%% file: main.tex
\documentclass{article}

\pdfoutput=1

\usepackage{arxiv}

\usepackage[utf8]{inputenc} 
\usepackage[T1]{fontenc}    
\usepackage{hyperref}       
\usepackage{url}            
\usepackage{booktabs}       
\usepackage{amsfonts}       
\usepackage{nicefrac}       
\usepackage{microtype}      
\usepackage{lipsum}		
\usepackage{graphicx}
\usepackage[numbers]{natbib}
\usepackage{doi}
\usepackage{xspace}
\usepackage{listings}

\usepackage{algorithm}
\usepackage{algpseudocode}
\usepackage{amsmath}
\usepackage{mathrsfs}  
\usepackage{enumitem}

\usepackage{wrapfig}
\usepackage{lipsum}
\usepackage{multirow}
\usepackage{graphicx}
\usepackage{float} 
\usepackage{threeparttable}
\usepackage{makecell}
\usepackage{colortbl}
\usepackage{listings}
\usepackage{xcolor}
\usepackage{subcaption}
\definecolor{backcolour}{rgb}{0.95,0.95,0.92}
\definecolor{codegray}{rgb}{0.5,0.5,0.5}
\definecolor{delim}{RGB}{20,105,176}
\definecolor{numb}{RGB}{106, 109, 32}
\definecolor{string}{rgb}{0.64,0.08,0.08}

\lstdefinestyle{code}{
	backgroundcolor=\color{backcolour},
        keywords={context, self, and, or, forAll, implies, xor, not},
        literate=
               {=:}{$\rightarrow{}$}{1},
	keywordstyle=\color{blue},
        morekeywords=[2]{inv, def},
	keywordstyle=[2]\bfseries,
	numberstyle=\tiny\color{codegray},
	basicstyle=\footnotesize,
	breakatwhitespace=false,
	breaklines=true,
	captionpos=b,
	keepspaces=true,
	numbers=none,
	showspaces=false,
	showstringspaces=false,
	showtabs=false,
	tabsize=1,
    xleftmargin=0em
}
\newcommand{\avut}{AVUT\xspace}
\newcommand{\deepcollision}{\textit{DeepCollision}\xspace}
\newcommand{\deepqtest}{\textit{DeepQTest}\xspace}
\newcommand{\randomStrategy}{\textit{RS}\xspace}
\newcommand{\greedyStrategy}{\textit{GS}\xspace}
\newcommand{\realisticConstraints}{\textit{realistic constraints}\xspace}

\newcommand{\metricReward}{\textit{Safety and Comfort Metrics}\xspace}

\newcommand{\metricDiversity}{\textit{Diversity Metrics}\xspace}

\newcommand{\metricTTC}{\textit{TTC}\xspace}
\newcommand{\metricDTO}{\textit{DTO}\xspace}
\newcommand{\metricJerk}{\textit{Jerk}\xspace}
\newcommand{\metricCollisonNum}{\#\textit{Collision}\xspace}
\newcommand{\metricCollisionTime}{\textit{CollisionTime}\xspace}

\newcommand{\metricRealism}{\textit{Realism Metrics}\xspace}
\newcommand{\metricRealismRCS}{\textit{RCS}\xspace}
\newcommand{\metricRealismRCTRCS}{$CT_{RCS}$\xspace}
\newcommand{\metricRealismUCS}{\textit{UCS}\xspace}
\newcommand{\metricRealismRNS}{\textit{RNS}\xspace}
\newcommand{\metricRealismUNS}{\textit{UNS}\xspace}
\newcommand{\metricRealismTC}{\#\textit{TS}\xspace}

\newcommand{\metricDiversityAPI}{$Div_{API}$\xspace}
\newcommand{\metricDiversityScenario}{$Div_{Scenario}$\xspace}

\newcommand{\rewardTTC}{$DeepQTest_{TTC}$\xspace}
\newcommand{\rewardDTO}{$DeepQTest_{DTO}$\xspace}
\newcommand{\rewardJerk}{$DeepQTest_{Jerk}$\xspace}

\newcommand{\rewardTTCRN}{$DeepQTest_{TTC_{RN}}$\xspace}
\newcommand{\rewardDTORN}{$DeepQTest_{DTO_{RN}}$\xspace}
\newcommand{\rewardJerkRN}{$DeepQTest_{Jerk_{RN}}$\xspace}

\newcommand{\rewardTTCSN}{$DeepQTest_{TTC_{SN}}$\xspace}
\newcommand{\rewardDTOSN}{$DeepQTest_{DTO_{SN}}$\xspace}
\newcommand{\rewardJerkSN}{$DeepQTest_{Jerk_{SN}}$\xspace}

\newcommand{\rewardTTCRD}{$DeepQTest_{TTC_{RD}}$\xspace}
\newcommand{\rewardDTORD}{$DeepQTest_{DTO_{RD}}$\xspace}
\newcommand{\rewardJerkRD}{$DeepQTest_{Jerk_{RD}}$\xspace}

\newcommand{\rewardTTCSD}{$DeepQTest_{TTC_{SD}}$\xspace}
\newcommand{\rewardDTOSD}{$DeepQTest_{DTO_{SD}}$\xspace}
\newcommand{\rewardJerkSD}{$DeepQTest_{Jerk_{SD}}$\xspace}

\newcommand{\unrealisticUTC}{{\textit{Unrealistic Time Changes (UTC)}}\xspace}
\newcommand{\unrealisticUWC}{{\textit{Unrealistic Weather Changes (UWC)}}\xspace}
\newcommand{\unrealisticVSD}{{\textit{Violation of Safety Distance (VSD)}}\xspace}
\newcommand{\unrealisticOA}{{\textit{Overlapping Areas (OA)}}\xspace}

\title{DeepQTest: Testing Autonomous Driving Systems with Reinforcement Learning and Real-world Weather Data}

\date{} 					

\author{
Chengjie~Lu
\\
Simula Research Laboratory and\\
University of Oslo\\
Oslo, Norway \\
\texttt{chengjielu@simula.no} \\
\And
 Tao~Yue
\\
Simula Research Laboratory\\
Oslo, Norway \\
\texttt{tao@simula.no} \\
\And
 Man~Zhang
\thanks{Corresponding author}
\\
Kristiania University College\\
Oslo, Norway \\
\texttt{man.zhang@kristiania.no} \\
\And
Shaukat~Ali
\\
Simula Research Laboratory and\\
Oslo Metropolitan University\\
Oslo, Norway \\
\texttt{shaukat@simula.no} \\
}

\renewcommand{\shorttitle}{DeepQTest}

\hypersetup{
pdftitle={DeepQTest: Testing Autonomous Driving Systems with Reinforcement Learning and Real-world Weather Data},
pdfsubject={Software Engineering},
pdfauthor={Chengjie~Lu, Tao~Yue, Man~Zhang, Shaukat~Ali},
pdfkeywords={Autonomous driving system testing, Test scenario generation, Reinforcement learning, Software testing}
}

\begin{document}
\maketitle
\begin{abstract}
Autonomous driving systems (ADSs) are capable of sensing the environment and making driving decisions autonomously. These systems are safety-critical, and testing them is one of the important approaches to ensure their safety. However, due to the inherent complexity of ADSs and the high dimensionality of their operating environment, the number of possible test scenarios for ADSs is infinite. Besides, the operating environment of ADSs is dynamic, continuously evolving, and full of uncertainties, which requires a testing approach adaptive to the environment. In addition, existing ADS testing techniques have limited effectiveness in ensuring the realism of test scenarios, especially the realism of weather conditions and their changes over time. 
Recently, reinforcement learning (RL) has demonstrated great potential in addressing challenging problems, especially those requiring constant adaptations to dynamic environments. To this end, we present \deepqtest, a novel ADS testing approach that uses RL to learn environment configurations with a high chance of revealing abnormal ADS behaviors. Specifically, \deepqtest employs Deep Q-Learning and adopts three safety and comfort measures to construct the reward functions. To ensure the realism of generated scenarios, \deepqtest defines a set of realistic constraints and introduces real-world weather conditions into the simulated environment. We employed three comparison baselines, i.e., random, greedy, and a state-of-the-art RL-based approach \deepcollision, for evaluating \deepqtest on an industrial-scale ADS. Evaluation results show that \deepqtest demonstrated significantly better effectiveness in terms of generating scenarios leading to collisions and ensuring the scenario realism compared with the baselines. In addition, among the three reward functions implemented in \deepqtest, \textit{Time-To-Collision} is recommended as the best design according to our study.
\end{abstract}


\section{Introduction}\label{sec:introduction}
Autonomous driving systems (ADSs), as cyber-physical systems, sense their operating environment and then autonomously make decisions based on sensed information from the environment. 
Different from advanced driver-assistance systems (ADASs), which are designed to assist/alert the driver in dealing with dangerous situations, ADSs are designed with the ultimate goal of operating vehicles without human intervention~\cite{shaout2011advanced}.
Testing ADSs has attracted a significant level of attention from both the industry and academia, in recent years. However, testing ADSs is still challenging because of the inherent complexity of ADSs themselves and their operating environment, which is dynamic, high-dimensional, and high-uncertain. In principle, there are an infinite number of driving (and therefore testing) scenarios, and it is critical to ensure that ADSs operate safely under any possible driving scenarios.

Typically, there are two ways of testing ADSs, Offline testing~\cite{codevilla2018offline} and online testing~\cite{haq2020comparing}. Offline testing, also called \textit{model-level} testing, does not test ADSs as a whole but focuses on testing the decision-making deep neural networks (DNNs) by considering DNNs as standalone components without involving the operating environment of ADSs. Several Offline testing approaches~\cite{zhang2018Deeproad,tian2018deeptest} have been proposed to test ADSs by generating adversarial test inputs/scenes (e.g., images) for DNNs. Such inputs are typically individual inputs that are expected to find prediction errors of DNNs (e.g., steering angle errors).
On the contrary, online testing (or \textit{system-level} testing) aims to identify system-level failures by generating critical test scenarios, in which an ADS under test is deployed in either a simulated or physical operating environment, and the ADS is tested as a whole when it interacts with the environment. 
Online testing approaches~\cite{ben2016testing,abdessalem2018ASE,abdessalem2018testing,lu2022learning} have demonstrated promising performance in identifying system-level failures, but it remains challenging to generate safety-critical test scenarios in a cost-effective way due to the huge time overhead caused by the use of simulated/physical environment. 

Existing online testing approaches usually apply techniques such as search algorithms and machine learning. For example, search-based testing (SBT) techniques~\cite{ben2016testing,abdessalem2018ASE} have been widely used to generate test scenarios by formulating a testing problem (e.g., searching critical configurations of driving scenarios) as an optimization problem, which can be addressed with meta-heuristic optimization algorithms, such as Non-dominated Sorting Genetic Algorithm II (NSGA-II)~\cite{deb2002fast}. However, given the complexity and uncertainty of the ADS operating environment, as well as the complexity of driving decision-making, the application of these approaches are frequently restricted by limited time, cost, and computational resource, or show limited effectiveness in testing a long-term decision-making task 
under a continuously-changing environment~\cite{sutton2018reinforcement}. 
Recently, reinforcement learning (RL) has shown great potential in addressing various challenging problems, where an agent learns optimal behavior strategies through constantly interacting with its environment without prior knowledge of the environment and the need for labeled training data~\cite{kiran2020deep}. Several RL-based testing approaches~\cite{lu2022learning,corso2019adaptive} have been proposed and exhibit good performance in dealing with continuous system and environment status changes. However, such an approach has limited effectiveness in ensuring the realism of test scenarios, especially the realism of weather conditions and their changes over time. Furthermore, there is a lack of empirical studies investigating the effectiveness of different reward functions.

Motivated by the above challenges and the successful applications of RL in solving complicated problems, in this paper, we propose \deepqtest, an online ADS testing approach that considers the test scenario generation problem as an environment configuration problem and formulates the sequential interaction between the environment configuration agent and the operating environment as an RL problem.
\deepqtest configures the operating environment and tests ADSs throughout driving tasks; in each discrete instant of driving in the environment, the RL agent decides an \textit{action} to take on how to configure the environment based on its observation of the environment (\textit{state}), and gets feedback (\textit{reward}) from the environment afterwards.
\deepqtest adopts \textit{Deep Q-Learning} (DQN)~\cite{mnih2015human}, a classical RL algorithm, which has demonstrated excellent performance in addressing various autonomous driving problems, such as lane following~\cite{ruiming2018end} and motion planning~\cite{chen2020conditional}. To encode the environment states and extract features from the high-dimensional operating environment, \deepqtest adopts multi-modal sensor fusion~\cite{feng2020deep} as its state encoding strategy. 
To ensure the realism of scenarios, we define a set of realistic constraints on configurable parameters (e.g., positions of objects and environmental parameters). We also develop a real-world weather generator that introduces real-world weather data in simulation.
Considering the importance of reward functions in RL~\cite{kiran2021deep}, we carefully design three reward functions: $TTC_{Reward}$, $DTO_{Reward}$ and $Jerk_{Reward}$, based on three metrics: time to collision (\textit{TTC}), distance to obstacles (\textit{DTO}), and the changing rate of acceleration (\textit{Jerk}). 
Corresponding to the three reward functions, we design three approaches: \rewardTTC, \rewardDTO, and \rewardJerk.

We trained 12 \deepqtest models using four real-world weather conditions
in San Francisco, and the three selected reward functions. The trained models are evaluated with an industrial-level ADS (i.e., Apollo~\cite{fan2018baidu}) and a commonly used autonomous driving simulator (i.e., SVL~\cite{rong2020lgsvl}). We perform an extensive empirical study with four driving roads and compare \deepqtest with three baseline strategies: random, greedy, and the state-of-the-art approach \deepcollision~\cite{lu2022learning}. Evaluation results show that 1) \deepqtest outperformed random and greedy strategies in terms of collision generation and time cost for enabling the occurrence of collisions; 2) \rewardTTC achieved the overall best performance among all the three reward functions; 3) \rewardTTC outperformed \deepcollision in terms of generating more realistic collision scenarios with less time, and it also achieved better performance than \deepcollision in introducing diverse environment configurations and generating diverse scenarios.
Based on the results on the four roads under the four weather conditions with 20 times repetition, \deepqtest is able to trigger a realistic collision scenario within 10.57 seconds on average.

The rest of the paper is organized as follows. Section~\ref{sec:backgroud} introduces RL, especially the DQN algorithm. Section~\ref{sec:relatedwork} surveys the existing research on testing ADSs. We present \deepqtest in Section~\ref{sec:deepqtest} and the evaluation design in Section~\ref{sec:evaluation}. In Section~\ref{sec:results}, we report results and analyses of the empirical study, followed by discussions in Section~\ref{sec:discussion}, and conclude the paper in Section~\ref{sec:conclusion}.

\section{Background}\label{sec:backgroud}
In the section, we provide the required background information about DQN and multi-modal sensor fusion, which are the key techniques applied in \deepqtest.

\subsection{Deep Q-Learning (DQN)}\label{subsec:dqn}
Reinforcement Learning (RL) is essentially about agents learning decision-making strategies while interacting with environments, which might not be known beforehand. More specifically, during a learning process, based on each observed \textit{state} of the environment, the agent decides an \textit{action} to take, and then the performance of the agent after taking the action is evaluated with a \textit{reward}. The ultimate goal is to maximize the cumulative reward of a long-term decision-making process.

When using Markov Decision Process (MDP) to formulate an RL problem, MDP can be represented as a 5-tuple <$\mathcal{S, A, R, P, \gamma}$>, representing a set of states $\mathcal{S}$, a set of actions $\mathcal{A}$, a reward function $\mathcal{R}(s_t, a_t)$, a probability distribution $\mathcal{P}(s_{t+1}|s_t, a_t)$, and a discount factor $\mathcal{\gamma} \in [0, 1]$. A stochastic policy $\pi: \mathcal{S} \times \mathcal{A} \rightarrow [0, 1]$ maps the state space to a probability distribution over the action space, and $\pi(a_t|s_t)$ represents the probability of choosing action $a_t$ at state $s_t$. 
The optimal policy $\pi^\ast$, which represents the goal for an RL agent, is expected to achieve the highest expected cumulative reward: $R_t = \sum_{t>0}\gamma^{t-1}r_t$, where future rewards are discounted by discount factor $\gamma$. Discount factor $\gamma$ controls how an agent regards future rewards; a lower value of $\gamma$ encourages the agent to focus more on short-term rewards, whereas a higher $\gamma$ value encourages the agent to be more concentrated on a long-term perspective and maximize long-term rewards. The learning process of an RL agent is that it first observes a state $s_t \in \mathcal{S}$, then based on the observation (i.e., $s_t$), the agent selects and executes an action $a_t \in \mathcal{A}$. After the execution of $a_t$, an immediate reward $r_t \sim \mathcal{R}(s_t, a_t)$ is fed back to the agent, along with a newly observed state $s_{t+1} \sim \mathcal{P}(s_{t+1}|s_t, a_t)$.

Many types of RL solutions have been proposed and demonstrate good performance in solving practical problems. One classical method is value-based RL, which uses an action-value function to estimate the performance of an action taken at a given state. Q-learning~\cite{watkins1992q} is a typical value-based method, which employs Q-value as the measure of the expected reward in a state-action pair: $(s_t, a_t)$. And the action-value function in Q-Learning is called the Q-value function $Q^\pi(s, a): (\mathcal{S, A}) \rightarrow \mathbb{R}$, which is the discounted expected return of rewards given the state, action, and policy represented by the equation:

\begin{equation}
    \label{equ:policy}
    Q^\pi(s_t, a_t) = \mathbb{E}_{\pi}[R_t|s_t, a_t].
\end{equation}

The Q-value of a given state-action pair $(s_t, a_t)$, i.e., $\textit{Q}^\pi(s_t, a_t)$, is an estimation of the expected future reward obtained from pair $(s_t, a_t)$ with policy $\pi$. The optimal action-value function $\textit{Q}^\ast(s_t, a_t)$ estimates the Q-value of possible actions for a given state and selects the action with the highest Q-value, based on the Bellman equation:

\begin{equation}
\label{equ:bellman_equation}
Q^\ast(s_t, a_t) = \mathbb{E}_{\pi^{\ast}}[r_t + \gamma \max_{a_{t+1}}Q^\ast(s_{t+1}, a_{t+1})].
\end{equation}

In Q-Learning, Q-values are stored in a Q-table, which represents the Q-value function for all state-action pairs and updates based on the Bellman function. Such a table is arbitrarily initialized and updated with data representing the agent's experience at each step using the equation:

\begin{equation}
    \label{equ:update_function}
    Q(s_t, a_t) \leftarrow Q(s_t, a_t) + \alpha(r_t + \gamma\max_{a_{t+1}}Q(s_{t+1}, a_{t+1}) - Q(s_t, a_t)),
\end{equation}
where, $0 < \alpha \leq 1$ is the learning rate. A larger learning rate leads to a stronger influence of new data on an update.

One drawback of a Q-table is that when the action space is large and the number of possible states of the environment is infinite, the memory required to store the Q-table, as well as the computation resource required to search from the Q-table, is huge. Therefore, it is very challenging to maintain a separate Q-table for each $\mathcal{S \times A}$~\cite{chen2019model}. Instead, a function is often used to approximate Q-values. To overcome the above challenges, Deep Q-Learning (DQN)~\cite{mnih2015human} was proposed, which approximates and stores Q-values with a neural network (i.e., Q-Network) parameterized by weights and biases, together denoted as $\theta$: $Q(s, a;\theta) \approx Q^\ast(s, a)$.
DQN employs several mechanisms to ensure the stability of the training process: it first uses a replay memory to store a certain number of state transitions $<s_t, a_t, r_t, s_{t+1}>$ and randomly samples data for training the Q-Network. This step aims to reduce correlations among samples and thereby increase sample efficiency through the reuse of data. Second, to ensure the stability of the training process, a separate target network $\hat{Q}(s, a;\theta^-)$ parameterized with $\theta^-$ is used for generating target Q-values in Q-Learning updates. $\hat{Q}$ is identical to the main network $Q$, except that its parameters $\theta^-$ are cloned from $\theta$ every $\textit{C}$ update. In this case, the Q-value function is updated as:

\begin{equation}
    \label{equ:q_update}
    \theta \leftarrow \theta + \alpha(y_{target} - Q(s_t, a_t;\theta))\nabla_\theta Q(s_t, a_t;\theta),
\end{equation}
where $y_{target}$ is the target value computed as:

\begin{equation}
    \label{equ:compute_target}
    y_{target} = r_t + \gamma\max_{a_{t+1}}\hat{Q}(s_{t+1}, a_{t+1};\theta^-)
\end{equation}
Finally, at each training iteration, a batch of memory is sampled uniformly from the replay memory for the Q-Network training. The training is based on target Q-values from $\hat{Q}$. The loss function of Q-Network in DQN is defined as:

\begin{equation}
    \label{equ:loss_function}
    \mathcal{L}(\theta) = \mathbb{E}[(y_{target} - Q(s_t, a_t;\theta))^2].
\end{equation}

\subsection{Multi-modal Sensor Fusion}\label{subsec:sensorfusion}
Perception is an essential task of ADSs. To gain robust and reliable environment comprehension, autonomous vehicles are usually equipped with multiple sensors, such as cameras (visual cameras, thermal cameras), LiDARs, and radars~\cite{feng2020deep}. Multi-modal sensor fusion is then a crucial method for perception tasks in autonomous driving. 
In multi-modal sensor fusion, DNNs are often employed to serve as the environmental perception module by taking a multi-modal representation of a driving environment (high dimensional) as the input and generating a low dimensional feature representation. The input consists of sensor data and vehicle state measurements. As for sensor data, LiDAR point clouds and camera images are often considered. LiDAR point clouds provide both depth and reflectance information of the driving environment and are encoded into different representations such as a bird's eye view~\cite{chen2017multi} and a front view~\cite{li2016vehicle}. Camera RGB images and depth images are also commonly used sensor data to provide rich texture information (e.g., color) of the driving environment. Vehicle state measurements are vehicle dynamics parameters such as speed and acceleration.

By combining LiDAR cloud points, camera images, vehicle measurements, and other sensor data, multi-modal representation can be encoded and inputted into a perception module for feature extraction. To construct a perception module, various neural networks will be adopted. For example, convolutional neural networks (CNNs), which are neural networks with multiple layers, are often used in image or object recognition and classification systems. A typical CNN usually has three types of layers, i.e., convolutional, pooling, and fully connected layers. Convolutional layers are the building block of a CNN, which carry the main computation responsibility, such as identifying edges of images for feature extraction; pooling layers then mitigate the dimensionality of extracted features; fully connected layers, also named output layers, recognize features of an image by using back-propagation algorithms. Due to its exclusive features, such as local connectivity and shared weights, CNN can achieve high accuracy and good performance in various tasks, such as image classification and object detection~\cite{dargan2020survey}. 

Another commonly applied network is the Long Short-Term Memory (LSTM)~\cite{hochreiter1997long}, which is capable of learning long-term dependencies and time series prediction. The key of LSTM is the cell state, which serves as a memory unit where the information can remain unchanged for a sufficient time. This helps the unit to memorize the last calculated value. There are three types of gates in an LSTM memory unit: forget gate, input gate, and output gate, where the forget gate decides which information needs attention and which to ignore, and the input gate updates the memory by controlling the flow of new information into the memory, and the output gate determines the value of the next hidden state, which contains information on previous inputs.

In multi-modal sensor fusion, DNNs represent features hierarchically and offer different schemes to multi-modal sensor fusion, including early, middle, and late fusion schemes. In the early fusion scheme, the multiple representations from raw data are joined together, then the DNN learns the joint features of multiple modalities at an early stage. Early fusion can fully exploit the information of the raw data because data from different sources are directly joined without losing the original information. In the middle fusion scheme, feature representations from different sensing modalities are combined at intermediate layers of a DNN, which enables it to learn cross-modalities with different feature representations. The late fusion scheme uses separate networks for feature extraction of each single sensing modality (e.g., LiDAR point clouds, camera images), and the outputs are further concatenated and processed by a network to get the final feature representation.

Evidence shows that multi-modal fusion outperforms single-sensor data perception in autonomous driving~\cite{feng2020deep}. Among different fusion schemes, middle fusion has demonstrated promising performance in terms of leading to more accurate object detection~\cite{schneider2017multimodal,ha2017mfnet} and improving the accuracy of geometry and semantics of the resulting representation~\cite{piewak2018improved}. Thus, we opt for the middle fusion scheme for multi-modal sensor fusion in our approach.

\section{Related Work}\label{sec:relatedwork}

As discussed in Section \ref{sec:introduction}, existing ADS testing techniques typically can be classified into \textit{offine} ADS testing (Section \ref{subsec:offlinetesting}) and online ADS testing (Section \ref{subsec:onlinetesting}). 

\subsection{Offline ADS Testing}\label{subsec:offlinetesting}
Offline testing primarily focuses on only testing decision-making DNNs embedded in ADSs without involving an operating environment. There are a large number of offline testing approaches that have been proposed by using machine learning (ML)-based testing~\cite{zhang2018Deeproad,tian2018deeptest,zhou2020deepbillboard}, fuzz testing~\cite{Deephunter,guo2018dlfuzz} and mutation testing~\cite{ma2018deepmutation,hu2019deepmutation++} techniques to generate adversarial inputs for DNNs in ADSs. To measure test adequacy, various metrics have been proposed, such as neuron coverage~\cite{pei2017deepxplore,tian2018deeptest} and surprise coverage~\cite{kim2019guiding}.

Several approaches leverage real-world changes in driving conditions (e.g., heavy rain, heavy snow, or adverse lighting), and synthesize adversarial driving conditions using ML techniques. Zhang et al.~\cite{zhang2018Deeproad} proposed DeepRoad, an Generative Adversarial Network (GAN)-based unsupervised learning framework, for identifying inconsistent behaviors of DNN-based ADSs under various weather conditions. DeepRoad synthesizes driving scenes with various weather conditions and utilizes metamorphic testing techniques for checking the consistency of such systems under synthetic driving scenes.
Tian et al.~\cite{tian2018deeptest} proposed DeepTest, a systematic testing tool focusing on generating realistic synthetic images by applying image transformations and mimicking different real-world phenomena (e.g., weather conditions, object movements). 
To explore possibilities for testing ADSs in the physical world, Guo et al.~\cite{zhou2020deepbillboard} proposed a systematic physical-world adversarial inputs generation approach, named DeepBillboard, which targets a specific driving scenario: drive-by billboards. DeepBillboard generates adversarial perturbations that can be added to roadside billboards in either a simulated/physical world and mislead CNN-based steering models.

Several approaches focusing on deep learning (DL) testing also demonstrate the potential for testing DNNs in ADSs. Guo et al.~\cite{guo2018dlfuzz} proposed DLFuzz, a differential fuzzing testing framework, to guide DL systems exposing incorrect behaviors. Xie et al.~\cite{Deephunter} proposed DeepHunter, a coverage-guided fuzz testing framework, for testing DNNs. DeepHunter detects potential defects of general-purpose DNNs by generating new semantically-preserved test inputs. Ma et al.~\cite{ma2018deepmutation} proposed a mutation testing framework, named DeepMutation, which directly injects faults into DL models through mutation operators. Later on, Hu et al.~\cite{hu2019deepmutation++} extended DeepMutation and hence proposed DeepMutation++ to cover more types of DNN models, such as feed-forward neural networks (FNNs) and stateful recurrent neural networks (RNNs).

To measure the test adequacy for DNNs, various metrics~\cite{pei2017deepxplore,hu2019deepmutation++,peng2020first,kim2019guiding} for evaluating the test quality have been proposed. For example, Pei et al.~\cite{pei2017deepxplore} proposed DeepXplore, which introduces the neuron coverage for measuring the number of rules in a DNN that have been exercised by a set of inputs, to maximize both the neuron coverage and the number of potentially erroneous behaviors without requiring manual labels. Kim et al.~\cite{kim2019guiding} proposed SADL, which introduces surprise adequacy and surprise coverage to measure the test adequacy for DNNs. Such metrics have been applied for guiding the testing of DNNs of ADSs~\cite{pei2017deepxplore,tian2018deeptest} to achieve higher coverages of activated neurons. However, due to an ADS being more complicated than a single DNN model, such DNN testing methods exhibit limitations when being applied to test an ADS with multiple DNN models. Moreover, it has been argued whether metrics such as neuron coverage are meaningful for testing a real-world DNN in ADSs~\cite{harel2020neuron}.

Offline testing techniques tend to be non-adaptive to contexts where a prediction model needs to adapt to continuous system status changes, which is exactly the case for ADSs' operating environment. This is mainly because such testing techniques test DNNs independent of their ADSs, and based on individual inputs without considering that a 
a single DNN prediction error may have an influence on future predictions and driving decisions~\cite{stocco2022mind}, leading to system failures. Therefore, although offline testing can detect single prediction errors in DNNs, it shows limitations in identifying system-level failures.

\subsection{Online ADS Testing}\label{subsec:onlinetesting}
Online testing approaches have been proposed to test ADSs in a simulated/physical operating environment. Various techniques have been applied in these approaches, including search algorithms and RL. 

Search algorithms have been typically applied for generating test scenarios for ADSs and ADASs~\cite{abdessalem2018testing,abdessalem2018ASE,buhler2008evolutionary,ben2016testing,calo2020generating}.
These approaches mostly focus on the automated test generation with optimization objectives such as minimizing distances to obstacles or unsafe area~\cite{abdessalem2018testing}, maximizing the speed of a vehicle at the time of collision~\cite{abdessalem2018testing}, violating safety requirements~\cite{abdessalem2018ASE}, and minimizing time to collision~\cite{buhler2008evolutionary,ben2016testing}.
Some works apply multi-objective search algorithms to select and/or generate safety-critical scenarios. Abdessalem et al.~\cite{ben2016testing} proposed NSGAII-SM, which combines multi-objective search algorithm NSGA-II~\cite{deb2002fast} with surrogate models~\cite{jin2011surrogate} to test an ADAS with three objectives.
Abdessalem et al.~\cite{abdessalem2018testing} also proposed another approach named NSGAII-DT, which considers two objectives (i.e, speed at the time of collision, distance to obstacles) to generate critical scenarios for vision-based control systems by combining NSGA-II with decision tree classification models.
To detect failures caused by feature interactions, FITEST~\cite{abdessalem2018ASE} integrates a set of hybrid test objectives designed based on distance functions measuring how far of violating system safety requirements (e.g., no collision with pedestrians, stopping at a stop sign). Such interactions involve features like automated emergency braking (AEB), adaptive cruise control (ACC), and traffic sign recognition (TSR). For instance, a feature interaction may arise when a braking command issued by ABE is overridden by ACC's command of maintaining the same speed as that of the front vehicle.
Cal\`o et al.~\cite{calo2020generating} adapted multi-objective search algorithms (i.e., NSGA-II) to search for collisions and ADS configurations that can avoid such collisions.

While SBT approaches show promising results in supporting online testing of ADSs, they introduce practical challenges in terms of time and computational overhead to evaluate test scenarios~\cite{konak2006multi}. 
Moreover, another challenge of SBT approaches is that their ability to deal with dynamic and continuously-changing operating environments is limited, which is mainly because they do not take advantage of the interaction between individuals (often encoding test scenarios) with the environment when searching critical scenarios.
However, this interaction information has been proven to be efficient for decision-making in RL, especially in an open and dynamic environment under uncertainty~\cite{dm2022,sallab2017deep}.

Several RL-based ADS online testing approaches have been proposed~\cite{chen2021adversarial,haq2022many,lu2022learning}. Chen et al.~\cite{chen2021adversarial} developed an RL-based adaptive testing framework to generate time-sequential adversarial environments specific to lane-changing driving models. Haq et al.~\cite{haq2022many} proposed MORLOT, an online testing approach, by combining RL and multi-objective search. MORLOT uses RL to adaptively generate sequences of environmental changes that can cause requirement violations and adopts multi-objective search to determine the changes that can cover as many requirements as possible. Lu et al.~\cite{lu2022learning} proposed DeepCollision, an RL-based approach, which generates safety-critical testing scenarios by configuring the operating environment of an ADS. 
Despite the promising results of the current RL-based approach, there is no work that uses real-world weather data when generating test scenarios. Besides, the literature mainly uses a few parameters to encode RL states, which may have limitations when representing a high dimensional state space of an autonomous driving environment. In addition, designing a proper reward function is vital to an RL-based testing approach, which requires further empirical studies.

To compare with the literature, \deepqtest, we propose in this paper, is different from existing RL-based approaches because 1) \deepqtest does not try to control the changes of weather conditions, while it introduces real-world weather data to the simulation, which can improve the realism of weather conditions and their changes over time in simulation and therefore improves the realism of test scenarios; 2) \deepqtest encodes RL \textit{state} with multi-modal sensor data, which can better represent the environment and system states; and 3) \deepqtest integrates three different safety/comfort-related reward functions and evaluates their effectiveness from various perspectives.

\section{DeepQTest Methodology}\label{sec:deepqtest}
In Section~\ref{subsec:overview}, we present the overview of \deepqtest, followed by a domain model describing all the configurable environment parameters and their relationships. We then describe how we introduce real-world weather data to the simulator (Section~\ref{subsec:weathertime}) and mathematically formalize the environment configuration problem as an MDP (Section~\ref{subsec:formulationMDP}). In Section~\ref{preprocessing}, we present the preprocessing of sensor data for encoding multimodal states and the Q-network architecture in Section~\ref{networkandTraining}.

\subsection{Overview}\label{subsec:overview}

\begin{figure}[ht]
    \centering
    \includegraphics[width=\textwidth]{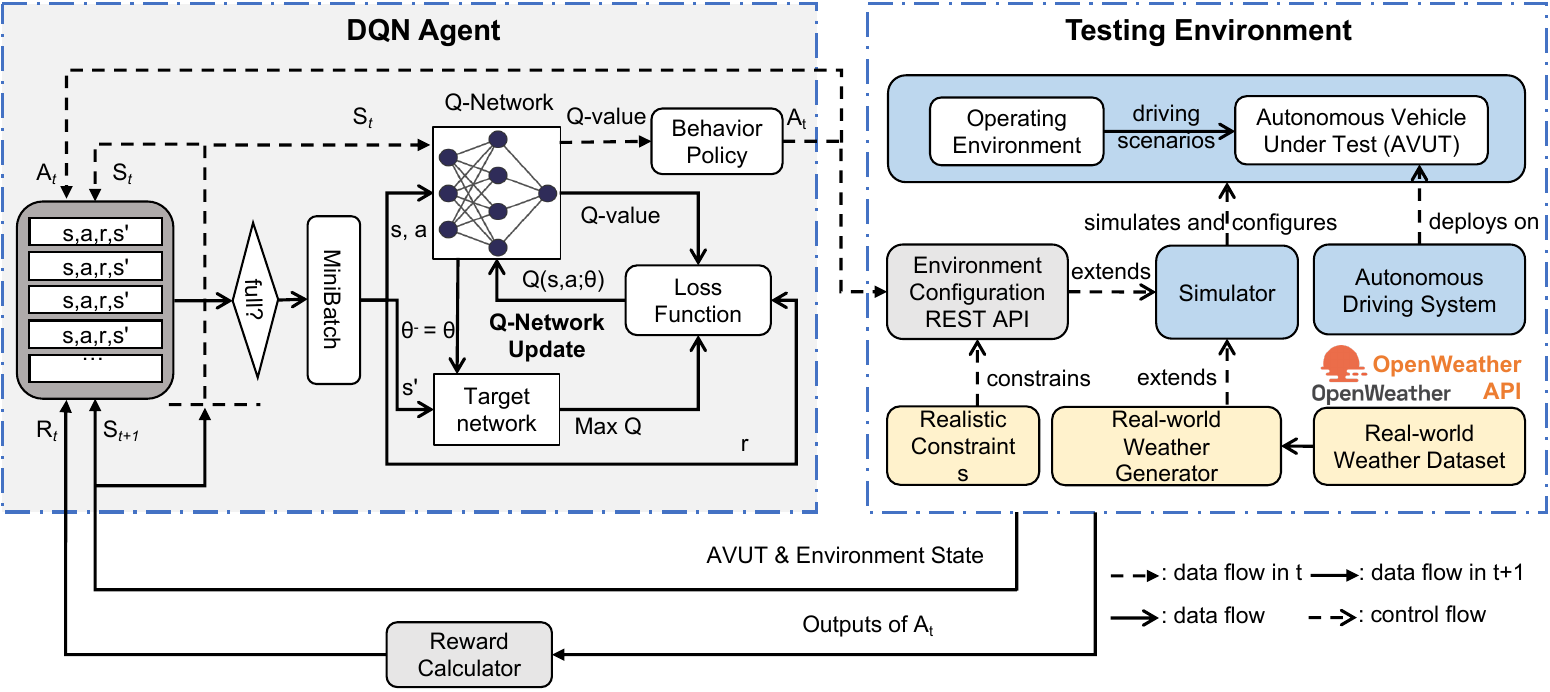}
    \caption{Overview of DeepQTest}
    \label{fig:overviewfig}
\end{figure}

\deepqtest learns environment configurations with RL, with the ultimate goal of identifying test scenarios for effectively testing autonomous vehicles, i.e., Autonomous Vehicles Under Test (\avut). As shown in Figure~\ref{fig:overviewfig}, \deepqtest employs a \textit{Simulator} (e.g., LGSVL \cite{rong2020lgsvl}) to simulate and configure a \textit{Testing Environment}, in which the (virtual) \avut drives in its simulated \textit{Operating Environment}. \deepqtest also integrates with an \textit{Autonomous Driving System} (e.g., the Baidu Apollo \cite{fan2018baidu}), which is deployed on the \avut to enable its autonomous driving.

\deepqtest utilizes DQN to generate actions to configure the operating environment of the \avut, e.g., adding a pedestrian crossing the road.
As Figure~\ref{fig:overviewfig} shows, during the configuration process, the DQN agent observes a state $S_t$ describing the current states of the \avut and its operating environment. With the state, \deepqtest decides an action $A_t$ based on the Q-Network and the behavior policy.
With our developed \textit{Environment Configuration REST API}, such an action $A_t$ is realized as an HTTP request for accessing the simulator to introduce environment configurations. The realism of the newly configured environment is ensured by \textit{Realistic Constraints} (defined on environment configuration parameters) and \textit{Real-world Weather Generator}, which simulates real-world weather conditions. 

After the \avut drives in the newly configured environment for a fixed time period, i.e., at $t+1$, both the \avut and its operating environment enter a new state observed as $S_{t+1}$, and action $A_t$ produces a set of outputs.
Based on these outputs, \textit{Reward Calculator} calculates reward $R_{t}$ for $A_t$ and $S_{t}$ at $t+1$. Then the DQN agent stores them (as $<S_t$, $A_t$, $R_t$, $S_{t+1}>$) into the replay memory buffer. 
Once the replay memory is full, the Q-Network is updated as $Q(s, a; \theta)$, using the loss function by a mini-batch randomly sampled from the updated replay memory. Meanwhile, the target network parameters $\theta ^-$ are updated with Q-Network parameters $\theta$ after a fixed number of steps and remain unchanged between two updates.
In addition, with $S_{t+1}$, the (updated) Q-Network with behavior policy decides the next action: $A_{t+1}$.

In \deepqtest, an episode is finished once the \avut arrives at its destination, the \avut cannot move for a defined duration, or a collision happens.
At each configuration step, information about the \avut (e.g., its driving and collision status) and its operating environment (e.g., its status and driving scenarios) are stored as \textit{Environment Configuration Logs} for further analyses and replaying.

\subsection{Configurable Environment Parameters}\label{subsec:configurableparameters} 
\begin{figure}[ht]
    \centering
    \includegraphics[width=0.95\textwidth]{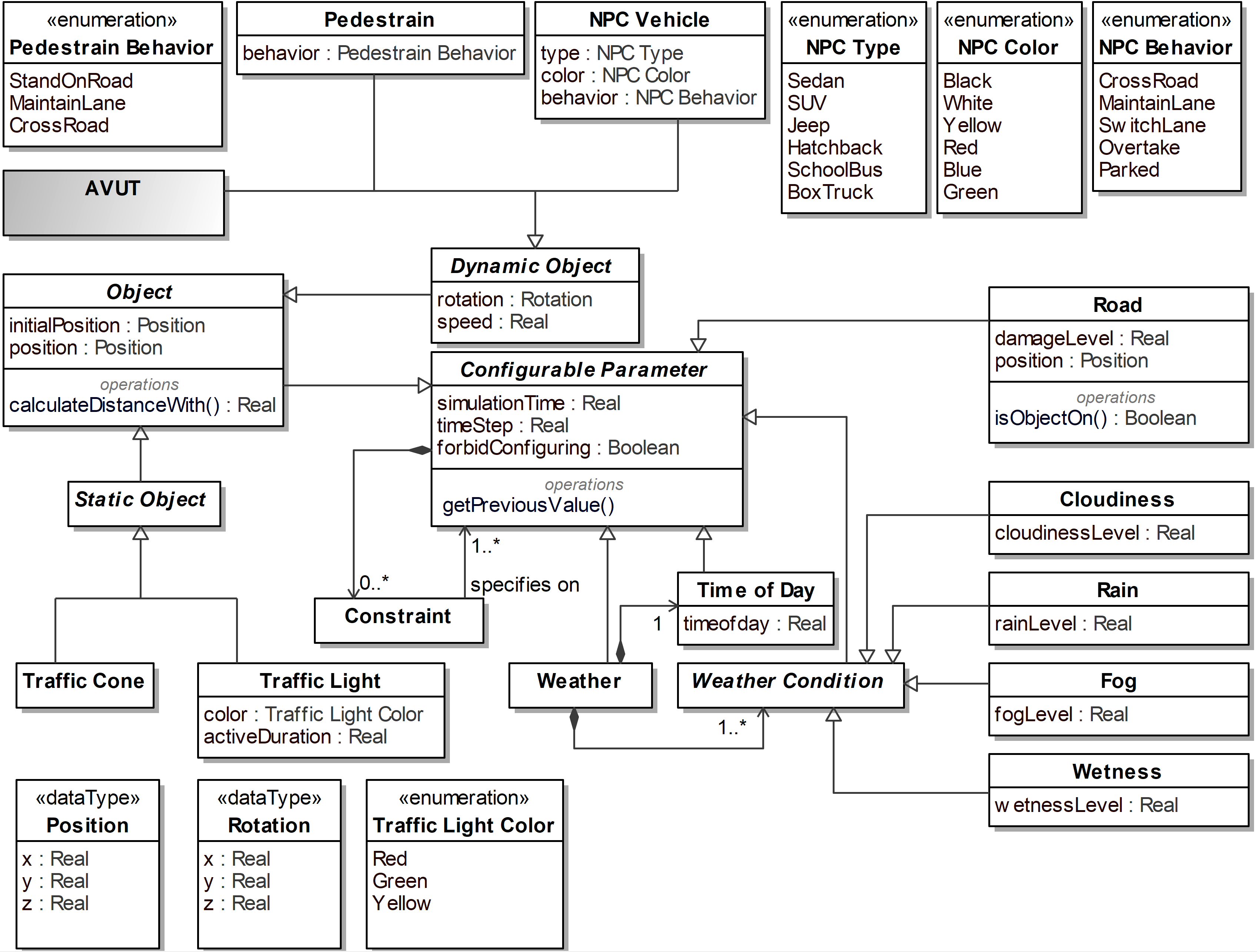}
    \caption{Domain model of the configurable parameters of the ADS operating environment}
    \label{fig:configurable_parameters}
\end{figure}

As part of this work, we developed a collection of environment configuration actions as REST API endpoints for manipulating the ADS operating environment by configuring parameters in simulation. These actions are utilized by \deepqtest for enabling testing under various operating environment configurations. Figure \ref{fig:configurable_parameters} presents a domain model (in the UML class diagram notations) capturing the\textit{Configurable Parameters} of the operating environment, based on which, \textit{Constraint}s can be specified to ensure (to a certain extent) the realism of configured operating environment.

\noindent\textbf{\textit{Configurable Parameters}}. In reality, the operating environment of ADSs constantly changes. However, when testing ADSs in a simulated environment, in practice, the number of parameters that can be manipulated is limited and dependent on the capability of the simulator being used. These configurable parameters and their valid configuration values form the \textit{action space}. Based on the types of objects being configured and properties of effect to create, we classify these parameters into three categories: parameters for configuring \textit{Object}s (\textit{Static Object}s, \textit{Dynamic Object}s), 
\textit{Road} and \textit{Weather and Time}.

(1) \textit {Configurable Parameters for Objects}. In the simulated environment, \textit{Object} can be manipulated by configuring its initial position (i.e., \textit{initialPosition} typed with datatype \textit{Position}), retrieving its current position (i.e., \textit{position}), and get a distance with other \textit{Object} with \textit{caclulateDistanceWith}.
\textit{initialPosition} can be configured only when the \textit{Object} is introduced into the simulated environment. Such introductions of \textit{Object}s can be performed during testing setup or during testing by an automated testing approach (such as \deepqtest). We further classified \textit{Object} into \textit{Static Object} and \textit{Dynamic Object}. 

For \textit{Static Object}s, configuration values of 
the current position (i.e., \textit{position}) remain unchanged within the specified \textit{Simulation Time}. We involve \textit{Traffic Cone} and \textit{Traffic Light} as \textit{Static Object}s in our ADS testing as shown in Figure~\ref{fig:configurable_parameters}. To test ADS, we allow introducing \textit{Traffic Cone} during testing (e.g., a road accident or a traffic accident could happen when driving), while \textit{Traffic Light} can only be introduced by setup. During testing, configurable parameters of \textit{Traffic Light} include its color (i.e., \textit{color} typed with enumeration \textit{Traffic Light Color}) and duration when the corresponding color is active (\textit{activeDuration} typed with \textit{Real}).


In a simulated environment, \textit{Dynamic Objects} include \avut, pedestrians and NPC (Non-Player Character) vehicles, and values of their properties (e.g., \textit{position}) change over time as shown by classes \textit{Dynamic Object} (with properties \textit{position} inherited from \textit{Object}, \textit{rotation} and \textit{speed}). 
\avut can be configured only in a test setup for a certain objective of testing ADSs, then denoted as grey background.
Pedestrians and NPC vehicles can be generated during testing. In addition to the three properties inherited from \textit{Dynamic Object}, a pedestrian has one additional property characterizing itself, i.e., its \textit{behavior} (typed with enumeration \textit{Pedestrian Behavior}). An NPC vehicle has three additional properties: its \textit{behavior} (typed with enumeration \textit{NPC Behavior}), \textit{color} (typed with \textit{NPC Color}), and \textit{type} (typed with enumeration \textit{NPC Type}).

Since pedestrians and NPC vehicles can have a lot of behaviors in real driving environments, the literature mainly focuses on safety-critical behaviors, such as those related to inevitable collisions or traffic rule violations~\cite{calo2020generating,li2020av}. Such behaviors mainly include overtaking, lane maintenance, lane switching, road crossing, etc. We, consequently, classify behaviors of pedestrians and NPC vehicles and specify them as two enumerations shown in Figure~\ref{fig:configurable_parameters}: \textit{Pedestrian Behavior} and \textit{NPC Behavior}.

\textbf{\textit{Realism.}} To ensure the realism of generations of \textit{Object}s, we defined constraints specified with Object Constraint Language~\cite{ocl} based on the configurable parameters.
When introducing a new \textit{Object} into the simulated environment, 
the \textit{Object} should keep a safe distance from the \avut and other objects, restricting where the \textit{Object} can be generated with \textit{initalPosition}.
Any violation of the safe distance constraint might result in unrealistic scenarios, e.g., colliding with a traffic cone that is just introduced in front of the \avut, heavy traffic congestion due to generated \textit{NPC Vehicle}s which are close to each other.
According to Ro et al.~\cite{ro2020new}, a vehicle should keep a safety distance of at least 5 meters from its surrounding objects,
however, the volume of \textit{Object} and potential uncertainty in the simulator need to be considered.
Then, based on a pilot study, we define the safety distance as 8 meters for \textit{Traffic Cone}, \textit{Pedestrian}, and \textit{NPC Vehicle} with its type of \textit{Sedan}, \textit{SUV}, \textit{Jeep}, and \textit{Hatchback}, and 10 metrics for \textit{NPC Vehicle} with its type of  \textit{SchoolBus} and \textit{BoxTruck}. The OCL constraint is defined as:
\begin{lstlisting}
context Object
inv: Object.allInstances =: excluding(self) =: forAll{o| ((self.oclIsKindOf(TrafficCone) or self.oclIsKindOf(Pedestrian) or (self.oclIsKindOf(NPCVehicle) and Set{NPCType::Sedan, NPCType::SUV, NPCType::Jeep, NPCType::Hatchback} =: includes(self.asTypeOf(NPCVehicle).type)) implies self.calculateDistanceWith(o) >= 8) or ((self.oclIsKindOf(NPCVehicle) and Set{NPCType::SchoolBus, NPCType::BoxTruck} =: includes(self.asTypeOf(NPCVehicle).type)) implies self.calculateDistanceWith(o) >= 10)}
\end{lstlisting}

For \textit{Traffic Light}, its color changes should also remain realistic. We restrict that the color of traffic lights can be changed only by following the order, i.e., red, green, yellow, then red again. 
Regarding the active duration for each color, we follow the traffic signal timing manual of the United States Department of Transportation~\cite{koonce2008traffic}, 
i.e., red 24 seconds, green 30 seconds, and yellow 6 seconds. Then we define the constraint as:

\begin{lstlisting}
context TrafficLight
inv:  (self.color =TrafficLightColor::Red implies (self.getPreviousValue().color =TrafficLightColor::Yellow and self.activeDuration=24)) or (self.color =TrafficLightColor::Green implies (self.getPreviousValue().color =TrafficLightColor::Red and self.activeDuration=30)) or (self.color =TrafficLightColor::Yellow implies (self.getPreviousValue().color =TrafficLightColor::Green and self.activeDuration=6))
\end{lstlisting}



(2) \textit{Configurable Parameters for Road}. 
\textit{Road} can be configured in terms of damage conditions, such conditions would impact ADS that need to be considered in 
autonomous driving testing~\cite{maeda2018road,guo2019safe}. Road damage can be classified into four levels according to Ferlisi et al.~\cite{ferlisi2021quantitative}: negligible (\textit{damageLevel} $\in$ [0\%, 25\%]), low (\textit{damageLevel} $\in$ (25\%, 50\%]), moderate (\textit{damageLevel} $\in$ (50\%, 75\%]), and severe (\textit{damageLevel} $\in$ (75\%, 100\%]).

\textbf{\textit{Realism.}} Regarding \textit{Road}, its \textit{damageLevel} cannot be configured (i.e., \textit{forbidConfiging} is \textit{true}) when \avut is driving on it. Then, we define a constraint on \textit{Road} as:

\begin{lstlisting}
context Road
inv: (not self.isObjectOn(AVUT)) xor self.forbidConfiguring
\end{lstlisting}

(3) \textit{Configurable Parameters for Weather and Time}. 
As shown in Figure~\ref{fig:configurable_parameters}, \textit{Weather} can be configured with various \textit{Weather Condition} and \textit{Time of Day}.
Configuring the time of day has a great impact on illumination conditions such as shadows, direct sunlight, or over/underexposed, which can lead to poor performance of vision-based modules of an ADS~\cite{zhou2019automated}, thereby important for testing. Similarly, weather conditions are one of the important factors causing accidents for both normal driving and autonomous driving~\cite{zhang2018Deeproad,bijelic2018benchmarking}. To test ADSs, we consider four \textit{Weather Condition}s, i.e., \textit{Cloudiness}, \textit{Rain}, \textit{Fog} and \textit{Wetness}, and define different levels for each condition based on real weather dataset as:
\begin{itemize}

    \item  Configuring \textit{Cloudiness} impacts how much the sky in the simulated environment is covered by clouds. To test ADSs with various degree of \textit{Cloudiness}, we defined five levels, i.e., clear sky (\textit{cloudinessLevel} $\in$ [0\%, 11\%]), few clouds (\textit{cloudinessLevel} $\in$ (11\%, 25\%]), scattered clouds (\textit{cloudinessLevel} $\in$ (25\%, 51\%]), broken clouds (\textit{cloudinessLevel} $\in$ (51\%, 84\%]), overcast clouds (\textit{cloudinessLevel} $\in$ (84\%, 100\%]).

    \item Configuring \textit{Rain} impacts how heavy the rain should fall in the simulated environment. Five levels are defined for \textit{Rain}, i.e., light rain (\textit{rainLevel} $\in$ [0\%, 20\%]), moderate rain (\textit{rainLevel} $\in$ (20\%, 40\%]), heavy intensity rain(\textit{rainLevel} $\in$ (40\%, 60\%]), very heavy rain(\textit{rainLevel} $\in$ (60\%, 80\%]), and extreme rain(\textit{rainLevel} $\in$ (80\%, 100\%]).

    \item Configuring \textit{Fog} impacts how tick the fog is in the simulated environment. Five levels are defined for \textit{Fog}, i.e., light fog (\textit{fogLevel} $\in$ [0\%, 20\%]), moderate fog (\textit{fogLevel} $\in$ (20\%, 40\%]), heavy fog (\textit{fogLevel} $\in$ (40\%, 60\%]), very heavy fog (\textit{fogLevel} $\in$ (60\%, 80\%]), and extreme heavy fog (\textit{fogLevel} $\in$ (80\%, 100\%]).

    \item Configuring \textit{Wetness} impacts how much the road in the simulated environment is covered by water in our context. Five levels are defined for \textit{Wetness}, i.e., light wetness (\textit{wetnessLevel} $\in$ [0\%, 20\%]), moderate wetness (\textit{wetnessLevel} $\in$ (20\%, 40\%]), heavy wetness (\textit{wetnessLevel} $\in$ (40\%, 60\%]), very heavy wetness (\textit{wetnessLevel} $\in$ (60\%, 80\%]), and extreme heavy wetness (\textit{wetnessLevel} $\in$ (80\%, 100\%]).
\end{itemize}

\textbf{\textit{Realism.}} To ensure the realism of weather and time, in Section~\ref{subsec:weathertime}, we design a mechanism that enables to introduce weather conditions based on real-world historical weather dataset
with the corresponding time of the day in the simulated environment.

The literature has shown that existing RL techniques cannot handle a large number of discrete actions, which may easily lead to scalability issues~\cite{dulac2019challenges,dulac2015deep}. Moreover, since the agent needs to evaluate all possible actions at each step, the larger the \textit{action space}, the larger the training cost (e.g., time) would be. For this reason, we specify valid values of the configurable parameters as enumeration literals, except for the time of day and the speed of any dynamic object, which takes a real value. The overall aim is to reduce the size of the \textit{action space}.

Eventually, invocations of configurable environment parameters are realized via \deepqtest APIs we implemented. For this purpose, we implemented in total 142 REST API endpoints~\cite{rodriguez2016rest}, which forms the entire \textit{action space} in our current implementation of \deepqtest.

\subsection{Real-world Weather Generation}\label{subsec:weathertime}
As shown in Figure~\ref{fig:configurable_parameters}, there are various ways of changing weather conditions, such as the time of day, whether it is raining at one moment, and what is the current fog level. In the current design of \deepqtest, we decide not to control the change of weather conditions since it is hard to simulate realistic and ever-changing weather conditions. Note that it's also hard to predict changes in weather conditions over time, even in the real world~\cite{sillmann2017understanding}. Existing works on testing ADSs mainly focus on simulating a fixed weather condition~\cite{ben2016testing,abdessalem2018testing} or applying Generative Adversarial Networks (GAN) to generate static weather condition inputs (e.g., GAN-generated images)~\cite{zhang2018Deeproad,tian2018deeptest}. These methods, however, cannot generate weather conditions that realistically change over time during a simulation.

Considering the above-mentioned limitations of existing methods, \deepqtest directly builds a mapping of weather conditions of the real world to its simulated world. To realize this, we opt for OpenWeather, an open-source online weather database, which provides rich information on history, and current, short-term, and long-term weather forecasts at any location of the globe, and this information can be accessed via APIs. 

To map real-world weather conditions to a simulated world, we consider weather data sources from two domains: $\mathcal{W_R}$, characterized with real-world weather parameters $\{w_r^0, w_r^1, w_r^2, ...\}$ (e.g., humidity, rain, wind), and $\mathcal{W_S}$, described with simulated-world weather parameters $\{w_s^0, w_s^1, w_s^2, ...\}$ (e.g., rain, wetness from a simulator), where $\mathcal{R}$ (or $r$) and $\mathcal{S}$ (or $s$) denote the real-world and simulated world, respectively. Given a specific city, since the weather condition in real-world changes over time in a day, we then first map the date and time of day to the corresponding real-world weather condition in the city, $\mathcal{G_R}: (coor; d, t)_R \rightarrow \mathcal{W}^{(coor; d, t)_R}_R$, where $coor$ denotes the coordinates of the given city, $d$ denotes a specific day (in the format of 2021-08-07, for instance), and $t$ specifies the time in a day (e.g., 20:00:00). Given a specific timestamp $(d, t)_R$ of a given city (e.g., San Francisco), the weather condition in real-world can be determined as $\mathcal{W}^{(coor; d, t)_R}_R$. To build a mapping of weather conditions of the real world to its simulated world, we first map the coordinates of the given city in the real world to the simulated world, $\mathcal{M_{COOR}}: coor_\mathcal{R} \rightarrow coor_\mathcal{S}$. Then, the timestamp $(d, t)_R$ in the real-world is mapped to the simulated world, $\mathcal{M_T}: (d, t)_R \rightarrow (d, t)_S$. Hence, a weather condition of a specific city in the real world with a timestamp $(d, t)_R$ can be mapped as, $\mathcal{M_{COOR, T}}: \mathcal{W}^{(coor; d, t)_R}_R \rightarrow \mathcal{W}^{(coor; d, t)_S}_S$. Thus, with $\mathcal{W}^{(coor; d, t)_S}_S$, we can simulate the weather condition of a specific timestamp of a specific city of the real world in a simulator.

Below we describe how we build $\mathcal{G_R}$, $\mathcal{M_{COOR}}$, $\mathcal{M_T}$ and $\mathcal{M_{COOR, T}}$ in detail.

$\mathcal{G_R}$ is built using the OpenWeather APIs. The coordinates $coor$ of a city in the real world are determined by the latitude and longitude of the city: $(latitude, longitude)$. The date and time $(d, t)$ is transformed to a Unix timestamp\footnote{The Unix timestamp is a way to track time as a running total of seconds. This count starts at the Unix Epoch on January 1st, 1970 at UTC.} (e.g., 2021-08-07 20:00:00 is represented as 1628337600 in the Unix timestamp format). Then, with the coordinates $coor$ and timestamp $(d, t)$, we can get access to corresponding weather condition data of the real world via the OpenWeather APIs.

$\mathcal{M_{COOR}}$ and $\mathcal{M_T}$ map the real-world coordinates and timestamp to the simulated world. For $\mathcal{M_{COOR}}$, \deepqtest supports importing High-Definition (HD) maps for testing ADSs in different cities, a HD map contains details including information about coordinates, road shape, and road marking, which are not normally presented on traditional maps~\cite{vardhan2017hd}. Thus, with the help of HD maps, we can easily map coordinates from the real world to the simulated world. The timestamp mapping $\mathcal{M_T}$ is already supported by the simulator (i.e., LGSVL) itself.

Weather conditions of the real world obtained via the OpenWeather APIs $\mathcal{W_R}$ are characterized by a variety of parameters that can be observed from the real world, including temperature, pressure, humidity, wetness, wind, cloudiness, rain, snow, fog and etc. However, due to the limitations of the simulator we currently use, we can only manipulate 4 types of weather parameters (Figure~\ref{fig:configurable_parameters}) to simulate weather conditions in the simulated world, which are cloudiness, rain, fog, and wetness. Thus, $\mathcal{W_S}$ is characterized with the parameters below:

\begin{equation*}
    \mathcal{W_S} = \{w_s^0, w_s^1, w_s^2, w_s^3\} = \{cloudiness_s, rain_s, fog_s, wetness_s\}
\end{equation*}
Moreover, we select four real-world parameters corresponding to the parameters in the simulator to define $\mathcal{W_R}$:

\begin{equation*}
    \mathcal{W_R} = \{w_r^0, w_r^1, w_r^2, w_r^3\} = \{cloudiness_r, rain_r, fog_r, wetness_r\}
\end{equation*}

Since weather parameters in the real world are described with real values, we correspondingly define real values (shown in Figure~\ref{fig:configurable_parameters}) for mapping real-world weather conditions to the simulated world. For example, the \textit{cloudiness} level in the real world varies from 0\% to 100\%. In contrast, in the simulator we use, the \textit{cloudiness} level ranges from 0 to 1, where, 1 is the maximum intensity of \textit{cloudiness}, which is equal to 100\% in the real world.

We acknowledge that our current mapping strategy has limitations in terms of not being able to cover a full list of parameters (e.g., \textit{temperature, pressure, wind}) due to the limitation of the simulator. However, these covered parameters have been proven to be effective in testing ADSs~\cite{zhang2018Deeproad,bijelic2018benchmarking}. Moreover, \deepqtest can be easily adapted, in future work, to cover more parameters for mapping as long as the simulator being used allows.

\subsection{Formulating Environment Configuration Learning as an MDP}\label{subsec:formulationMDP}
\subsubsection{State Encoding}\label{state} 
In the context of autonomous driving, perception of the driving environment and encoding it to extract features are the key to constructing safety ADSs. In recent years, prior works adopt different approaches to encode the operating environment of an ADS for constructing an object detector~\cite{pfeuffer2018optimal,chen2017multi} or training an end-to-end autonomous driving model~\cite{chen2019model,codevilla2018end}. For instance, Codevilla et al.~\cite{codevilla2018end} proposed a multi-modal perception method for constructing end-to-end autonomous driving models, and their experiment results indicate that an end-to-end driving model can be improved with multi-modal sensor data (Section~\ref{sec:backgroud}), instead of just relying on a single modality. To this end, in \deepqtest, we adopt multi-modal sensor fusion as the overall strategy of our state encoding. Specifically, a state inputted to the RL model is composed of the following three parts:

\begin{itemize}
    \item{Camera RGB Image}: a front view image containing information on the road structure and obstacles ahead;
    \item{Lidar Bird's Eye View}: an eye view representation encoded with height, intensity, and density, containing information of the surrounding environment of the vehicle;
    \item{Vehicle's State Measurement}: measuring the driving state of the vehicle (e.g., speed, acceleration, jerk).
\end{itemize}

After a state is observed and inputted into the RL model, features will be extracted, based on which the RL agent consequently determines which environment configuration action to take. Details about how a state is processed by the RL model will be presented in Section~\ref{networkandTraining}.

\subsubsection{Action Space}\label{action} 
The \textit{action space} in \deepqtest is a set of actions with discrete integer values representing the actions' IDs. These actions are used to configure the driving environment of an ADS, which forms test scenarios for testing the ADS. Each action is associated with a \textit{simulation time} denoted by \textit{T} (in seconds) and a \textit{time step} denoted by \textit{t} (in seconds). The \textit{simulation time T} indicates the time we let the simulation run after invoking an action. The \textit{simulation time} is divided into time steps with equal time duration \textit{t}. Additionally, an action is associated with several outputs, defined as \textit{Environment Configuration Output}.

\noindent\textbf{\textit{Outputs of Environment Configuration Actions}}. In our context, invoking an environment configuration action produces the following three types of outputs. 
\begin{itemize}
    \item \textit{Time-To-Collision (TTC)} measures the time left for a vehicle to collide with obstacles. In our current design of \deepqtest, for calculating \textit{TTC}, we adapted the prediction-based method from the Self-Driving Cars Specialization online course~\cite{UniversityofTorontoTTC} provided by the University of Toronto. The pseudocode of the calculation is presented in Algorithm~\ref{alg::ttc_calculation}.
    \item \textit{Distance-To-Obstacles (DTO)} measures the distance between \avut and the obstacles, which is closely related to driving safety~\cite{furda2011enabling}.
    \item \textit{JERK} measures the change rate of the acceleration of \avut, which is one of the important indicators of the degree of comfort of passengers~\cite{itkonen2017trade}. 
\end{itemize}
In \deepqtest, values of \textit{TTC}, \textit{DTO} and \textit{JERK} are collected every \textit{time step t}, simultaneously, and after a \textit{simulation time T} of an action being invoked, three lists ($TTC_{buff}$, $DTO_{buff}$, $JERK_{buff}$) containing the corresponding outputs are returned.

\begin{algorithm}[ht]
    \small
  \caption{Prediction-Based Time-To-Collision Calculation}
  \label{alg::ttc_calculation}
  \input{algos/algo-ttc_calculation}
\end{algorithm}

Since \deepqtest uses RL with a proper reward function to find test scenarios in a cost-effective manner, we design the reward function based on the outputs of environment configuration actions. Prior works on ADS testing have shown that using \textit{Time-To-Collision}, \textit{Distance-To-Obstacles} as objectives can guide the algorithms to generate safety-critical scenarios~\cite{ben2016testing,abdessalem2018testing}, and in~\cite{bae2020self} it is demonstrated that \textit{jerk} is an important factor to passengers' riding experience. We however acknowledge that there might exist other types of outputs, which can be investigated in the future. 

\subsubsection{Reward Function}\label{subsubsec:rewardfunctions}
The reward function is defined based on the environment configuration outputs (Section \ref{action}). More specifically, to calculate reward, we define three rewards corresponding to the three outputs, denoted as: \textit{Time-To-Collision Reward} ($TTC_{Reward}$), \textit{Distance-To-Obstacles Reward} ($DTO_{Reward}$), and \textit{JERK Reward} ($JERK_{Reward}$), respectively.

\noindent\textbf{\textit{Time-To-Collision Reward}}, $TTC_{Reward}$. A smaller \textit{TTC} indicates a higher collision risk and vice versa~\cite{vogel2003comparison}. In the context of testing ADSs, we aim to put the \avut at risk, meaning that smaller \textit{TTC} values lead to higher reward values. 

As shown in Algorithm~\ref{alg::ttc_calculation}, after an environment configuration action is executed, a buffer of \textit{TTC} (i.e., $TTC_{buff}$) will be returned as an output, and the algorithm then takes the minimum \textit{TTC} in $TTC_{buff}$ for calculating $TTC_{Reward}$. To apply \textit{TTC} as a metric indicating the risk to collisions, a \textit{TTC} threshold should be determined, denoted as $ttc_{thre}$, which is configured as 7 seconds, in the current design of \deepqtest, based on the guideline from~\cite{zhu2020safe}.
The calculation of $TTC_{Reward}$ is shown in Equation~\ref{equ:ttc_reward}. If a \textit{TTC} value is less than or equal to $ttc_{thre}$, the reward will be positive; otherwise, a punishment being -1 will be returned.

\begin{equation}
    \label{equ:ttc_reward}
    TTC_{Reward} = \left\{
    \begin{aligned}
    & -\ln{\dfrac{nor(TTC)}{nor(ttc_{thre})}}, & 0 < TTC \leq ttc_{thre}\\
    & -1, & TTC > ttc_{thre}
    \end{aligned}
    \right.
\end{equation}

\textbf{\textit{Distance-To-Obstacles Reward}}, $DTO_{Reward}$. Prior works have shown that different distances to obstacles during driving will cause risks of varying severity, and a smaller distance to obstacles will lead to a higher risk of collision, especially when the safety distance is violated~\cite{ro2020new}. In \deepqtest, we take the minimum $DTO$ value from the $DTO_{buff}$ to calculate $DTO_{Reward}$. The equation for calculating $DTO_{Reward}$ is similar to the calculation of $TTC_{Reward}$, shown in Equation~\ref{equ:dto_reward}, where, $dto_{thre}$ is the safety distance indicating the minimum distance by which the \avut avoids a collision with the obstacles. $dto_{thre}$ is related to the kinematics parameters (e.g., speed, acceleration) of the \avut and obstacles, and in \deepqtest, we take the default value from Berkeley algorithm~\cite{bella2011collision} which is 10 meters.

\begin{equation}
    \label{equ:dto_reward}
    DTO_{Reward} = \left\{
    \begin{aligned}
    & -\ln{\dfrac{nor(DTO)}{nor(dto_{thre})}}, & 0 < DTO \leq dto_{thre}\\
    & -1, & dto > dto_{thre}
    \end{aligned}
    \right.
\end{equation}

\textbf{\textit{JERK Reward}}, $JERK_{Reward}$. Since the jerk of the \avut has a strong influence on the comfort of passengers; the higher the jerk, the less comfortable the passengers~\cite{bae2020self}. When a list of jerks ($JERK_{buff}$) is returned after an environment configuration action is executed, we then take the maximum $JERK$ value in $JERK_{buff}$ for calculating $JERK_{Reward}$. We also need a \textit{JREK} threshold ($jerk_{thre}$) in the calculation, and according to our preliminary study, we found that for normal driving, the jerk values are lower than $5m/s^3$. Therefore, we set the threshold as $5m/s^3$ since we want to maximize $JERK$.

\begin{equation}
    \label{equ:jerk_reward}
    JERK_{Reward} = \left\{
    \begin{aligned}
    & e^{\dfrac{nor(JERK)}{nor(jerk_{thre})}} - 1, & JERK \geq jerk_{thre}\\
    & -1,  & 0 \leq JERK < jerk_{thre}
    \end{aligned}
    \right.
\end{equation}

Considering that environment configuration outputs (\textit{TTC}, \textit{DTO}, and \textit{JERK}) may be not comparable, in the above calculations, we apply the normalization function~\cite{greer2004software}, $nor(F_x) = (F_x-F_{min}) / (F_{max}-F_{min})$, to ensure all values fall into $[0, 1]$, where, $F_x$ is an environment configuration output to be normalized; \textit{max} and \textit{min} represent the maximum value and minimum value of the environment configuration output, respectively.

\subsection{Sensor Data Preprocessing}\label{preprocessing}

Some preprocessing steps for camera images and lidar point cloud data in multimodal state encoding are necessary because working directly with raw data can be computation and memory demanding~\cite{mnih2015human}. In \deepqtest, a camera image is an RGB image of $1920 \times 1080$ pixel frames at each color channel, we crop off the top rectangle area of the image so that the picture in front of the vehicle is centered in the image,
then resize the cropped image to $160 \times 376$ pixel frames at each color channel while keeping its aspect ratio constant.

A lidar sensor generates a point cloud \textit{P} containing $N$ points: $P=\{p_1, p_2, ..., p_N\}$, $p_i = (x, y, z, r)^T$, i $\in \{1, 2, ..., N\}$, where $(x, y, z)$ denotes a 3D position in space and $r$ denotes the reflectance of point $p_i$. To get a bird's eye view representation from the point cloud, we follow the steps from MV3D~\cite{chen2017multi} to encode a bird's eye view representation with height, intensity, and density. Hence, the point cloud is discretized into a 2D grid with a resolution of 0.1 meters. The height of each cell in the grid is calculated as the maximum height of the points. The intensity is the reflectance value of the maximum height point. The density is computed using the number of the points in the cell ($n$) as $min(1.0, \dfrac{log(n+1)}{log(64)})$. To encode more detailed height information, the point cloud is divided equally into $M$ slices. Finally, the bird's eye view is encoded as $(M + 2)$-channel features. In \deepqtest, we are using a lidar sensor with 128 laser beams, which can generate 9391 points in one point cloud. Thus, in total the bird's eye view is a 15-channel feature with size $300 \times 300$ at each channel.

\subsection{Network Architecture and Training Algorithm}\label{networkandTraining}

\subsubsection{Network Architecture}

Deep Q-Learning (Section~\ref{subsec:dqn}) utilizes a neural network (Deep Q-Network) to approximate and store Q-values and the design of the Q-Network plays a key role in learning tasks. In the context of autonomous driving, a good network architecture is conducive to the learning tasks, such as the success in image classification (e.g., VGGNet~\cite{simonyan2014very}, Alexnet~\cite{krizhevsky2012imagenet}), object detection (e.g., MV3D~\cite{chen2017multi}) and pedestrian prediction~\cite{keller2013will}. Focusing on multimodality and sensor data fusion, various network architectures have been proposed as well as various fusion schemes. As discussed in Section~\ref{subsec:sensorfusion}, the middle fusion scheme has demonstrated promising performance in various autonomous driving perception tasks, which gives us the confidence to adopt the middle fusion scheme to construct the Q-network in \deepqtest.

\begin{figure}[ht]
    \centering
    \includegraphics[width=\textwidth]{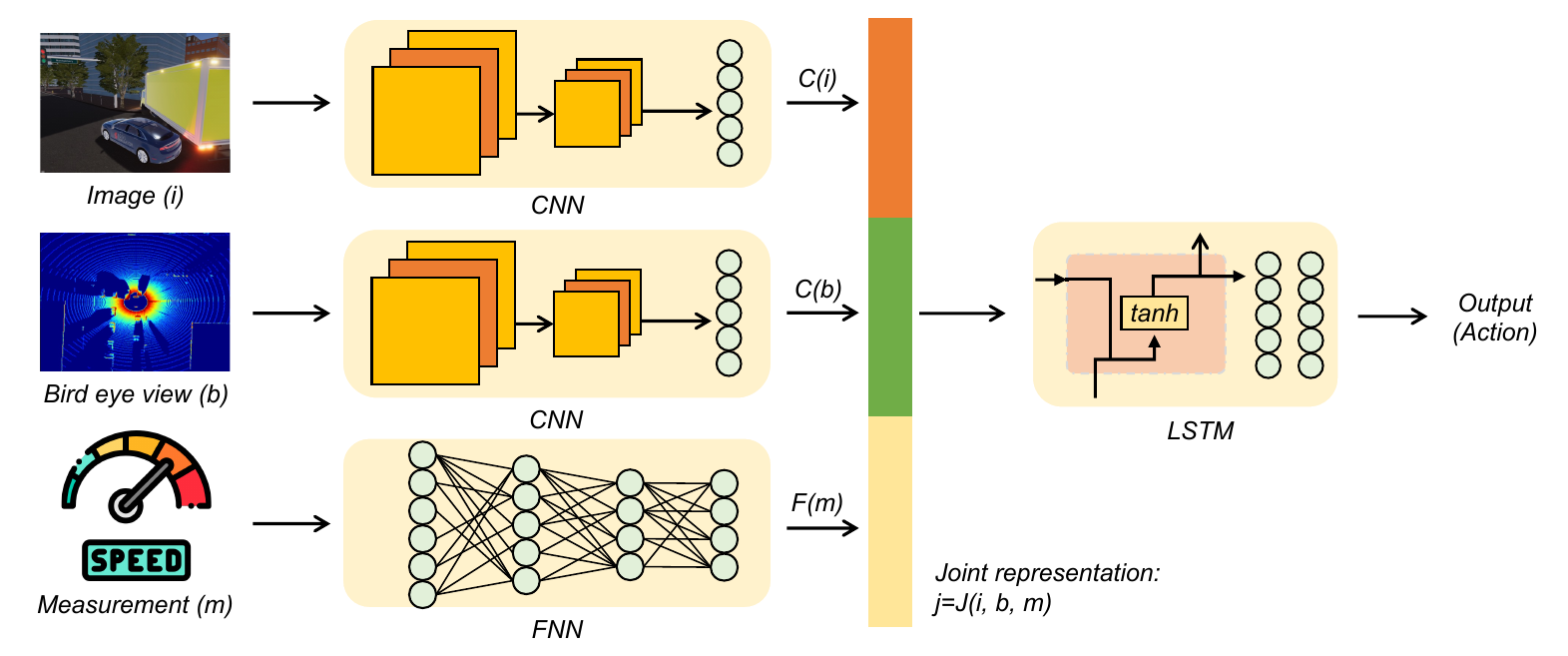}
    \caption{\textbf{Q-Network Architecture.} The RGB image, bird's eye view, and vehicle status measurement are the inputs of Q-Network. The output is the environment configuration action.
    }
    \label{fig:network_architecture}
\end{figure}

Assume that each state s = <\textbf{i}, \textbf{b}, \textbf{m}> comprises an image \textbf{i}, a lidar bird's eye view \textbf{b}, and a low-dimensional vector \textbf{m} referring to vehicle state measurements. The deep neural network (DNN) in a DQN algorithm takes a state as the input and produces an action with the maximum Q-value as its output. 
As shown in Figure~\ref{fig:network_architecture}, the network includes an image-processing CNN module, taking image \textbf{i} as the input to extract image features \textit{C}(\textbf{i}); a lidar-processing CNN module which takes lidar bird's eye view \textbf{b} as the input to extract lidar bird's eye view features, \textit{C}(\textbf{b}); and a module of fully-connected neural network (FNN) which takes vehicle state measurements as the input to extract measurement features, \textit{F}(\textbf{m}). After a state is processed by these three modules, the output features (i.e., \textit{C}(\textbf{i}), \textit{C}(\textbf{b}), \textit{F}(\textbf{m})) are concatenated into a joint representation:

\begin{equation}
    j = J(i, b, m) = <C(i), C(b), F(m)>
\end{equation}
The final decision module, implemented as an LSTM~\cite{hochreiter1997long}, followed by two fully-connected layers, takes the joint representation \textbf{j} as input and outputs an action with the maximum Q-value.

More specifically, the image-processing CNN module consists of 4 convolutional layers and 2 fully-connected layers. The input \textbf{i} consists of a $160 \times 376 \times 3$ image. The first hidden layer convolves 32 filters of $5 \times 5$ with stride 4. The second hidden layer convolves 64 filters of $3 \times 3$ with stride 2. The third hidden layer convolves 128 filters of $3 \times 3$ with stride 2, followed by the final convolutional layer that convolves 256 filters of $3 \times 3$ with stride 2. This is followed by a hidden and fully-connected layer with its neuron number (128 neurons) equal to the output of the last convolutional layer of the module.
The lidar-processing CNN module has the same design as the image-processing CNN module, but with its input \textbf{b} being a $300 \times 300 \times 15$ image.
The FNN module with its input as measurement \textbf{m} is designed as a four-layer fully-connected neural network with the architecture of an input layer with 1 neuron, 200 hidden layers with 200 neurons for each, and one output layer with 64 neurons.
The final decision module consists of an LSTM with 704 neurons as the input, 64 hidden neurons, and 2 recurrent layers, which is followed by a fully-connected hidden layer with 64 neurons and another fully-connected layer for outputting an action with the maximum Q-value.
To ensure stability, prevent overfitting, and accelerate the convergence of the network training process,
we perform batch normalization after each convolutional layer, then apply the Rectified Linear Unit (ReLU) nonlinearity after each normalization. We also apply the ReLU nonlinearity after each fully-connected layer and use 10\% dropout after LSTM.


\section{Experiment Design}\label{sec:evaluation}

\subsection{Research Questions}\label{subsec:RQs}
To assess our approach,
we carried out an empirical study to answer the following three main research questions (RQs):

\textbf{RQ1}:
Does \deepqtest achieve better performance compared to \textit{Random} and \textit{Greedy Strategies}?
RQ1 aims to do a sanity check that the problem we address is complex and it is hard to be solved by naive
\textit{Random} and \textit{Greedy Strategies}.

\textbf{RQ2}: Among the three reward functions, which one performs the best?
RQ2 aims to identify the best design(s) (Section~\ref{subsubsec:rewardfunctions}), which can guide \deepqtest to generate more effective environment configurations for testing \avut.

\textbf{RQ3}: How does the \deepqtest 
perform compared to the start-of-the-art technique?
RQ3 aims to demonstrate the effectiveness of \deepqtest compared to the start-of-the-art technique, i.e., \deepcollision.

Table~\ref{tab:RQ_descriptions} shows an overview of 
our experiment design. In the rest of this section, we present comparison baselines (Section~\ref{subsec:baseline}), experiment settings (Section~\ref{subsec:experimentSettings}), evaluation metrics (Section~\ref{subsec:metrics}), and statistical test (Section~\ref{subsec:statisticaltest}).

\begin{table}
    \centering
    \caption{
    Description of tasks, employed approaches, road settings, weather condition settings, metrics, and statistical test for each RQ
    }
    \resizebox{\textwidth}{!}{

\input{tables/RQ_descriptions.tex}
    }
    \label{tab:RQ_descriptions}
\end{table}

\subsection{Comparison Baselines}\label{subsec:baseline}
To evaluate \deepqtest, we selected three baseline techniques, i.e., 
a random strategy (i.e., \randomStrategy), a greedy strategy (i.e., \greedyStrategy), and \deepcollision~\cite{lu2022learning}.
\randomStrategy is commonly used as the baseline for a sanity check on whether an optimization problem is complex enough. 
\greedyStrategy has been used as the comparison baseline in the literature for several RL-based approaches, including RL-based decision-making for autonomous lane changing~\cite{mukadam2017tactical} and RL-based navigation optimization~\cite{abedin2020data}. 
\deepcollision is a recent study that is an RL-based ADS testing strategy proposed by Lu~\cite{lu2022learning} for testing \avut with various environment configurations.
Detailed implementations and setup for each baseline are:
\begin{enumerate}
    \item 
\textbf{\randomStrategy} (Algorithm~\ref{alg::baseline_random}):
\randomStrategy randomly selects an action to configure the environment at each environment configuration step.
    \item
    \textbf{\greedyStrategy} (Algorithm~\ref{alg::baseline_greedy}): \textit{GS} was implemented to select the best configuration at each step. 
To determine the next action to be executed, \greedyStrategy executes all the actions (Lines 7--14), then selects an action that achieves the best performance based on a defined reward (Line 15). For \textit{GS}, we implemented three rewards as \deepqtest (i.e., \metricTTC, \metricDTO, and \metricJerk), and comparison was conducted by employing the same reward (e.g., \rewardTTC \textit{vs.} $GS_{TTC}$). To ensure that the comparison among actions is fair, we implemented a \textit{rollback} mechanism (see Line 13) that facilitates, at each iteration, executing each action from the same state of the testing environment. A sequence of the best actions is returned at the end of the search.
\item
\textbf{\deepcollision}~\cite{lu2022learning}: \deepcollision learns environment configurations with RL guided by the collision probability to maximize collision occurrences.
There are four models trained in \deepcollision, and we selected the recommended one~\cite{lu2022learning}, i.e., \textit{M6} model, in our experiment.
In \deepcollision, a state of the driving environment is encoded based on 8 environmental parameters, while \deepqtest adopts multi-modal sensor fusion to encode the state.
Compared to \deepcollision, we further consider realism by defining realistic constraints in \deepqtest to ensure that \avut can be tested with realistic scenarios.
To better explore testing of ADS with various environment configurations, in \deepqtest, we involved 90 more environment configurations APIs than \deepcollision, i.e., 142 APIs in \deepqtest \textit{vs.} 52 APIs in \deepcollision.

\end{enumerate}

\begin{algorithm}[htbp]
    \small
  \caption{RS-based Environment Configuration}
  \label{alg::baseline_random}
  \input{algos/algo-RS}
\end{algorithm}

\begin{algorithm}[htbp]
    \small
  \caption{GS-based Environment Configuration}
  \label{alg::baseline_greedy}

\input{algos/algo-GS}
\end{algorithm}

\subsection{Experiment Settings}\label{subsec:experimentSettings}
\subsubsection{Subject System and Simulator}

To evaluate \deepqtest, we adopt Baidu's open-source autonomous driving platform Apollo~\cite{apollo} as the subject system under test, which is a high-performance ADS with various modules responsible for various tasks, such as traffic light recognition, obstacle perception and avoidance, and trajectory planning and routing. Apollo is a flexible architecture with high-driving automation, aiming to support the development, testing, and deployment of autonomous vehicles. Concretely, we used the Apollo Open Platform 5.0, which can operate without human interactions in most circumstances and handle complex road scenarios with enhanced perception deep learning models.

We adopt LGSVL Simulator~\cite{rong2020lgsvl} to simulate the autonomous vehicle and its operating environment. Specifically, LGSVL is an open-source, high-fidelity autonomous driving simulation platform that provides an essential vehicle dynamic model for vehicle simulation and supports a set of sensors, including cameras, LiDAR, GPS, and Radar. LGSVL can easily connect to autopilot platforms such as Apollo and Autoware~\cite{kato2018autoware}, enabling us to perform software-in-the-loop and hardware-in-the-loop ADS testing. In our experiment, we employed LGSVL Simulator 5.0. We selected the San Francisco HD map, 
a digital re-creation of a section of the South of Market Street (SoMa) in San Francisco with various driving environment characteristics, e.g., traffic light intersections and multi-lane streets. Regarding the \avut, we used Lincoln2017MKZ, a four-door Sedan car equipped with various sensor configurations.

\subsubsection{Model Training} We selected 4 different types of real-world weather conditions
(shown in Table~\ref{tab:real-world-weather}) in San Francisco from the OpenWeather dataset. Next, we trained 12 (3 reward settings $\times$ 4 real-world weather conditions) \deepqtest models. Accordingly, we defined 12 \randomStrategy and 12 \greedyStrategy strategies. We use convention $\deepqtest_{[reward][real-world-weather]}$ to denote each trained model, and use $\greedyStrategy_{[reward][real-world-weather]}$, $\randomStrategy_{[reward][real-world-weather]}$ to denote each baseline strategy (Section~\ref{subsec:baseline}). For example, $\deepqtest_{TTC_{RD}}$ denotes the model trained in the rainy day (RD) with reward function $TTC_{Reward}$.

All models were trained with the same network architecture, learning algorithm, reward design, and hyperparameter settings to ensure fair comparisons.
For training each model, we used the Adam algorithm~\cite{zhang2018improved} with mini-batches of size 64 as the adaptive optimization algorithm. To manage the trade-off between exploration and exploitation, we applied $\epsilon$-greedy as the behavior policy during training with $\epsilon$ annealed from 1.0 to 0.2 over the first 10000 observed states and fixed at 0.2 thereafter. Such a setting of behavior policy lets the agent explore more at the beginning of the training process when little is known about the problem environment, and as the training progresses, the agent gradually conducts more exploitation than exploration. We trained for a total of 40000 observed states and used a replay memory with a size of 6000.

\begin{table}[htbp]
    \centering
    \small
    \caption{Real-world weather conditions applied in the experiment}
    \resizebox{\textwidth}{!}{
        \input{tables/real-world-effect}
    }
    \label{tab:real-world-weather}
\end{table}

Based on our pilot study, we found that directly training on a huge HD map with many roads is not easy. Because we are trying to generate complex driving scenarios for an \avut driving from origin to destination, thus if the driving path is too long, then it will take much more time for one episode. Moreover, as the training progresses, the environment will be more complex for the \avut arriving at the destination, therefore, collisions tend to occur early in the driving task.
Thus, we divided the map into road segments with lengths of about 1km to 2km. Next, we trained each model on each road segment to gain a final model. This allows the agent to finish one episode within a reasonable time and ensures that the whole road is fully explored and exploited.

The values of all hyperparameters are determined based on the parameter settings from~\cite{mnih2015human,paszke2019pytorch}. 
We provide the settings of all hyperparameters in our online repository (see Section~\ref{subsec:data}).

\subsubsection{Experiment Design and Execution}\label{subsubsec:experimentdesign}
\begin{figure*}[htbp] 
	\centering
        \input{figs/fig_4-roads}
	\caption{Graphical representations of the roads from ~\cite{lu2022learning}}
	\label{fig:roads}
\end{figure*}
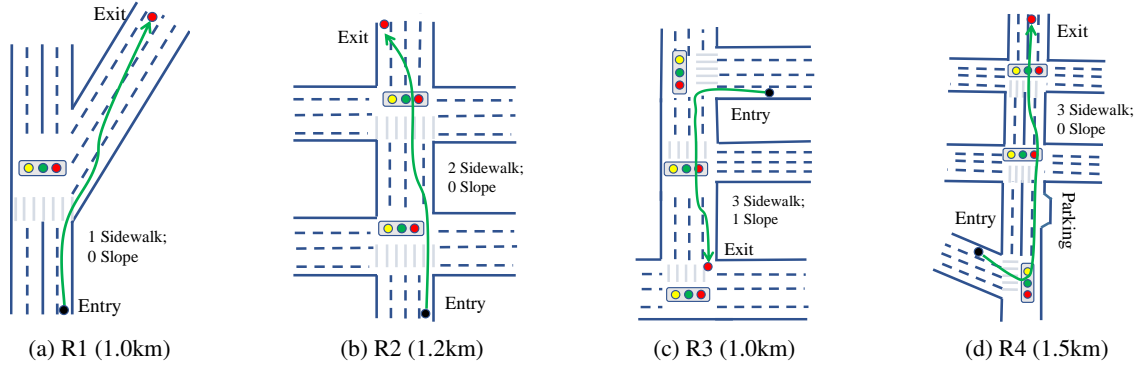

\textit{\textbf{Roads}}. To ensure that the experiments are run in the same environment, we used the same roads adopted in \deepcollision~\cite{lu2022learning}, which are selected based on the road structure definition in~\cite{czarnecki2018operational}. Figure~\ref{fig:roads} graphically depicts their characteristics.

\textit{\textbf{Driving Scenarios}}. A driving scenario \textit{S} describes the temporal development between several scenes, where a scene describes a snapshot of the environment including the \avut, static and dynamic obstacles, and environmental conditions such as weather and traffic rules~\cite{ulbrich2015defining}. We, therefore, formally define a driving scenario as a tuple with scenes: \textit{S}=$<scene^1, scene^2, ..., scene^n, ST>$, where \textit{ST} is the time period that the scenario spans and \textit{n} denotes the number of scenes in the scenario. A driving scene is a 3-tuple <\textit{\avut Operation}, \textit{\avut Speed}, \textit{Environment State}>, denoting \avut operations such as cruise, a speed level of the \avut, e.g., zero or slow, and 3) environment state about weather condition, time of date, traffic rules, obstacles, and their behaviors, etc. More specifically, we use 11 properties (column 3, Table~\ref{tab:scene}) to characterize \avut’s states (rows 1 and 2, Table~\ref{tab:scene}) and the environment (rows 3 to 11, Table~\ref{tab:scene}). Note that NPC vehicles, Pedestrians, Static obstacles, Traffic Lights, and Sidewalks are collected within the sensing range of the radar deployed on the \avut. Moreover, in our current design, the time period that the scenario spans (i.e., \textit{ST}) is set to 3 seconds.

\begin{table}[htbp]
    \centering
        \caption{Properties characterizing scenes in the formal experiment}
    \resizebox{\textwidth}{!}{
        \input{tables/scene_definition}}
    \label{tab:scene}
\end{table}

\textit{\textbf{Execution}}. To answer the RQs (Section~\ref{subsec:RQs}), each trained \deepqtest model, \randomStrategy, \greedyStrategy, and \deepcollision were run 20 times on each of the four roads (R1...R4). The trained \deepqtest models were evaluated with an $\epsilon$-greedy behavior policy with $\epsilon$=0.05. As suggested by Mnih~\cite{mnih2015human}, this setting is adopted to minimize the possibility of overfitting during evaluation. In total, we obtained results of 2880 executions (20 runs $\times$ 4 roads $\times$ 3 strategies $\times$ 3 reward functions $\times$ 4 Real-World Effects). All the experiments were executed on three machines with identical configurations with a 2.4GHz Intel Xeon E5-2680 v4 CPU, an 11GB NVIDIA GeForce GTX 1080 Ti GPU, and 32GB RAM. The operating systems are all Ubuntu 18.04.6 LTS.

\subsection{Evaluation Metrics}\label{subsec:metrics}
In this section, we present the definition of the evaluation metrics: \metricReward, \metricRealism, and \metricDiversity. As shown in Table~\ref{tab:RQ_descriptions}, for RQ1, we use \metricReward to evaluate the performance of \deepqtest when comparing with \randomStrategy and \greedyStrategy. We adopt \metricReward, and \metricDiversity to evaluate the performance of the three reward functions of \deepqtest (RQ2). Regarding the comparison with \deepcollision (RQ3), we use \metricRealism to evaluate the realism of the generated scenarios because both \deepqtest and \deepcollision have different mechanisms to ensure the scenario realism, and \metricDiversity is used for evaluating the diversity of environment configurations and scenarios of the two approaches.

\subsubsection{\metricReward}
In \deepqtest, we defined three reward functions based on \metricTTC, \metricDTO, and \metricJerk, respectively, which are related to driving safety or comfort. In our evaluation, we employed them to study the performance of the reward functions and their potential correlations. These metrics are formally defined as Equation~\ref{eq:rewardmetric}, where $k$ denotes the $k$th execution, and $i$ denotes the $i$th \textit{simulation time T} (see Section~\ref{action}) in the execution.

\begin{equation}\label{eq:rewardmetric}
    SCM_k = \dfrac{\sum_{i=1}^{n}SCM^{i}_k}{n}, SCM \in \{TTC, DTO, Jerk\}, k= 1...20
\end{equation}

In addition, during each execution $k$, we collected the number of collisions $\#Collision_k$. $\#Collision_k$ could be 0 if no collision occurred. For \deepqtest, $\#Collision_k \in \{0, 1\}$ as it terminates the execution once a collision is identified.
For \deepcollision, $\#Collision_k$ could be more than one, because \deepcollision allows the \avut to continue to move forward after the occurrence of a collision. We also calculated $CollisionTime_k$, which measures how long a collision can be observed on average.

\subsubsection{\metricRealism}

To evaluate \deepqtest and \deepcollision in terms of whether they generate realistic scenarios, in addition to the \realisticConstraints on \textit{Object}s and \textit{Road}s (Section~\ref{subsec:configurableparameters}), we further define a set of \realisticConstraints on the parameters for weather and time. Specifically, for weather and its change over time, we ensure its realism based on four weather-related measurements which can be observed in the simulator, i.e., rain, cloudiness, fog, and wetness. By analyzing real-world historical weather data from an open weather dataset, i.e., OpenWeather\footnote{\url{https://openweathermap.org/}}, we calculate the range of realistic changes over time of these four weather-related measurements and then define the following constraints for ensuring the realism of the weather and its change. 


\begin{itemize}
    \item \textit{Rain} reflects the intensity of the rainfall, to be realistic, the rain level change within 1 hour should be less than 20\%.
\begin{lstlisting}
context Rain
inv: self.rainLevel <= (self.getPreviousValue(1h).rainLevel+0.2).max(1.0) and self.rainLevel >= (self.getPreviousValue(1h).rainLevel-0.2).min(0.0)
\end{lstlisting}
    \item \textit{Cloudiness} refers to the extent to which the atmosphere is covered by clouds (measured by \textit{cloudinessLevel} in Figure~\ref{fig:configurable_parameters}).
    For instance, cloudiness which is less than 30\% represents a clear sky while near 100\% cloudiness represents an overcast sky. To be realistic, the cloudiness level change within an hour should be less than 25\%. For example, if the current cloudiness is 25\%, the realistic cloudiness within one hour should be between 0\% and 50\%.
\begin{lstlisting}
context Cloudiness
inv: self.cloudinessLevel <= (self.getPreviousValue(1h).cloudinessLevel+0.25).max(1.0) and self.cloudinessLevel >= (self.getPreviousValue(1h).cloudinessLevel-0.25).min(0.0)
\end{lstlisting}

    \item \textit{Fog} impacts the visibility of the vehicle, according to the analysis of its changes based on real-world weather data, we consider it realistic that the change of fog level within one hour is within 10\%.
\begin{lstlisting}
context Fog
inv: self.fogLevel <= (self.getPreviousValue(1h).fogLevel+0.1).max(1.0) and self.fogLevel >= (self.getPreviousValue(1h).fogLevel-0.1).min(0.0)
\end{lstlisting}
    \item \textit{Wetness} (expressed as a percentage) reflects how the road surfaces should be. The realistic wetness level changes within an hour should be less than 5\%. For example, if the current wetness level is 80\%, the realistic wetness level in one hour should be between 75\% and 85\%.
\begin{lstlisting}
context Wetness
inv: self.wetnessLevel <= (self.getPreviousValue(1h).wetnessLevel+0.05).max(1.0) and self.wetnessLevel >= (self.getPreviousValue(1h.wetnessLevel)-0.05).min(0.0)
\end{lstlisting}
    \item Regarding \textit{time of day}, its changes should follow the passage of time in the real world. However, in the simulated world, the simulator has its own clock. 
    So, we consider the passage of time to be realistic if the passage of time follows or is slower than the simulation time.
\begin{lstlisting}
context TimeOfDay
def: t = secondinsimulation
inv: self.timeofday <= self.getPreviousValue(t).timeofday + t
\end{lstlisting}
    
\end{itemize}

By validating if generated scenarios satisfy the above constraints in terms of realism,
we further classified the generated scenarios into four types by considering collisions, i.e., \textit{Realistic Collision Scenarios} (\metricRealismRCS), \textit{Unrealistic Collision Scenarios} (\metricRealismUCS), \textit{Realistic Non-Collision Scenarios} (\metricRealismRNS), and \textit{Unrealistic Non-Collision Scenarios} (\metricRealismUNS).
For each execution $k$, we collected the total number of the generated test scenarios as $\metricRealismTC_{k}$, the number of these four types of scenarios as $\#\metricRealismRCS_{k}$, $\#\metricRealismUCS_{k}$, $\#\metricRealismRNS_k$, and $\#\metricRealismUNS_k$, and proportion of the four types out of all scenarios as $RCS\%_k$, $UCS\%_k$, $RNS\%_k$, and $UNS\%_k$. 
In addition, we also calculated the average time it takes for a realistic collision scenario (i.e., $\metricRealismRCS_{k}$) to happen as \textit{Realistic Collision Time} ($RCT_{{RCS}_{k}}$).

\subsubsection{\metricDiversity}
To measure \textit{Diversity} in one execution, we defined two metrics: \textit{API Diversity} (i.e., \metricDiversityAPI), and \textit{Scenario Diversity} (i.e., \metricDiversityScenario).

\textit{API Diversity} (\metricDiversityAPI) measures the diversity of the environment configuration APIs invoked. \metricDiversityAPI is calculated for each execution $k$ as ${Div_{API}}_k=\dfrac{\#UniqueAPI_k}{\#InvokedAPI_k}$, where $\#UniqueAPI_k$ represents the number of the unique environment configuration APIs invoked in execution $k$ and $\#InvokedAPI_k$ represents the total number of the invoked APIs in execution $k$.

\begin{figure}
    \centering
    \includegraphics[width=\textwidth]{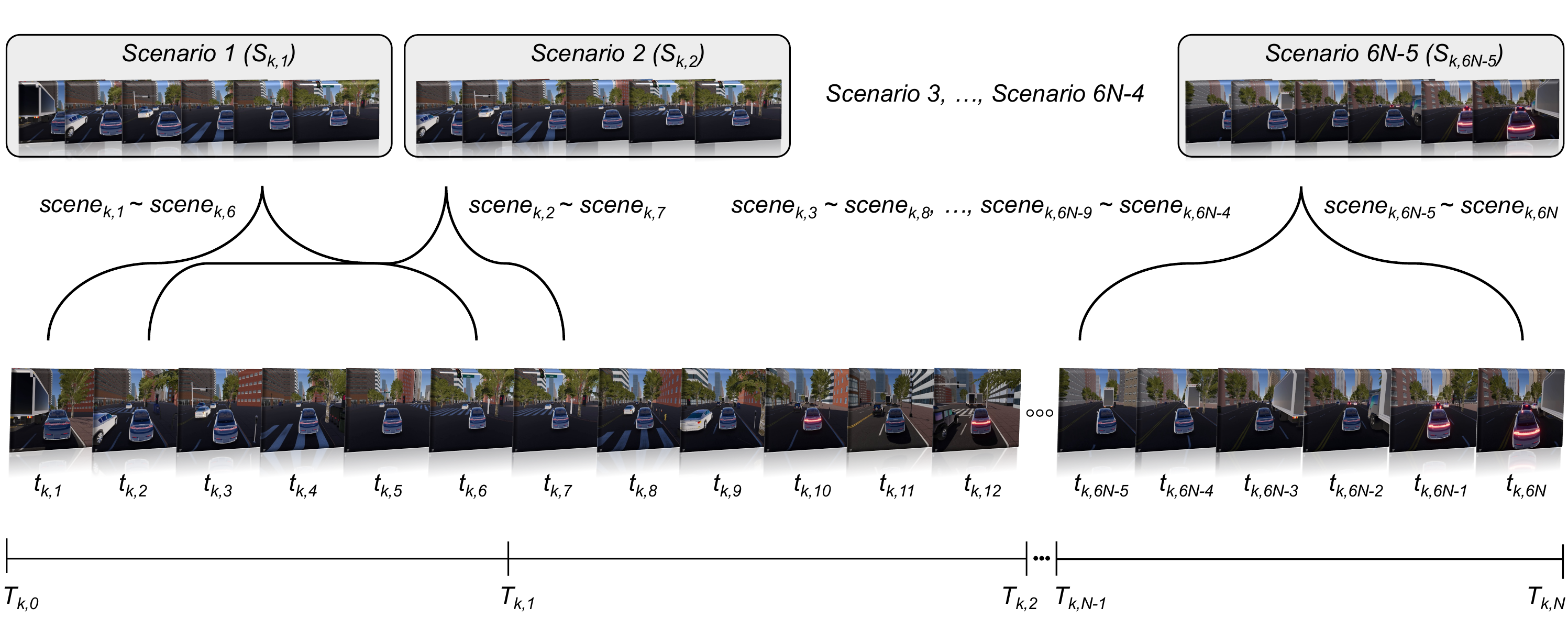}
    \caption{\textbf{The process for scenario identification in the $k$th execution.} $k$ denotes the $k$th execution, and $T$ is the simulation time after an environment configuration action is invoked. $scene_{k,x}$ is the $xth$ scene captured in the $k$th execution.}
    \label{fig:div_scenario}
\end{figure}

\textit{Scenario Diversity} (\metricDiversityScenario) measures the diversity of generated driving scenarios.
A driving scenario (see Section~\ref{subsubsec:experimentdesign}) is regarded as a sequence of scenes over a time period, i.e., 
\begin{equation}\label{eq:scenario}
    S_{k,i}=<scene_{k,i}^{1}, scene_{k,i+1}^{2},...,scene_{k,i+n-1}^{n}, ST>
\end{equation}
where $k$ is the $k$th execution, $i$ is the $i$th scenario that occurred in the execution, $scene_{k,x}$ is the $xth$ scene captured in the execution, $n$ is the number of scenes in the scenario, and $ST$ is the time period that the scenario spans. In our current design, we set $ST$ as 3 seconds, which is the same as the simulation time (i.e., \textit{T}) after invoking an environment configuration action, and capture a scene every 0.5 seconds; therefore, each scenario contains 6 scenes, i.e., $n = 6$. Figure~\ref{fig:div_scenario} shows an example of how scenarios are identified in the $k$th execution.
For an execution $k$, \metricDiversityScenario is calculated by comparing occurred scenarios based on the 11 properties characterizing scenes (Table~\ref{tab:scene}). For each pair of scenarios (i.e., $S_{k,i}$, $S_{k,j}$ and $i \neq j$) in the $k$th execution, we first calculate $similarity$ between $S_{k,i}$ and $S_{k,j}$ based on Algorithm~\ref{alg::calculate_diversity}, which is inspired by the Ratcliff-Obershelp similarity algorithm~\cite{string_sim} which is originally designed
for measuring the similarity of two strings
~\cite{kolchin2014template}. Algorithm~\ref{alg::calculate_diversity} calculates $Sim_{S_{k,i}, S_{k,j}}$ by finding a set of consecutive scenes that can maximize the similarity between $S_{k,i}$ and $S_{k,j}$:
\begin{equation}
    Sim_{S_{k,i}, S_{k,j}} = \textsc{scenarioSimilarity}(S_{k,i}, S_{k,j}), Sim_{S_{k,i}, S_{k,j}} \in [0,1];
\end{equation}
then, we define their $diversity$ ($Div_{S_{k,i}, S_{k,j}}$) based on the $similarity$
as:
\begin{equation}
    Div_{S_{k,i}, S_{k,j}} = 1 - Sim_{S_{k,i}, S_{k,j}};
    \label{equ:div}
\end{equation}
%
\textit{Scenario Diversity} is calculated based on all such pairwise $diversity$ of the scenarios in the $k$th execution as:
\begin{equation}
    {Div_{Scenario}}_{k} = \dfrac{\sum_{i=1}^{m-1}\sum_{j=i+1}^{m}Div_{S_{k,i}, S_{k,j}}}{m},
\end{equation}
where $m$ is the number of scenarios generated in execution $k$.

\subsection{Statistical Test}\label{subsec:statisticaltest}
To account for the randomness caused by the RL algorithm (i.e., DQN) and the ADS itself, 
it is essential to use statistical tests to compare the performance of different strategies~\cite{arcuri2011practical,harman2010search}. The selection of appropriate statistical techniques is vital to answer the RQs correctly. As suggested by Arcuri and Briand~\cite{arcuri2011practical}, we first checked if our samples are normally distributed using the Kolmogorov-Smirnov test~\cite{KVtest}, and results show that they are not normally distributed. Thus, we applied the non-parametric statistical techniques to compare samples generated by \deepqtest and baseline strategies. 

Specifically, we use the Mann and Whitney U test~\cite{mann1947test} for assessing the statistical significance and calculate the effect size with Vargha and Delaney metric $\hat{A}_{12}$. 
Given an evaluation metric, $\mathcal{M}$, $\hat{A}_{12}$ was used to compare the probability (i.e., how often) of yielding higher values of the metric $\mathcal{M}$ for two strategies, A and B. If $\hat{A}_{12}$ is 0.5, then they are equal. If $\hat{A}_{12}$ is greater than 0.5, then A has a higher chance to obtain higher values of $\mathcal{M}$ than B, and vice versa. The Mann–Whitney U test calculates a $p$-value to determine if the performance difference between A and B is significant. A $p$-value less than 0.05 indicates the significant difference between A and B. In addition, by following the guidelines proposed by Kitchenham et al.~\cite{RobustTest}, we divide the Vargha and Delaney effect size magnitude of $\hat{A}_{12}$ into four levels:
\textit{negligible} ($\hat{A}_{12}$ $\in$ (0.444, 0.556)), \textit{small} ($\hat{A}_{12}$ $\in$ [0.556, 0.638) or (0.362, 0.444]), \textit{medium} ($\hat{A}_{12}$ $\in$ [0.638, 0.714) or (0.286, 0.362]), \textit{large} ($\hat{A}_{12}$ $\in$ [0.714, 1.0] or [0, 0.286]).

In addition to investigating the performance of different strategies, we also want to investigate the correlation among the three selected measures, i.e., \metricTTC, \metricDTO, \metricJerk. We choose to use Spearman's rank correlation ($\rho$) test, which is a non-parametric, rank-based assessment of correlation test. The $\rho$ value ranges from $-1.0$ to 1.0, i.e., there is a positive correlation if $\rho$ is close to 1.0 and a negative correlation when $\rho$ closes to $-1.0$. If $\rho$ is closer to 0, then there is no correlation between the two measures. We also reported the significance of the correlation using a $p$-value, i.e., a $p$-value less than 0.05 tells that the correlation is statistically significant.

\section{Experiment Results and Analysis}\label{sec:results}

\subsection{Results for RQ1}\label{subsubsec:rq1_results}
To answer RQ1, we compare each of the three reward settings of \deepqtest 
(i.e., \rewardTTC, \rewardDTO, \rewardJerk)
with the \randomStrategy and \greedyStrategy strategies.
We repeated the experiment 20 times
under all real-world weather conditions on all roads (i.e., 20 runs $\times$ 4 real-world weather conditions $\times$ 4 roads = 320 samples for each of the three strategies).
The comparison analysis was performed with \metricReward using the Mann and Whitney U test and Vargha and Delaney effect size.
Results are summarized in Table~\ref{tab:RQ1}, and the detailed statistical results on each road for the three reward settings can be seen in Table~\ref{tab:Appendix_RQ1_TTCReward}, ~\ref{tab:Appendix_RQ1_DistanceReward}, and ~\ref{tab:Appendix_RQ1_JerkReward} in Appendix~\ref{Appendix}.

As can be seen from Table~\ref{tab:RQ1}, \deepqtest with all settings significantly outperformed \randomStrategy and \greedyStrategy in terms of all metrics.
\begin{table}
    \centering
    \small
    \caption{Results of pair comparison of \deepqtest and RS/GS in terms of \metricReward using the Vargha and Delaney statistics and the Mann–Whitney U test - RQ1}
    \begin{threeparttable}
    \resizebox{.9\textwidth}{!}{
\input{tables/RQ1_Roads_Wea}
    }
        \begin{tablenotes}
            \scriptsize
            \item[$^{*}$] $p$ is \textit{p-value}; a bold $\hat{A}_{12}$ with a $p < 0.05$ implies that \deepqtest is significantly better than \textit{GS/RS}. 
            $\hat{A}_{12}$ magnitude: \textit{negligible} ($\hat{A}_{12}$ $\in$ (0.444, 0.556)), \textit{small} ($\hat{A}_{12}$ $\in$ [0.556, 0.638) or (0.362, 0.444]), \textit{medium} ($\hat{A}_{12}$ $\in$ [0.638, 0.714) or (0.286, 0.362]), \textit{large} ($\hat{A}_{12}$ $\in$ [0.714, 1.0] or [0, 0.286]).
        \end{tablenotes}
    \end{threeparttable}
    \label{tab:RQ1}
\end{table}
In terms of $\hat{A}_{12}$, we found that for \metricTTC, 
and \metricCollisionTime
, $\hat{A}_{12}$ is at \textit{large} magnitude level (i.e., $\geq$ 0.714 or $\leq$ 0.286) for all the cases. For \metricDTO, $\hat{A}_{12}$ is at the \textit{large} magnitude level (i.e., $\leq$ 0.286) except for \rewardJerk with which we observed the \textit{small} magnitude level results. For \metricJerk, $\hat{A}_{12}$ is at least at the \textit{large} magnitude level (i.e., $\geq$ 0.714) for \rewardTTC, the \textit{medium} magnitude level (i.e., $\geq$ 0.638) for \rewardDTO, and the \textit{small} magnitude level for \rewardJerk (i.e., $\geq$ 0.556). For the \metricCollisonNum metric, we can observe that $\hat{A}_{12}$ is at the \textit{large} magnitude level (i.e., $\geq$ 0.714) for \rewardTTC and \rewardDTO, and at \textit{medium} magnitude level (i.e., $\geq$ 0.638) for \rewardJerk.

\begin{figure}
    \centering
    \includegraphics[width=\textwidth]{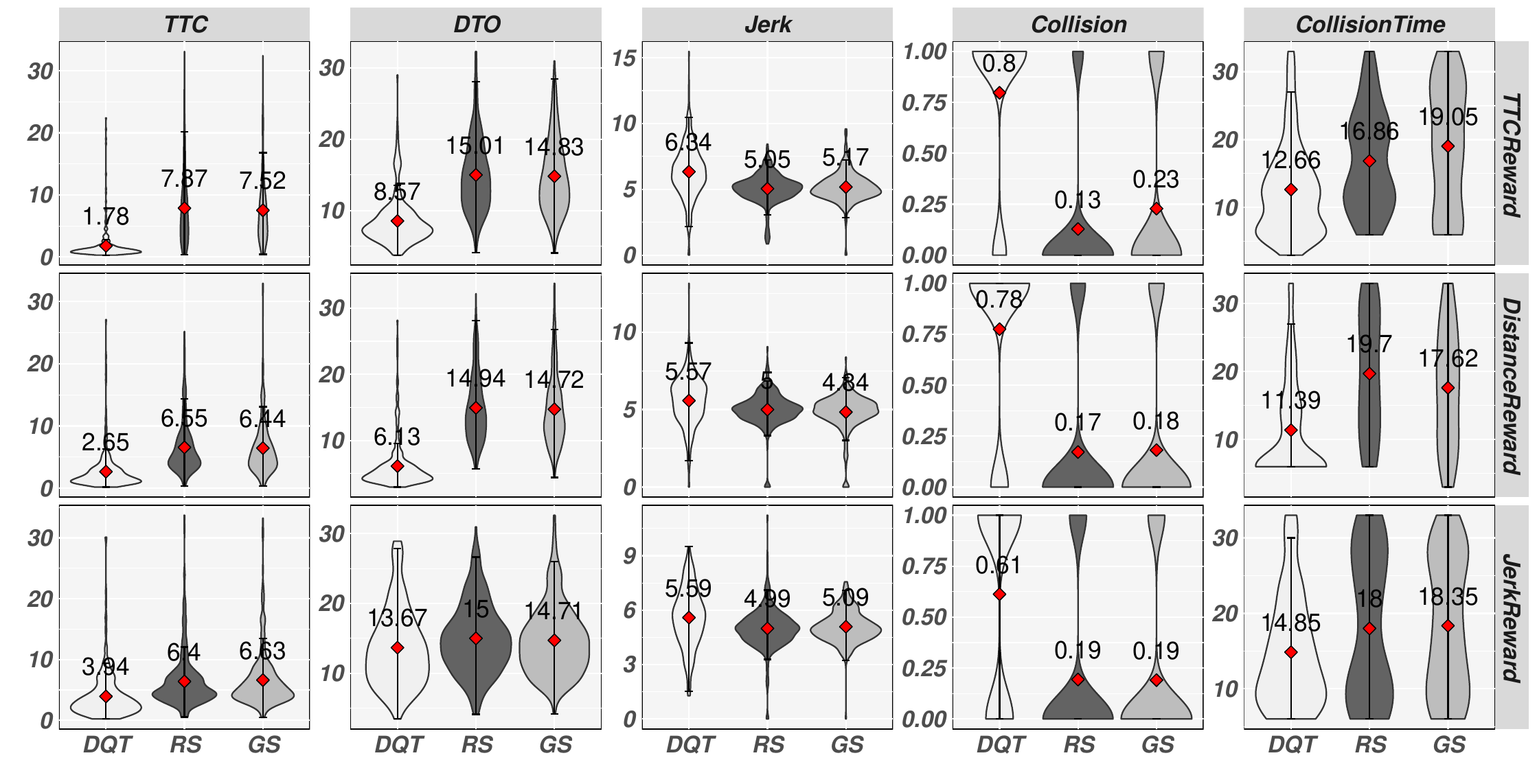}
        \scriptsize {* $DQT$: \deepqtest; \randomStrategy: \textit{Random Strategy}; \greedyStrategy: \textit{Greedy Strategy}. 
        }
    \caption{Descriptive statistics of the evaluation metrics achieved by \deepqtest, \randomStrategy and \greedyStrategy -- RQ1}
    \label{fig:RQ1}
\end{figure}

In addition, to study the distributions of each metric, we present violin plots for \deepqtest, \randomStrategy, and \greedyStrategy for each reward setting in Figure~\ref{fig:RQ1}. From the figure, we can observe that all settings of \deepqtest outperformed \randomStrategy and \greedyStrategy in terms of the average values (red dots in Figure~\ref{fig:RQ1}) of all evaluation metrics. 
We looking at the distributions, we can see that \deepqtest achieved less variability, except for \metricJerk, implying that \deepqtest performed more stably in terms of \metricTTC, \metricDTO, \metricCollisonNum, and \metricCollisionTime. For \metricJerk, although \deepqtest performed less stably, it still achieved higher \metricJerk values on average. One plausible explanation is that \rewardJerk may not be as effective as \rewardTTC, \rewardDTO) in terms of training a stable \deepqtest model. More discussions about this observation will be provided in Section~\ref{subsubsec:rq2_results}. 

\begin{center}
    \fcolorbox{black}{gray!10}{
    \parbox{\textwidth}{
    Conclusion for RQ1: 
    Compared to \randomStrategy and \greedyStrategy, \deepqtest achieved significant improvements in terms of all \metricReward.
    This represents that \deepqtest is more effective in generating environment configurations that lead to shorter time to collide with obstacles (\metricTTC), shorter distance to obstacles (\metricDTO), a lower degree of passenger comfort (\metricJerk), and more collision occurrences (\metricCollisonNum), and shorter (\metricCollisionTime). In summary, the problem we address is complex enough, which cannot effectively be addressed with \randomStrategy and \greedyStrategy. 
    }}
\end{center}

\subsection{Results for RQ2}\label{subsubsec:rq2_results}

To answer RQ2, we compare the performance achieved by each of the models under three reward function settings (i.e., \rewardTTC, \rewardDTO, \rewardJerk) on all four roads for each weather condition (i.e., \textit{RD}, \textit{RN}, \textit{SD}, \textit{SN}).
We repeated the experiment 20 times. Thus, for each weather condition, there are in total of 80 (i.e., 20 runs $\times$ 4 Roads) samples for each reward function.
Table~\ref{tab:RQ2} reports the results of pair-wise comparisons of the three reward functions in terms of \metricReward and \metricDiversity for each weather condition. More detailed results for each road can be found in Table~\ref{tab:RQ2_Appendix} in Appendix.

Regarding \metricReward, for all the weather conditions, we found that \rewardTTC and \rewardDTO consistently achieved the best performance; \rewardTTC is the most effective in generating environment configurations that lead to less time for \avut to collide with obstacles (\metricTTC) while \rewardDTO is the most effective in generating environment configurations that lead to less distance between \avut and obstacles (\metricDTO), as expected. 
For \metricJerk, it is surprising that the best results are achieved by \rewardTTC but not \rewardJerk. 
Recall that \metricJerk is calculated as the rate of changes of acceleration, therefore, to figure out the reason that \rewardTTC outperformed \rewardJerk, we further analyzed the changes in acceleration and the corresponding \metricJerk values by checking scenarios generated by \rewardJerk and \rewardTTC. We found that \rewardJerk generates scenarios that result in higher \metricJerk over a period of time. For instance, a sequence of \textit{acceleration ($m/s^2$)}/\textit{Jerk} values over a period of time can be: 

{\centering $T_0$: 2.70/2.44, $T_1$: 0.70/3.82, $T_2$: -2.22/6.12, $T_3$: 1.73/10.42.\par}

\noindent However, \rewardTTC could generate scenarios where \avut collides with obstacles in a less period of time,
and such scenarios would likely result in a higher \metricJerk value, i.e., a lower degree of comfort for passengers. For instance, in a run of \rewardTTC, a sequence of \textit{acceleration ($m/s^2$)}/\metricJerk values over three steps is: 

{\centering $T_0$: 3.30/4.64, $T_1$: -0.91/8.58, $T_2$: 2.18/8.02.\par}

\noindent There might exist correlations among reward functions based on their corresponding metrics.
To study the correlations, for each reward function, we calculated \metricTTC, \metricDTO, \metricJerk at each step for all 20 runs on all of the four roads, then analyzed results using Spearman's rank correlation ($\rho$) test.
Results show a significantly positive correlation between \metricTTC and \metricDTO (i.e., $p < 0.05$ and $\rho=0.423$), and significantly negative correlations observed in \metricTTC-\metricJerk (i.e., $p < 0.05$ and $\rho=-0.254$) and \metricDTO-\metricJerk (i.e., $p < 0.05$ and $\rho=-0.292$). 
It indicates that a lower \metricTTC (less time to collide with obstacles) would lead to a lower \metricDTO (i.e., less distance to obstacles) and a higher \metricJerk (i.e., a lower degree of comfort). 

In addition, we observed similar pair comparison results among reward functions for metrics \metricCollisonNum and \metricCollisionTime. No significant difference is observed between \rewardTTC and \rewardDTO under all four kinds of weather conditions. 
Both \rewardTTC and \rewardDTO outperformed \rewardJerk consistently on the four weather conditions, and the differences are significant on \textit{RN}, \textit{SD}, and \textit{SN}.
It indicates that reward functions based on \metricTTC and \metricDTO might have more chances to guide RL to introduce environment configurations that result in collisions.

In Table~\ref{tab:RQ2}, we also analyzed the diversity among the scenarios generated by each reward function in each weather condition. Results show that \rewardJerk performed the best in terms of both diversity-related metrics, i.e., invoking more diverse APIs (\metricDiversityAPI) and generating more diverse scenarios (\metricDiversityScenario). More specifically, \rewardJerk outperformed \rewardTTC and \rewardDTO in terms of \metricDiversityAPI and \metricDiversityScenario in all weather conditions, except \metricDiversityAPI in \textit{RD}. As for \metricDiversityAPI in \textit{RD}, \rewardJerk outperformed \rewardDTO, and \rewardJerk and \rewardTTC performed equivalently in statistics. Regarding \rewardTTC and \rewardDTO, there is no significant difference in terms of \metricDiversityAPI and \metricDiversityScenario. But in terms of \metricDiversityAPI, \rewardTTC is slightly better than \rewardDTO as $\hat{A}_{12}$ of comparing \rewardTTC with \rewardTTC is greater 0.5 in all weather conditions.

\begin{table}
    \centering
        \caption{Results of pair comparisons among the three reward functions in terms of \metricReward and \metricDiversity using the Vargha and Delaney statistics and the Mann–Whitney U test - RQ2}
    \begin{threeparttable}
    \resizebox{\textwidth}{!}{
\input{tables/RQ2_1_MergeRoad}
}
        \begin{tablenotes}
            \scriptsize
            \item[$^{*}$] \textit{RD}: \textit{Rainy Day}; \textit{RN}: \textit{Rainy Night}; \textit{SD}: \textit{Sunny Day}; \textit{SN}: \textit{Sunny Night}; 
            $p$ is \textit{p-value}; a bold $\hat{A}_{12}$ with a $p < 0.05$ implies that the former model is significantly \\better than the latter, whereas a $\hat{A}_{12}$ decorated by symbol "×" indicates the former model performed significantly worse than the latter; a $p > 0.5$ means no\\ significant difference between two models.
            $\hat{A}_{12}$ magnitude: \textit{negligible} ($\hat{A}_{12}$ $\in$ (0.444, 0.556)), \textit{small} ($\hat{A}_{12}$ $\in$ [0.556, 0.638) or (0.362, 0.444]), \textit{medium} \\($\hat{A}_{12}$ $\in$ [0.638, 0.714) or (0.286, 0.362]), \textit{large} ($\hat{A}_{12}$ $\in$ [0.714, 1.0] or [0, 0.286]).
        \end{tablenotes}
    \end{threeparttable}
    \label{tab:RQ2}
\end{table}

\begin{table}[htbp]
    \small
    \centering
        \caption{Ranking of \rewardTTC, \rewardDTO, and \rewardJerk in terms of metrics and real-world weather conditions - RQ2}
    \begin{threeparttable}
    \resizebox{\textwidth}{!}{

\input{tables/RQ2_Ranking.tex}
        }
        \begin{tablenotes}
            \scriptsize
            \item[$^{*}$] $DQT_T$: \rewardTTC, $DQT_D$: \rewardDTO, $DQT_J$: \rewardJerk. \textit{RD}: \textit{Rainy Day}; \textit{RN}: \textit{Rainy Night}; \textit{SD}: \textit{Sunny Day}; \textit{SN}: \textit{Sunny Night}; \\a reward function before "," is statistically significantly better than the reward function after ","; a "/" means two reward functions have no statistically significant \\difference, and a reward function before "/" achieved better (or the same) performance than the reward function after "/" in terms of mean metric values. $Rec_M$:\\ recommended by metric.
        \end{tablenotes}
    \end{threeparttable}
    \label{tab:RQ2_ranking}
\end{table}

Based on Table~\ref{tab:RQ2}, we summarized the results obtained by the three reward functions in each weather condition for each metric using ranking and provided a recommendation of reward function for each metric in Table~\ref{tab:RQ2_ranking}.
For all metrics, we found that the ranking results of the three reward functions do not vary much in weather conditions. 
\rewardTTC achieved the overall best performance in terms of all metrics.

\begin{center}
    \fcolorbox{black}{gray!10}{\parbox{\textwidth}{Conclusion for 
    RQ2: Among all three reward functions (i.e., \rewardTTC, \rewardDTO, and \rewardJerk), \rewardTTC achieved the overall best performance. Regarding \metricReward, \rewardTTC performed the best in terms of \metricTTC and \metricJerk, and the second best in terms of \metricDTO. As for \metricCollisonNum and \metricCollisionTime, \rewardTTC and \rewardDTO both performed the best. Regarding \metricDiversity, \rewardJerk performed the best, followed by \rewardTTC.
    }}
\end{center}

\subsection{Results for RQ3} 

To answer RQ3, we compared \deepqtest with \deepcollision~\cite{lu2022learning} regarding \metricRealism and \metricDiversity. Table~\ref{tab:RQ3.1_realistic-unrealistic} reports results of scenarios in terms of realism along with the occurrence of collisions, achieved by the best reward function (i.e., \rewardTTC) and \deepcollision on all of the four roads (i.e., \textit{R1}--\textit{R4}) under the four weather conditions (i.e., \rewardTTCRD, \rewardTTCRN, \rewardTTCSD, \rewardTTCSN). 
Table~\ref{tab:RQ3.1_statistical_realistic-unrealistic} represents the results of pair comparisons of \rewardTTC and \deepcollision using Vargha and Delaney statistics ($\hat{A}_{12}$) and the Mann–Whitney U test ($p$-value).

The termination criterion of \deepcollision is either to reach the specified destination or to run out of the specified time budget, which allows \avut to continue driving even if a collision happens. But, for \deepqtest, besides reaching the destination or running out of the budget, it also terminates testing once a collision occurs. Thus, compared to \deepqtest, \deepcollision generates more test scenarios, as shown in column \metricRealismTC) in Table~\ref{tab:RQ3.1_realistic-unrealistic}.

As for \metricRealism, based on results shown in Table~\ref{tab:RQ3.1_realistic-unrealistic}, all scenarios generated by \rewardTTC are realistic scenarios (with \metricRealismRCS\% and \metricRealismRNS\% being 100\%, and unrealistic scenarios with \metricRealismUCS\% and \metricRealismUNS\% being 0\%). This is because \deepqtest is designed by considering realistic constraints (Section~\ref{subsec:configurableparameters}), and all scenarios generated by \deepqtest conform to these constraints. Instead, \deepcollision achieved a limited performance in generating realistic scenarios on R1, R3, and R4: the percentages of generated realistic scenarios (\metricRealismRCS\% and \metricRealismRNS\%) are 40.93\% (0.55\% + 40.38\%), 3.49\% (0.38\% + 3.44\%) and 2.68\% (0.04\% + 2.64\%), respectively. On \textit{R2}, \deepcollision generated 89.36\% realistic scenarios out of the 3440 scenarios. 

Regarding realistic collision scenarios, \rewardTTC generated more realistic collision scenarios (\#\metricRealismRCS) than \deepcollision on \textit{R2}, \textit{R3} and \textit{R4}. Especially for \textit{R3} and \textit{R4}, with 20 runs, \rewardTTC generated a minimum of 13 and a maximum of 18 realistic collision scenarios, while \deepcollision produced only 2 realistic collision scenarios. On \textit{R1}, although \deepcollision generated 19 realistic collision scenarios, the percentage of realistic collision scenarios (i.e., \metricRealismRCS\% = 0.55\%) is much smaller than that for \deepqtest (i.e., at least 2.02\%), implying that on R1, \deepcollision needed to generate a higher number of scenarios (3440) to get a comparable number of realistic collision scenarios as \deepqtest. Besides, the proportions of collision scenarios among all realistic scenarios achieved by \deepcollision are all lower than \deepqtest. 
In addition, the significant outperformance of \deepqtest over \deepcollision, with large effect size magnitude, can be observed in Table~\ref{tab:RQ3.1_statistical_realistic-unrealistic}, indicated as $p$-$value<0.05$ and $\hat{A}_{12} > 0.825$. On \textit{R1}, \deepcollision generated the most realistic collision scenarios (i.e., 19), but the difference is modest (i.e., 19 vs. 16) and not significant (i.e., $p$-$value >0.05$ and $\hat{A}_{12}\in [0.425, 0.450]$). 
Note that the low percentage of realistic collision scenarios (\metricRealismRCS\%) by \rewardTTC (i.e., the maximum is 3.44\% with \rewardTTCSD on \textit{R3}) relates to how we identify and collect test scenarios. 
As a change in the operating environment can impact ADS continuously in the real world, we identify a test scenario with a time span \textit{ST} that is composed of a successive sequence of \textit{scenes} (see Equation~\ref{eq:scenario}), then collect test scenarios every time step (\textit{t}) in one execution (see an example in Figure~\ref{fig:div_scenario}).
Such identification and collection aim to depict the test scenarios over time in detail, but it increases the number of collected test scenarios.
Thus, it might result in low \metricRealismRCS\% as one execution by \deepqtest can observe at maximum one collision scenario. 

We further analyzed the average realistic collision time (i.e., \metricRealismRCTRCS). Based on \metricRealismRCTRCS in Tables~\ref{tab:RQ3.1_realistic-unrealistic} and ~\ref{tab:RQ3.1_statistical_realistic-unrealistic}, \rewardTTC significantly outperformed \deepcollision on all four roads, i.e., \rewardTTC took less time to generate realistic scenarios that can lead to collisions than \deepcollision. For example, one can observe from Table~\ref{tab:RQ3.1_realistic-unrealistic} that on \textit{R1}, \deepcollision took on average 18 seconds to find a realistic collision scenario, while \deepqtest took 
on average 11.58 seconds (maximum 12.4 with \rewardTTCRD, and minimum 9.92 with \rewardTTCSD).

Regarding the effectiveness of \deepqtest in terms of triggering realistic collision scenarios, we found that, with the results on four roads under four weather conditions, \rewardTTC identified on average 16 realistic collision scenarios out of 20 runs (i.e., 16/20=80\% per run), and it took on average 10.58 seconds per realistic collision scenario.

\begin{table}
    \small
    \centering
    \caption{
    Results of a number of test scenarios (\metricRealismTC), a number/percentage of realistic collision scenarios (\#\metricRealismRCS/\%), average time spent on observing each realistic collision scenario (\metricRealismRCTRCS \textit{seconds}), a number/percentage of unrealistic collision scenarios (\#\metricRealismUNS/\%), a number/percentage of realistic non-collision scenarios (\#\metricRealismRNS/\%) and a number/percentage of unrealistic non-collision scenarios (\#\metricRealismRCS/\%) achieved by \rewardTTC and \deepcollision.}
    \begin{threeparttable}
    \resizebox{\textwidth}{!}{

\input{tables/RQ2.2_realistic-unrealistic.tex}
    }
    \begin{tablenotes}
    \scriptsize
    \item[$^{*}$] $R1$: $Road1$; $R2$: $Road2$; $R3$: $Road3$; $R4$: $Road4$. \textit{RD}: \textit{Rainy Day}; \textit{RN}: \textit{Rainy Night}; \textit{SD}: \textit{Sunny Day}; \textit{SN}: \textit{Sunny Night}.
    \metricRealismTC: \textit{Total Scenarios}; \metricRealismRCS: \\\textit{Realistic Collision Scenario}; \metricRealismRCTRCS: \textit{realistic collision time for} \metricRealismRCS, \metricRealismUCS: \textit{Unrealistic Collision Scenario}; \metricRealismRNS: \textit{Realistic Non-Collision Scenario}, \metricRealismUNS: \\\textit{Unrealistic Non-Collision Scenario}.
    \end{tablenotes}
    \end{threeparttable}
    \label{tab:RQ3.1_realistic-unrealistic}
\end{table}

\begin{table}
    \small
    \centering
    \caption{Results of pair comparisons of \rewardTTC and \deepcollision using Vargha and Delaney statistics ($\hat{A}_{12}$) and the Mann–Whitney U test ($p$-value) in terms of a number/percentage of realistic collision scenarios (\#\metricRealismRCS/\%), average realistic collision time (\metricRealismRCTRCS \textit{seconds}), a number/percentage of unrealistic collision scenarios (\#\metricRealismUNS/\%), a number/percentage of realistic non-collision scenarios (\#\metricRealismRNS/\%) and a number/percentage of unrealistic non-collision scenarios (\#\metricRealismRCS/\%).
    }
    \begin{threeparttable}
        \resizebox{\textwidth}{!}{
\input{tables/RQ2.2_statistical_realistic-unrealistic.tex}}
    \begin{tablenotes}
    \scriptsize
    \item[$^{*}$] $R1$: $Road1$; $R2$: $Road2$; $R3$: $Road3$; $R4$: $Road4$. \textit{RD}: \textit{Rainy Day}; \textit{RN}: \textit{Rainy Night}; \textit{SD}: \textit{Sunny Day}; \textit{SN}: \textit{Sunny Night}.
    \metricRealismRCS: \textit{Realistic Collision Scenario}; \\\metricRealismUCS: \textit{Unrealistic Collision Scenario}; \metricRealismRNS: \textit{Realistic Non-Collision Scenario}, \metricRealismUNS: \textit{Unrealistic Non-Collision Scenario}.
    $p$ is \textit{p-value}; a bold $\hat{A}_{12}$ with a $p < 0.05$ \\implies that $\deepqtest^*$ is significantly better than \deepcollision, whereas a $\hat{A}_{12}$ decorated by symbol "×" indicates $\deepqtest^*$ performed significantly worse \\than \deepcollision; a $p > 0.5$ means no significant difference between $\deepqtest^*$ and \deepcollision.
    $\hat{A}_{12}$ magnitude: \textit{negligible} ($\hat{A}_{12}$ $\in$ (0.444, 0.556)), \textit{small}\\($\hat{A}_{12}$ $\in$ [0.556, 0.638) or (0.362, 0.444]), \textit{medium} ($\hat{A}_{12}$ $\in$ [0.638, 0.714) or (0.286, 0.362]), \textit{large} ($\hat{A}_{12}$ $\in$ [0.714, 1.0] or [0, 0.286]).
    \end{tablenotes}
    \end{threeparttable}

    \label{tab:RQ3.1_statistical_realistic-unrealistic}
\end{table}

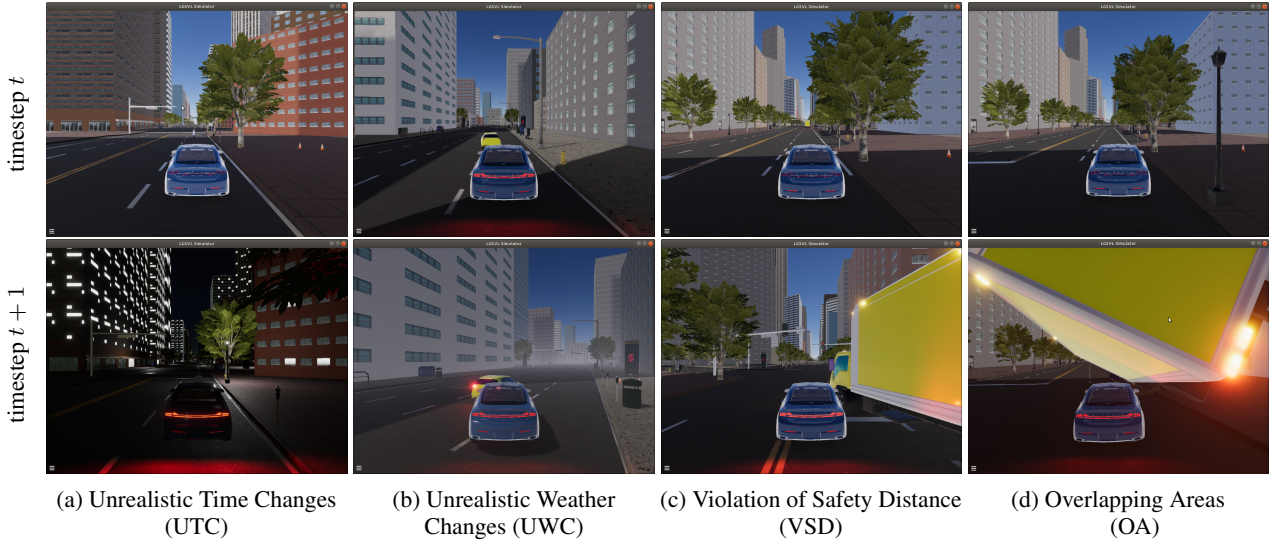
\begin{figure*}
	\flushright
        \resizebox{\textwidth}{!}{
            \input{figs/unrealistic.tex}
        }
        \put(-482, 140){\small \rotatebox{90}{timestep $t$}}
        \put(-482, 45){\small \rotatebox{90}{timestep $t + 1$}}
	\caption{\textbf{Examples of Unrealistic Environment Configurations and the Resulting Unrealistic Driving Scenarios - RQ3}. From left to right are: (a) The \avut was maintaining its lane and a pedestrian was standing on the sidewalk in front of the \avut at timestep $t$, and at timestep $t+1$, an environment configuration action changed the time from day to night and the \avut collided with the pedestrian; (b) The \avut and an NPC vehicle both were trying to change the left lane at timestep $t$, and at timestep $t+1$, an environment configuration action that changed the weather condition to moderate fog was invoked, the invocation of the action caused a collision between the \avut and the NPC vehicle; (c) The \avut was maintaining its lane at timestep $t$, and at timestep $t+1$, a truck was generated whose initial position violated the safety distance from the \avut and caused a collision; (d) The \avut was maintaining its lane at timestep $t$, and at timestep $t+1$, a truck was generated whose area overlaps with the \avut and directly caused a collision.
 }
\label{fig:RQ3.1_unrealistic_pictures}
\end{figure*}

Moreover, we studied unrealistic scenarios generated by \deepcollision by 
replaying them on the four roads.
We found that the generation of unrealistic scenarios is mainly caused by the four types of unrealistic environment configurations, i.e., \unrealisticUTC, \unrealisticUWC, \unrealisticVSD, and \unrealisticOA.
Examples of each type are shown in Figure~\ref{fig:RQ3.1_unrealistic_pictures}. We further explain the four types of unrealistic environment configurations as follows:

\begin{itemize}[noitemsep,topsep=0pt,parsep=0pt,partopsep=0pt]
\item \unrealisticUTC refers to the unrealistic manipulation of time changes. As Figure~\ref{fig:u1} shows, the time was manipulated and suddenly changed from the day (12 am) to the night (8 pm), which caused the \avut to collide with the pedestrian in front. 
\item \unrealisticUWC refers to unrealistic weather changes. As shown in Figure~\ref{fig:u5}, an environment configuration action could immediately shift the weather condition from sunny to foggy. As the foggy weather condition affects the visibility of \avut, such an instant change might leave less time for \avut to react and hence lead to collisions (e.g., colliding with the vehicle in the front in Figure~\ref{fig:u5}). Considering that the weather condition changes over a period of time, such a sudden change is unrealistic.
\item \unrealisticVSD happens when obstacles (e.g., NPC vehicles, pedestrians) are generated with initial positions violating the safety distances of \avut from the obstacles. For testing purposes, we could introduce new obstacles into the driving environment. But such an introduction should be in accord with real-world scenarios. For instance, it is impossible for a truck to suddenly appear beside \avut (Figure~\ref{fig:u9}).
In such a case, \avut did not have time to react and therefore collided with the truck.
As driving tasks on \textit{R3} and \textit{R4} need to go through a T-junction and then turn left from a four-lane road to another four-lane road at the beginning (see Figure~\ref{fig:roads}), we found that \deepcollision generated many collision scenarios by introducing new obstacles when \avut was turning left or entering the new lane. This is a reason why most of the collision scenarios on \textit{R3} and \textit{R4} are unrealistic (see Table~\ref{tab:RQ3.1_realistic-unrealistic}).
\item \unrealisticOA is the case that an obstacle is generated in an area overlapping the areas covered by the \avut and the newly introduced obstacles due to a lack of consideration of the volume of the new obstacles. Such introductions will directly lead to collisions. Figure~\ref{fig:u11} is an example of a collision caused by generating a truck in the same area of \avut. Such introductions of obstacles might mislead the approach by introducing large obstacles such as trucks. 
\end{itemize}

Furthermore, we compared \rewardTTC and \deepcollision in terms of \metricDiversity.
\deepqtest achieved an overall better performance than \deepcollision in terms of \metricDiversity. Results of the pair-wise comparison are shown in Table~\ref{tab:RQ3.2_Diversity_All_Realistic}. In terms of \metricDiversityAPI, we found that \rewardTTC achieved significantly better results than \deepcollision, which indicates that, compared to \deepcollision, \rewardTTC could test \avut under more diverse combinations of environment configurations in one run. In terms of \metricDiversityScenario, \rewardTTC significantly outperformed \deepcollision on \textit{R1} and \textit{R2}. On \textit{R3} and \textit{R4}, \rewardTTC is either significantly better than or statistically equivalent to \deepcollision. 
The results of \metricDiversityScenario indicate that \rewardTTC might have more chance to test \avut under more diverse test scenarios. Note that \metricDiversityAPI measures the diversity of applying environment configurations, while \metricDiversityScenario measures the diversity of test scenarios resulting from the environment configurations. 

\begin{center}
    \fcolorbox{black}{gray!10}{\parbox{\textwidth}{
    Conclusion for RQ3: \rewardTTC outperformed \deepcollision in terms of \metricRealism and \metricDiversity. Concretely, as for \metricRealism, \rewardTTC achieved a higher proportion of realistic scenarios among all the generated ones, and it also generated more realistic collision scenarios with less time.
    On the four roads under the four weather conditions, \rewardTTC identified on average 16 realistic collision scenarios out of 20 runs (i.e., 16/20=80\% per run), and it took on average 10.58 seconds to trigger a realistic collision scenario.
    In addition, 
    \deepqtest is effective in avoiding the generation of four kinds of unrealistic environment configurations.
    Regarding \metricDiversity, \deepqtest achieved an overall better performance than \deepcollision in introducing diverse environment configurations (i.e., \metricDiversityAPI) and generating diverse scenarios (i.e., \metricDiversityScenario).
    }}
\end{center}

\begin{table}
    \small
    \centering
    \caption{Results of pair-wise comparisons of \rewardTTC and \deepcollision in terms of \metricDiversity using the Vargha and Delaney statistics and the Mann–Whitney U test - RQ3}
    \begin{threeparttable}
        \resizebox{\textwidth}{!}{
\input{tables/RQ2.2_Diversity_All_Realistic.tex}
        }
        \begin{tablenotes}
        \scriptsize
        \item[$^{*}$] 
        $R1$: $Road1$; $R2$: $Road2$; $R3$: $Road3$; $R4$: $Road4$. \textit{RD}: \textit{Rainy Day}; \textit{RN}: \textit{Rainy Night}; \textit{SD}: \textit{Sunny Day}; \textit{SN}: \textit{Sunny Night}.
        $p$ is \textit{p-value}; a bold $\hat{A}_{12}$ with a \\$p < 0.05$ implies that \deepqtest is significantly better than \deepcollision, whereas a $\hat{A}_{12}$ decorated by symbol "×" indicates \deepqtest performed significantly \\worse than \deepcollision; a $p > 0.5$ means no significant difference between \deepqtest and \deepcollision.
        $\hat{A}_{12}$ magnitude: \textit{negligible} ($\hat{A}_{12}$ $\in$ (0.444, 0.556)), \\\textit{small} ($\hat{A}_{12}$ $\in$ [0.556, 0.638) or (0.362, 0.444]), \textit{medium} ($\hat{A}_{12}$ $\in$ [0.638, 0.714) or (0.286, 0.362]), \textit{large} ($\hat{A}_{12}$ $\in$ [0.714, 1.0] or [0, 0.286]).
        \end{tablenotes}
    \end{threeparttable}

    \label{tab:RQ3.2_Diversity_All_Realistic}
\end{table}

\subsection{Threats to Validity}\label{subsec:threats}
\textbf{\textit{Internal Validity}}. Threats to internal validity are concerned about the design factors that might have an impact on the conclusion. To migrate the potential threat to the internal validity caused by MDP formulation and DQN parameter settings, we trained the 12 \deepqtest models following identical state encoding, action space, Q-network architecture, learning algorithm, training details, and hyperparameter settings, to ensure fair comparison among them. Another threat to internal validity is related to the potential faults within our algorithm infrastructure. To control this threat, we used the open-source machine learning framework, PyTorch~\cite{paszke2019pytorch}, which has been widely applied and tested in both academics and industry. Furthermore, we realized that the hyperparameter tuning of DQN may have an influence on the performance of \deepqtest, therefore, we followed the default hyperparameter settings from Zhou~\cite{PyTorchTutorial} when training all of the 12 \deepqtest models. We understand that tuning hyperparameters of DQN might improve the performance of \deepqtest, which unfortunately cannot be covered in this paper because it requires extensive and dedicated empirical studies.

\textbf{\textit{External Validity}}. The threat is related to the currently applied subject system and simulator and the possible generalization to other case studies. In our evaluation, we conducted the empirical study with an industrial-scale ADS (i.e., Apollo) and a commonly applied simulator (i.e., LGSVL). We understand that conducting more case studies with different ADSs and simulators would help to better evaluate and generalize DeepTest, and using our already developed REST API endpoints, other ADSs, and simulators can be easily integrated with \deepqtest with minimum effort. However, it is important to mention that experiments on testing ADSs in the simulation are very computationally expensive, which limited the number of case studies we used in our experiment.

\textbf{\textit{Construct Validity}}. Threat to construct validity concerns whether the metrics for evaluation can precisely reflect the variables we want to measure in the experiment. As discussed in Section~\ref{subsec:metrics}, we present three types of metrics (i.e., \metricReward, \metricRealism, and \metricDiversity) including 16 comparable metrics (e.g., \metricTTC and \metricCollisonNum) to measure the effectiveness and efficiency of \deepqtest. 

\textbf{\textit{Conclusion Validity}}. The threat to conclusion validity is related to the use of appropriate analytical methods and the reliability of conclusions. To draw a reliable conclusion, we followed a rigorous procedure to collect the data and employed statistical methods to analyze the collected data. Specifically, 20 independent runs of each experiment setting were performed to account for randomness.
In addition, following the guideline by Arcuri and Briand~\cite{arcuri2011practical}, we performed pairwise comparisons with the Mann-Whitney U test and used the Vargha-Delaney $\hat{A}_{12}$ metric for effect size.

\subsection{Data Availability}\label{subsec:data}
To enable the reproducibility of our research findings, we provide the replication package in our GitHub online repository\footnote{\url{https://github.com/Simula-COMPLEX/DeepQTest}}. The replication package contains the following contents: 1) all related algorithms and source code for conducting the experiment; 2) All the raw data for the experiment results and analyses; and 3) The REST API endpoints for configuring the environment.

\section{Discussion}\label{sec:discussion}
\textbf{\textit{Recording and replaying scenarios for further diagnosis and analysis}}. Due to the complexity and uncertainty of the operating environment, as well as the complexity of the driving tasks, it is always very time-consuming to use simulations to evaluate scenarios of an ADS testing approach. Thus, it is necessary to record the generated scenarios to mitigate the cost for further possible analysis and usage. For example, the recorded scenarios could be utilized for supporting further diagnoses of ADSs, and can also be used for regression testing.

To facilitate the recording and reusing process of test scenarios, we first proposed a Domain-Specific Language (DSL) for scenario representation and evaluation, named \textit{DeepScenario}~\cite{10174023}, which is based on the \textit{Driving Scenario} definition in Section~\ref{sec:evaluation}. Additionally, to support automatically collecting and replaying driving scenarios when testing ADSs by \deepqtest, we developed an \textit{ScenarioCollector} and an \textit{ScenarioRunner} to easily collect and replay scenarios using \deepqtest. With \textit{ScenarioCollector} and \deepqtest, driving scenarios that have been used for testing ADSs can be automatically recorded for further usage. Meanwhile, by selecting and replaying specific types of driving scenarios (e.g., collision scenarios) through \textit{ScenarioRunner}, \deepqtest can better facilitate diagnoses of ADSs.

There are existing works in the literature that study scene representation and propose scene standards and support toolsets. For instance, GeoScenario~\cite{queiroz2019geoscenario} is an open DSL for autonomous driving scenario representation, and it has a toolset available that can support the design, evaluation, and execution of scenarios. ASAM OpenSCENARIO~\cite{openscenario} is a file format for describing the dynamic content of driving and traffic simulators. Scenarios defined following the OpenSCENARIO standard can be executed by Carla ScenarioRunner~\cite{dosovitskiy2017carla}. Both specifications can be used for scenario representation and execution, however, the process of scenario definition requires a lot of manual work, which is time-consuming. \deepqtest automates both the scenario collection and execution process using \textit{ScenarioCollector} and \textit{ScenarioRunner}.

Finally, the recorded driving scenarios can also be used for regression testing of advanced ADS versions. Specifically, each scenario is associated with several attributes characterizing its execution results. By using search-based techniques (e.g., genetic algorithms), these associated attributes can support for selection and prioritizing of scenarios for regression testing of ADSs.
The recorded driving scenarios and toolset for collecting and replaying these scenarios can be found in the replication package which is listed in Section~\ref{subsec:data}.

\textbf{\textit{Multi-objective DeepQTest}}. In the current design of \deepqtest, only one objective (i.e., reward) is used to guide the test scenarios generation process. In practice, however, multiple objectives are often considered due to the multi-objective nature of many sequential decision and adaptive optimization problems~\cite{liu2014multiobjective}. In addition, from the results of RQ2 in Section~\ref{subsubsec:rq2_results}, correlations can be observed among the three selected reward functions, i.e., positive correlation between \metricTTC and \metricDTO, negative correlation between \metricTTC (\metricDTO) and \metricJerk. This observation inspires us to investigate different designs of \deepqtest which can optimize multi-objectives at one time. Considering that the selected rewards are to some extent interrelated, one possible solution can be deriving one objective by combining the multiple objectives together. 
However, if the considered objectives conflict with each other, and we want to achieve a trade-off between the conflicting objectives, then multi-objective reinforcement learning (MORL) can be utilized to find a policy that can optimize multiple objectives simultaneously~\cite{vamplew2011empirical}. We think that there is a need to provide a multi-objective strategy for \deepqtest in the future.

\section{Conclusion and Future work}\label{sec:conclusion}
This paper proposed a reinforcement learning (RL)-based ADS testing approach, named \deepqtest, which adaptively learns critical configurations of the operating environment to generate test inputs for ADSs. \deepqtest employs Deep Q-Learning (DQN) as the RL solution and adopts three reward functions as the learning guidelines. To extract features from a high-dimensional operating environment, \deepqtest utilizes multi-modal sensor fusion as the state encoding. Furthermore, the environment configuration processes are realized through a list of REST API endpoints specifying which environment parameters will be configured and their values. In addition, the realism of the environment configurations is ensured with a list of predefined realistic constraints and the real-world effect generator, which maps real-world weather and time to simulation. To assess the cost-effectiveness of \deepqtest, we compared \deepqtest with three baselines, i.e., random, greedy, and a start-of-the-art RL-based approach namely \deepcollision. Results show that \deepqtest significantly outperformed random and greedy strategies in terms of all selected metrics. Compared to \deepcollision, \deepqtest is capable of generating more realistic scenarios leading to collisions for testing ADSs. We also compared different reward functions to study their performance, and \textit{Time-To-Collision} achieved the overall best results that is set as a recommended reward function in \deepqtest.

In the future, we plan to study the generalization of \deepqtest with more case studies and conduct additional experiments on more ADSs and simulators. In addition, investigating algorithm settings and tuning hyperparameter settings is one of our future works. We plan to employ automatic hyperparameter optimization frameworks (e.g., Optuna~\cite{akiba2019optuna} and Tune~\cite{liaw2018tune}) to facilitate this process. We also plan to study different RL solutions and the possibility of integrating these solutions with \deepqtest. Furthermore, environmental uncertainties are crucial factors for the safety and reliability of ADS, thus, systematically introducing and quantifying uncertainties in the operating environment is another future work. Finally, \deepqtest was evaluated with simulators and the generated driving scenarios tested ADS with the simulated vehicle in a simulated operating environment, while it is important to evaluate the effectiveness of generated scenarios in various testing contexts, e.g., hardware-in-the-loop and physical-world ADS testing. Studying the transferability of virtual testing to other ADS testing contexts is another work in the future.




\section*{ACKNOWLEDGMENTS}
This work is supported by the Co-evolver project (No. 286898/F20) funded by the Research Council of Norway.
Man Zhang is funded by the European Research Council (ERC) under the European Union’s Horizon 2020 research and innovation program (grant agreement No 864972).

\bibliographystyle{plain}
\bibliography{main}  

\clearpage
\section*{Appendix}\label{Appendix}

\begin{algorithm}[htbp]
\small
  \caption{Pseudo-Code of \deepqtest Environment Configuration Learning}
  \label{alg::baseline_deepqtest}
  \input{algos/algo-deepqtest.tex}
\end{algorithm}

\begin{algorithm}[htbp]
\small
  \caption{Pseudo-Code of \textit{Scenario Similarity} Calculation}
  \label{alg::calculate_diversity}
  \input{algos/algo-calculate_diversity.tex}
\end{algorithm}

\begin{table}[htbp]
    \centering
    \small
        \caption{Results of the Vargha and Delaney statistics and the Mann–Whitney U test ($TTC_{Reward}$) - RQ1}
\input{tables/Appendix_RQ1_TTCReward.tex}
    \label{tab:Appendix_RQ1_TTCReward}
\end{table}

\begin{table}[htbp]
    \centering
    \small
        \caption{Results of the Vargha and Delaney statistics and the Mann–Whitney U test ($DTO_{Reward}$) - RQ1}
\input{tables/Appendix_RQ1_DistanceReward.tex}
    \label{tab:Appendix_RQ1_DistanceReward}
\end{table}

\begin{table}[htbp]
    \centering
    \small
        \caption{Results of the Vargha and Delaney statistics and the Mann–Whitney U test ($Jerk_{Reward}$) - RQ1}
\input{tables/Appendix_RQ1_JerkReward.tex}
    \label{tab:Appendix_RQ1_JerkReward}
\end{table}

\begin{table}[htbp]
    \centering
        \caption{Results of the Vargha and Delaney statistics and the Mann–Whitney U test - RQ2}
    \resizebox{\textwidth}{!}{

\input{tables/RQ2_Appendix.tex}
    }
    \label{tab:RQ2_Appendix}
\end{table}

\end{document}

%% file: algos/algo-ttc_calculation.tex
  \begin{algorithmic}[1]
    \Require
      \Statex Let \textit{D} be a set of dynamic obstacles;
      \textit{T} be time period an environment configuration action will last;
      \textit{dt} be time step between simulation steps
    \Ensure
      \Statex A buffer of TTC between AVUT with all dynamic obstacles
    \Statex
    \State $t \leftarrow 0$
    \State $TTC_{buff} \leftarrow []$ \Comment{initialize an empty buffer to store TTC}
    \While {t \textless T}
      \State t = t + dt;
      \State $Pos_{avut}, Velo_{avut} \leftarrow getPosition(avut)$ \Comment{get position and velocity of avut}
      \For{each $d \in |D|$}
      \State $Pos_{d}, Velo_{d} \leftarrow getPosition(d)$ \algorithmiccomment{get position and velocity of dynamic obstacle d}
      \State $Pos_{coll} \leftarrow collisionPrediction(Pos_{avut}, Velo_{avut}, Pos_{d}, Velo_{d})$ 
      \Statex \algorithmiccomment{Predict the collision point between AVUT and d}
      \State $TTC_{avut,d} \leftarrow ttcEstimation(Pos_{avut}, Velo_{avut}, Pos_{d}, Velo_{d}, Pos_{coll})$ 
      \Statex \algorithmiccomment{Estimate time to collision between AVUT and d}
      \State Put $TTC_{avut,d}$ into $TTC_{buff}$
      \EndFor
    \EndWhile\\
    \Return $TTC_{buff}$
  \end{algorithmic}

%% file: tables/RQ_descriptions.tex
\begin{tabular}{lllllll}
\toprule
RQs & Tasks                                                                                                                                                                                                                          & Approaches                                                                                                                 & Roads                                                     & Weather                                                                                 & Metrics (Section~\ref{subsec:metrics})                                                                                                                                 & Statistical Test \\ \midrule
RQ1 & \begin{tabular}[c]{@{}l@{}}Compare \deepqtest with \randomStrategy and \greedyStrategy\end{tabular}                                                                             & \begin{tabular}[c]{@{}l@{}}\deepqtest\\ \randomStrategy, \greedyStrategy\end{tabular}                                                 &                                                           &                                                                                         & \begin{tabular}[c]{@{}l@{}}\metricReward\end{tabular}                                  & \begin{tabular}[c]{@{}l@{}}Mann-Whitney U test \\ Vargha-Delaney A measure\end{tabular} \\ \cmidrule{1-3} \cmidrule{6-7} 
RQ2 & Identify the best reward function of \deepqtest                                                                                                                                                                 & \begin{tabular}[c]{@{}l@{}}\rewardTTC\\ \rewardDTO\\ \rewardJerk\end{tabular} & \begin{tabular}[c]{@{}l@{}}\textit{R1}\\ \textit{R2}\\ \textit{R3}\\ \textit{R4}\end{tabular} & \begin{tabular}[c]{@{}l@{}}\textit{Rainy Day}\\ \textit{Rainy Night}\\ \textit{Sunny Day}\\ \textit{Rainy Day}\end{tabular} & \begin{tabular}[c]{@{}l@{}}\metricReward\\ \metricDiversity\end{tabular} & \begin{tabular}[c]{@{}l@{}}Mann-Whitney U test \\ Vargha-Delaney A measure \\Spearman's rank correlation \\coefficient
\end{tabular}  \\ \cmidrule{1-3} \cmidrule{6-7} 
RQ3 & \begin{tabular}[c]{@{}l@{}}Compare \deepqtest with the state-of-the-art \\in terms of the effectiveness of detecting unrealistic \\collision scenarios, and the diversity of generated \\scenarios.\end{tabular} & \begin{tabular}[c]{@{}l@{}}\rewardTTC\\ \deepcollision\end{tabular}                          &                                                           &                                                                                         & \begin{tabular}[c]{@{}l@{}}\metricRealism\\ \metricDiversity\end{tabular}                                  & \begin{tabular}[c]{@{}l@{}}Mann-Whitney U test \\ Vargha-Delaney A measure\end{tabular} \\ \bottomrule
\end{tabular}


%% file: algos/algo-RS.tex





\begin{algorithmic}[1]
\Require
  \Statex Let \textit{TEnv} be the test environment; \textit{AS} be the environment configuration action space; \textit{terminate} be the stopping criteria.
\Ensure
  \Statex $Act_{list}$: a list of actions selected from \textit{AS}
  \Statex

\State $Act_{list}$ $\leftarrow$ []
\State \textit{s} $\leftarrow$ observe a state from \textit{TEnv}

\While{$\neg$ \textit{terminate}}
\State $act$ $\leftarrow$ randomly select an action from $AS$
\State \textit{terminate}, \textit{reward}, $s'$ $\leftarrow$ step($act$, \textit{s})
\algorithmiccomment{execute the selected action in state \textit{s}}
\State Put $act$ into $Act_{list}$
\State $s \leftarrow s'$

\EndWhile
\State \Return $Act_{list}$

\end{algorithmic}

%% file: algos/algo-GS.tex
\begin{algorithmic}[1]
\Require
  \Statex Let \textit{TEnv} be the test environment; \textit{AS} be the environment configuration action space; \textit{terminate} be the stopping criteria.

\Ensure
  \Statex $Act_{list}$: a list of actions selected from \textit{AS}
  \Statex

\State $Act_{list}$ $\leftarrow$ []
\State \textit{s} $\leftarrow$ observe a state of \textit{TEnv}
\State $TEnv_{pre}$ $\leftarrow$ SaveEnv(\textit{TEnv, s}) \algorithmiccomment{save the \textit{TEnv} with state $s$}
\While{$\neg$ \textit{terminate}}
\State \textit{reward}$_{max}$ $\leftarrow$ 0
\State $act_{optimal}$ $\leftarrow$ 0
\For{each action \textit{act} in \textit{AS}} \algorithmiccomment{try each action in the action space}
\State \_, \textit{reward}, \_ $\leftarrow$ step(\textit{act, s})
\If{\textit{reward} \textgreater \textit{reward}$_{max}$}
\State \textit{reward}$_{max}$ $\leftarrow$ \textit{reward}
\State $act_{optimal}$ $\leftarrow$ \textit{act}
\EndIf
\State \textit{TEnv} $\leftarrow$ RollBackEnv($TEnv_{pre}$, $s$)
\EndFor
\State \textit{terminate}, $reward$, $s'$ $\leftarrow$ step($act_{optimal}$) \algorithmiccomment{execute the action that achieves the maximum reward}
\State Put $act_{optimal}$ into $Act_{list}$
\State $TEnv_{pre}$ $\leftarrow$ SaveEnv(\textit{TEnv, $s'$})
\State $s \leftarrow s'$ \algorithmiccomment{move to the next environment state}

\EndWhile
\State \Return $Act_{list}$

\end{algorithmic}

%% file: tables/real-world-effect.tex
\begin{tabular}{ccccc}
\toprule
        & Rainy Day (RD)            & Rainy Night (RN)         & Sunny Day (SD) & Sunny Night (SN)\\ \midrule
Date    & 2021-08-07              & 2021-08-07             & 2021-07-24   & 2021-07-24    \\ \midrule
Time    & 8:00:00              & 20:00:00             & 8:00:00   & 20:00:00    \\ \midrule
Description & Heavy intensity rain & Heavy intensity rain & Clear sky     & Clear sky \\ \bottomrule     
\end{tabular}

%% file: figs/fig_4-roads.tex
	\begin{subfigure}[b]{0.24\textwidth}
		\centering
		\includegraphics[width=0.6\textwidth]{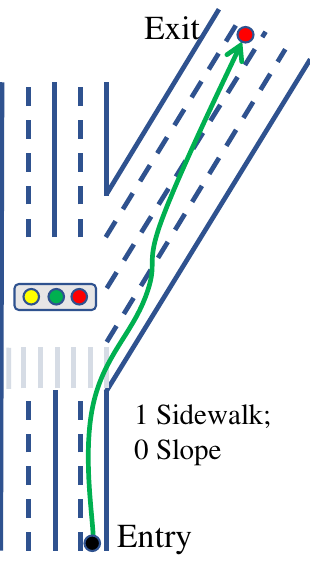}
		\caption{R1 (1.0km)}
		\label{fig:r1}
	\end{subfigure}
	\hfill
	\begin{subfigure}[b]{0.24\textwidth}
		\centering
		\includegraphics[width=0.8\textwidth]{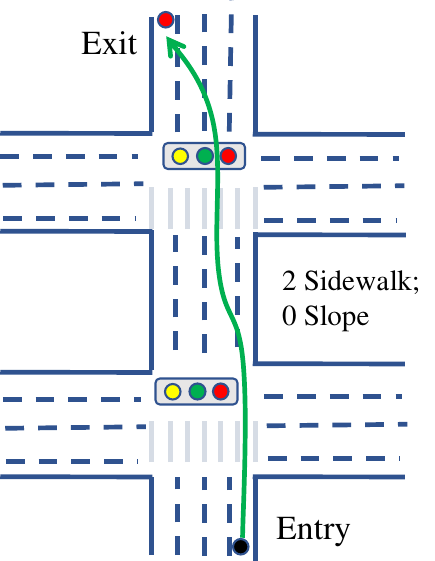}
		\caption{R2 (1.2km)}
		\label{fig:r2}
	\end{subfigure}
	\hfill
	\begin{subfigure}[b]{0.24\textwidth}
		\centering
		\includegraphics[width=0.6\textwidth]{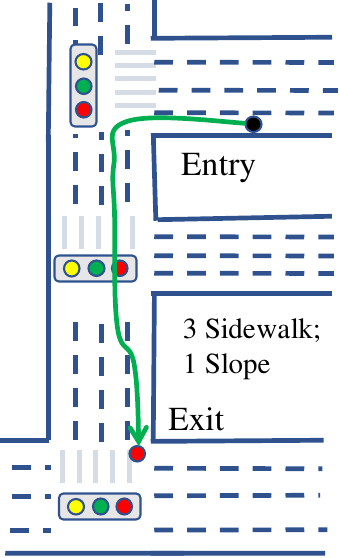}
		\caption{R3 (1.0km)}
		\label{fig:r3}
	\end{subfigure}
	\hfill
	\begin{subfigure}[b]{0.24\textwidth}
		\centering
		\includegraphics[width=0.7\textwidth]{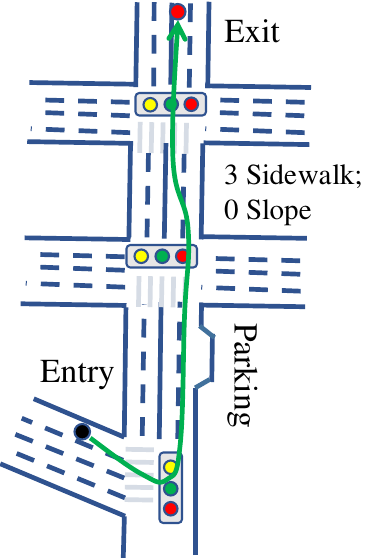}
		\caption{R4 (1.5km)}
		\label{fig:r4}
	\end{subfigure}

%% file: tables/scene_definition.tex
\begin{tabular}{llrl}
\toprule
\#                 &                                                                                                 & \multicolumn{1}{l}{Property} & Property Value                                                                                                                                                                                  \\ \midrule
1                  & \multirow{2}{*}{\begin{tabular}[c]{@{}l@{}}AVUT\end{tabular}} & Operation                      & Stop; Cruise; SpeedUp; SpeedCut; EmergencyBrake; SwitchLaneToRight; SwitchLaneToLeft; TurnLeft; TurnRight                                                                                            \\ \cmidrule{4-4}
2                  &                                                                                                 & Speed (m/s)                          & Zero (0 \textless speed  $\leq$ 1); Slow (1 \textless speed $\leq$ 5); Moderate (5 \textless speed $\leq$ 12); Fast (speed \textgreater 12)   \\  \cmidrule{2-4}
3                  & \multirow{13}{*}{Env}                                                                   & Weather: rain                  & None (rain$_{level}$ = 0); Light (0 \textless rain$_{level} \leq$ 0.2); Moderate (0.2 \textless rain$_{level} \leq$ 0.5); Heavy (rain$_{level} > 0.5$)   \\ \cmidrule{4-4}
4                  &                                                                                                 & Weather: fog                   & None (fog$_{level}$ = 0); Light (0 \textless fog$_{level} \leq$ 0.2); Moderate (0.2 \textless fog$_{level} \leq$ 0.5); Heavy (fog$_{level} > 0.5$)                                    \\ \cmidrule{4-4}
5                  &                                                                                                 & Weather: wetness               & None (wetness$_{level}$ = 0); Light (0 \textless wetness$_{level} \leq$ 0.2); Moderate (0.2 \textless wetness$_{level} \leq$ 0.5); Heavy (wetness$_{level} > 0.5$)                    \\ \cmidrule{4-4}
6                  &                                                                                                 & Time of the day                & Morning (10am); Noon (12am); Night (8pm)                                                                                                                                                     \\ \cmidrule{4-4}
\multirow{4}{*}{7} &                                                                                                 & \multirow{4}{*}{NPC vehicle*}  & \textit{Volume ($m^3$)}: Small (volume  $\leq$ 10); Medium (10 < volume $\leq$ 60); Large (volume \textgreater 60) \\
                   &                                                                                                 &                                & \textit{Operation}: Stop; SwitchLane; LeftLaneDriving; RightLaneDriving; CurrentLaneDriving   \\
                   &                                                                                                 &                                & \textit{Speed (m/s)}: Zero (0 \textless speed $\leq$ 1); Slow (1 \textless speed $\leq$ 5); Moderate (5 \textless Speed $\leq$ 12); Fast (speed \textgreater 12)   \\
                   &                                                                                                 &                                & \textit{Distance from AVUT (m)}: \makecell[l]{Zero ($\lvert d \rvert$ = 0); Very near (0 \textless $\lvert d \rvert$ $\leq$ 8); Near (8 \textless $\lvert d \rvert$ $\leq$ 18); Far (18 \textless $\lvert d \rvert$ $\leq$ 28); Very far ($\lvert d \rvert$ \textgreater 28)}                                                          \\ \cmidrule{4-4}
\multirow{3}{*}{8} &                                                                                                 & \multirow{3}{*}{Pedestrian*}   & \textit{Volume ($m^3$)}: Small (volume \textless 1)                                                                                                                               \\
                   &                                                                                                 &                                & \textit{Operation}: Stop; Crossing the road; FrontLaneWalking   \\
                   &                                                                                                 &                                & \textit{Distance from AVUT (m)}: \makecell[l]{Zero ($\lvert d \rvert$ = 0); Very near (0 \textless $\lvert d \rvert$ $\leq$ 8); Near (8 \textless $\lvert d \rvert$ $\leq$ 18); Far (18 \textless $\lvert d \rvert$ $\leq$ 28); Very far ($\lvert d \rvert$ \textgreater 28)}   \\ \cmidrule{4-4}
\multirow{2}{*}{9} &                                                                                                 & \multirow{2}{*}{Static obstacles*} & \textit{Volume ($m^3$)}: Small (volume $\leq$ 3); Medium (3 < volume $\leq$ 10); Large (volume \textgreater 10)                                                                                                                                                \\
                   &                                                                                                 &                                    & \textit{Distance from AVUT (m)}: \makecell[l]{Zero ($\lvert d \rvert$ = 0); Very near (0 \textless $\lvert d \rvert$ $\leq$ 8); Near (8 \textless $\lvert d \rvert$ $\leq$ 18); Far (18 \textless $\lvert d \rvert$ $\leq$ 28); Very far ($\lvert d \rvert$ \textgreater 28)}\\ \cmidrule{4-4}
10                 &                                                                                                 & Traffic Rule: traffic light    & None; Green (allow to pass but slow at intersection); Yellow (stop for a while); Red (do not run a red light)                                                           \\ \cmidrule{4-4}
11                 &                                                                                                 & Traffic Rule: side walk         & None; SlowDown                                                                                                                                                                                     \\ \bottomrule
\end{tabular}

%% file: tables/RQ1_Roads_Wea.tex
\begin{tabular}{ccccccccccccccc}
\toprule
\multicolumn{2}{l}{\multirow{2}{*}{\begin{tabular}[l]{@{}l@{}}\deepqtest vs. \\(\randomStrategy, \greedyStrategy)\end{tabular}}} & \multicolumn{2}{c}{\metricTTC}                   & \multicolumn{2}{c}{\metricDTO}              & \multicolumn{2}{c}{\metricJerk}                  & \multicolumn{2}{c}{\metricCollisonNum}             & \multicolumn{2}{c}{\metricCollisionTime}                \\
\multicolumn{2}{c}{}                                                                                  & $\hat{A}_{12}$            & $p$                        & $\hat{A}_{12}$            & $p$                        & $\hat{A}_{12}$            & $p$                        & $\hat{A}_{12}$            & $p$                        & $\hat{A}_{12}$            & $p$                          \\ \midrule
\multirow{2}{*}{\rewardTTC}                                        & \greedyStrategy                                      & \textbf{0.095} & \textbf{\textless{}0.05} & \textbf{0.117} & \textbf{\textless{}0.05} & \textbf{0.735} & \textbf{\textless{}0.05} & \textbf{0.834} & \textbf{\textless{}0.05} & \textbf{0.149} & \textbf{\textless{}0.05}   \\
                                                            & \randomStrategy                                      & \textbf{0.091} & \textbf{\textless{}0.05} & \textbf{0.125} & \textbf{\textless{}0.05} & \textbf{0.724} & \textbf{\textless{}0.05} & \textbf{0.784} & \textbf{\textless{}0.05} & \textbf{0.179} & \textbf{\textless{}0.05}\\ \midrule
\multirow{2}{*}{\rewardDTO}                                        & \greedyStrategy                                      & \textbf{0.172} & \textbf{\textless{}0.05} & \textbf{0.058} & \textbf{\textless{}0.05} & \textbf{0.639} & \textbf{\textless{}0.05} & \textbf{0.802} & \textbf{\textless{}0.05} & \textbf{0.162} & \textbf{\textless{}0.05}  \\
                                                            & \randomStrategy                                      & \textbf{0.186} & \textbf{\textless{}0.05} & \textbf{0.064} & \textbf{\textless{}0.05} & \textbf{0.661} & \textbf{\textless{}0.05} & \textbf{0.797} & \textbf{\textless{}0.05} & \textbf{0.176} & \textbf{\textless{}0.05}   \\ \midrule
\multirow{2}{*}{\rewardJerk}                                       & \greedyStrategy                                      & \textbf{0.285} & \textbf{\textless{}0.05} & \textbf{0.422} & \textbf{\textless{}0.05} & \textbf{0.635} & \textbf{\textless{}0.05} & \textbf{0.709} & \textbf{\textless{}0.05} & \textbf{0.279} & \textbf{\textless{}0.05}   \\
                                                            & \randomStrategy                                      & \textbf{0.269} & \textbf{\textless{}0.05} & \textbf{0.437} & \textbf{\textless{}0.05} & \textbf{0.619} & \textbf{\textless{}0.05} & \textbf{0.711} & \textbf{\textless{}0.05} & \textbf{0.274} & \textbf{\textless{}0.05}   \\ \bottomrule
\end{tabular}

%% file: tables/RQ2_1_MergeRoad.tex
\begin{tabular}{ccccccccccccccccc}
\toprule
\multicolumn{2}{c}{\multirow{2}{*}{\textit{Reward Comparison}}}       & \multicolumn{2}{c}{\metricTTC}                   & \multicolumn{2}{c}{\metricDTO}              & \multicolumn{2}{c}{\metricJerk}                  & \multicolumn{2}{c}{\metricCollisonNum}             & \multicolumn{2}{c}{\metricCollisionTime}   & \multicolumn{2}{c}{\metricDiversityAPI}    & \multicolumn{2}{c}{\metricDiversityScenario}    \\
  &       & $\hat{A}_{12}$            & \textit{p}                        & $\hat{A}_{12}$            & \textit{p}                        & $\hat{A}_{12}$            & \textit{p}                        & $\hat{A}_{12}$            & \textit{p}                        & $\hat{A}_{12}$            & \textit{p}                        & $\hat{A}_{12}$            & \textit{p}                        & $\hat{A}_{12}$    &\textit{p}               \\ \midrule
       
\multirow{2}{*}{\rewardTTCRD} & \rewardDTORD & \textbf{0.281} & \textbf{\textless{}0.05} & 0.789×         & \textless{}0.05          & \textbf{0.606} & \textbf{\textless{}0.05} & 0.525          & 0.475                    & 0.499          & 0.980                    & 0.530  & 0.503           & 0.461  & 0.475           \\
       & \rewardJerkRD& \textbf{0.210} & \textbf{\textless{}0.05} & \textbf{0.228} & \textbf{\textless{}0.05} & \textbf{0.621} & \textbf{\textless{}0.05} & 0.562          & 0.086                    & \textbf{0.409} & \textbf{\textless{}0.05}         & 0.420  & 0.079           & 0.374× & \textless{}0.05 \\
\rewardDTORD         & \rewardJerkRD & \textbf{0.332} & \textbf{\textless{}0.05} & \textbf{0.117} & \textbf{\textless{}0.05} & 0.529          & 0.533                    & 0.537          & 0.314                    & \textbf{0.411} & \textbf{\textless{}0.05}  & 0.388× & \textless{}0.05 & 0.412× & \textless{}0.05 \\ \midrule
\multirow{2}{*}{\rewardTTCRN} & \rewardDTORN & \textbf{0.297} & \textbf{\textless{}0.05} & 0.811×         & \textless{}0.05          & \textbf{0.625} & \textbf{\textless{}0.05} & 0.475          & 0.442                    & 0.571          & 0.111                 & 0.610  & \textless{}0.05 & 0.568  & 0.094 \\
       & \rewardJerkRN & \textbf{0.247} & \textbf{\textless{}0.05} & \textbf{0.177} & \textbf{\textless{}0.05} & \textbf{0.609} & \textbf{\textless{}0.05} & \textbf{0.600} & \textbf{\textless{}0.05} & \textbf{0.366} & \textbf{\textless{}0.05} & 0.409× & \textless{}0.05 & 0.378× & \textless{}0.05 \\
\rewardDTORN         & \rewardJerkRN & \textbf{0.432} & \textbf{\textless{}0.05} & \textbf{0.067} & \textbf{\textless{}0.05} & 0.501          & 0.984                    & \textbf{0.625} & \textbf{\textless{}0.05} & \textbf{0.309} & \textbf{\textless{}0.05}  & 0.306× & \textless{}0.05 & 0.317× & \textless{}0.05 \\ \midrule
\multirow{2}{*}{\rewardTTCSD} & \rewardDTOSD & \textbf{0.304} & \textbf{\textless{}0.05} & 0.826×         & \textless{}0.05          & \textbf{0.636} & \textbf{\textless{}0.05} & 0.525          & 0.421                    & 0.535          & 0.435                 & 0.553  & 0.235           & 0.527  & 0.639           \\
       & \rewardJerkSD& \textbf{0.185} & \textbf{\textless{}0.05} & \textbf{0.188} & \textbf{\textless{}0.05} & \textbf{0.650} & \textbf{\textless{}0.05} & \textbf{0.606} & \textbf{\textless{}0.05} & \textbf{0.344} & \textbf{\textless{}0.05} & 0.365× & \textless{}0.05 & 0.386× & \textless{}0.05 \\
\rewardDTOSD         & \rewardJerkSD & \textbf{0.346} & \textbf{\textless{}0.05} & \textbf{0.073} & \textbf{\textless{}0.05} & 0.524          & 0.604                    & \textbf{0.581} & \textbf{\textless{}0.05} & \textbf{0.325} & \textbf{\textless{}0.05} & 0.330× & \textless{}0.05 & 0.360× & \textless{}0.05 \\ \midrule
\multirow{2}{*}{\rewardTTCSN} & \rewardDTOSN & \textbf{0.371} & \textbf{\textless{}0.05} & 0.787×         & \textless{}0.05          & \textbf{0.625} & \textbf{\textless{}0.05} & 0.519          & 0.551                    & 0.536          & 0.417                    & 0.529  & 0.523           & 0.490  & 0.832           \\
       & \rewardJerkSN & \textbf{0.185} & \textbf{\textless{}0.05} & \textbf{0.182} & \textbf{\textless{}0.05} & 0.581          & 0.078                    & \textbf{0.600} & \textbf{\textless{}0.05} & \textbf{0.339} & \textbf{\textless{}0.05}   & 0.323× & \textless{}0.05 & 0.338× & \textless{}0.05 \\
\rewardDTOSN         & \rewardJerkSN & \textbf{0.347} & \textbf{\textless{}0.05} & \textbf{0.082} & \textbf{\textless{}0.05} & 0.457          & 0.352                    & \textbf{0.581} & \textbf{\textless{}0.05} & \textbf{0.335} & \textbf{\textless{}0.05}     & 0.333× & \textless{}0.05 & 0.349× & \textless{}0.05 \\ \bottomrule
\end{tabular}

%% file: tables/RQ2_Ranking.tex
\begin{tabular}{lrrrrrrr}
\toprule
   & $TTC$            & $DTO$            & $Jerk$          & $\#Collision$   & $CollisionTime$ & $DIV_{API}$ & $DIV_{Scenario}$       \\ \midrule
$RD$ & $DQT_T$, $DQT_D$, $DQT_J$ & $DQT_D$, $DQT_T$, $DQT_J$ & $DQT_T$, $DQT_D$/$DQT_J$ & $DQT_T$/$DQT_D$/$DQT_J$ & $DQT_D$/$DQT_T$, $DQT_J$ & $DQT_J$,$DQT_T$/$DQT_D$ & $DQT_J$,$DQT_D$/$DQT_T$\\
$RN$ & $DQT_T$, $DQT_D$, $DQT_J$ & $DQT_D$, $DQT_T$, $DQT_J$ & $DQT_T$, $DQT_J$/$DQT_D$ & $DQT_D$/$DQT_T$, $DQT_J$ & $DQT_T$/$DQT_D$, $DQT_J$ & $DQT_J$,$DQT_T$/$DQT_D$ & $DQT_J$,$DQT_T$/$DQT_D$\\
$SD$ & $DQT_T$, $DQT_D$, $DQT_J$ & $DQT_D$, $DQT_T$, $DQT_J$ & $DQT_T$, $DQT_D$/$DQT_J$ & $DQT_T$/$DQT_D$, $DQT_J$ & $DQT_T$/$DQT_D$, $DQT_J$ & $DQT_J$,$DQT_T$/$DQT_D$ & $DQT_J$,$DQT_T$/$DQT_D$ \\
$SN$ & $DQT_T$, $DQT_D$, $DQT_J$ & $DQT_D$, $DQT_T$, $DQT_J$ & $DQT_T$,$DQT_J$/$DQT_D$ & $DQT_T$/$DQT_D$, $DQT_J$ & $DQT_T$/$DQT_D$, $DQT_J$ & $DQT_J$,$DQT_T$/$DQT_D$ & $DQT_J$, $DQT_D$/$DQT_T$ \\ 
\midrule
$Rec_M$ & $DQT_T$ & $DQT_D$ & $DQT_T$ & $DQT_T$ & $DQT_T$ & $DQT_J$ & $DQT_J$ \\
\bottomrule
\end{tabular}

%% file: tables/RQ2.2_realistic-unrealistic.tex
\begin{tabular}{c|rrrrrr|rrrrrr}
\toprule
 \multirow{2}{*}{\begin{tabular}[l]{@{}l@{}}\deepqtest vs. \\ \deepcollision\end{tabular}} &         &         \multicolumn{3}{c}{\textit{Collision}}      &       \multicolumn{2}{c|}{\textit{Non-Collision}}               &           &          \multicolumn{3}{c}{\textit{Collision}}        &        \multicolumn{2}{c}{\textit{Non-Collision}}   \\  \cmidrule(r){3-5} \cmidrule(r){6-7}\cmidrule(r){9-11}\cmidrule(r){12-13}
 & \textit{\#TS} & \textit{\#\metricRealismRCS/\%}  & $CT_{RCS}$ & \textit{\#\metricRealismUCS/\%} & \textit{\#\metricRealismRNS/\%} & \textit{\#\metricRealismUNS/\%} & \textit{\#TS} & \textit{\#\metricRealismRCS/\%}  & $CT_{RCS}$ & \textit{\#\metricRealismUCS/\%} & \textit{\#\metricRealismRNS/\%} & \textit{\#\metricRealismUNS/\%} \\ \midrule
& \multicolumn{6}{c|}{\textit{R1}} & \multicolumn{6}{c}{\textit{R2}} \\                                                \cmidrule(r){1-1} \cmidrule(r){2-7} \cmidrule(r){8-13}
\rewardTTCRD                                   & 770           & 16/2.08\%    & 12.40     & 0/0\%             & 754/97.92\%       & 0/0\%             & 614           & 15/2.44\%     & 15.67    & 0/0\%             & 599/97.56\%       & 0/0\%             \\
\rewardTTCRN                                    & 794           & 16/2.02\%   & 12.00     & 0/0\%             & 778/97.98\%       & 0/0\%             & 548           & \textbf{17/3.10\%  }   & 8.20    & 0/0\%             & 531/96.90\%       & 0/0\%             \\
\rewardTTCSD                                   & 662           & 16/2.42\%    & 9.92      & 0/0\%             & 646/97.58\%       & 0/0\%             & 602           & 16/2.66\%      & 11.14      & 0/0\%          & 586/97.34\%       & 0/0\%             \\
\rewardTTCSN                                & 680           & \textbf{17/2.50\%} & 12.00        & 0/0\%             & 663/97.50\%       & 0/0\%             & 548           & \textbf{17/3.10\% }     & 11.44        & 0/0\%        & 531/96.90\%       & 0/0\%             \\
\deepcollision                                     & 3440          & 19/0.55\%   & 18.00      & 23/0.67\%         & 1389/40.38\%      & 2009/58.40\%      & 3440          & 13/0.38\%       & 18.00       & 9/0.26\%     & 3061/88.98\%      & 357/10.38\%        \\ 
\cmidrule(r){1-1} \cmidrule(r){2-7} \cmidrule(r){8-13}


& \multicolumn{6}{c|}{\textit{R3}} & \multicolumn{6}{c}{\textit{R4}} \\                                                \cmidrule(r){1-1} \cmidrule(r){2-7} \cmidrule(r){8-13}

\rewardTTCRD                                    & 770           & 14/1.82\%      & 8.67       & 0/0\%             & 756/98.18\%       & 0/0\%             & 662           & 17/2.57\%       & 9.60      & 0/0\%             & 645/97.43\%       & 0/0\%             \\
\rewardTTCRN                                 & 764           & 13/1.70\%       & 10.71      & 0/0\%             & 751/98.30\%       & 0/0\%             & 752           & 15/1.99\%        & 9.92     & 0/0\%             & 737/98.01\%       & 0/0\%             \\
\rewardTTCSD                                    & 524           & \textbf{18/3.44\%}       & 7.50    & 0/0\%             & 506/96.56\%       & 0/0\%             & 536           & \textbf{17/3.17\%}          & 9.00   & 0/0\%             & 519/96.83\%       & 0/0\%             \\
\rewardTTCSN                                   & 644           & 16/2.48\%        & 7.33     & 0/0\%             & 628/97.52\%       & 0/0\%             & 680           & 16/2.35\%        & 13.67     & 0/0\%             & 664/97.65\%       & 0/0\%             \\
\deepcollision                                  & 4220          & 2/0.05\%        & 21.00      & 94/2.23\%         & 145/3.44\%        & 3979/94.29\%      & 4700          & 2/0.04\%       & 30.00       & 145/3.09\%        & 124/2.64\%        & 4429/94.23\%      \\ \bottomrule
\end{tabular}

%% file: tables/RQ2.2_statistical_realistic-unrealistic.tex
\begin{tabular}{ccccccccccccccccccc}
\toprule
&  \multicolumn{8}{c}{\textit{Collision}} & \multicolumn{8}{c}{\textit{Non-Collision}} \\ \cmidrule(r){1-1} \cmidrule(r){2-11} \cmidrule(r){12-19}
\rewardTTC \textit{vs.} & \multicolumn{2}{c}{\#\metricRealismRCS}                 & \multicolumn{2}{c}{\metricRealismRCS\%}    & \multicolumn{2}{c}{$CT_{RCS}$}             & \multicolumn{2}{c}{\#\metricRealismUCS}                 & \multicolumn{2}{c}{\metricRealismUCS\%}                 & \multicolumn{2}{c}{\#\metricRealismRNS}                 & \multicolumn{2}{c}{\metricRealismRNS\%}                 & \multicolumn{2}{c}{\#\metricRealismUNS}                   & \multicolumn{2}{c}{\metricRealismRNS\%}                 \\
\deepcollision & $\hat{A}_{12}$            & $p$                        & $\hat{A}_{12}$            & $p$       & $\hat{A}_{12}$            & $p$                 & $\hat{A}_{12}$            & $p$                        & $\hat{A}_{12}$            & $p$                        & $\hat{A}_{12}$            & $p$                        & $\hat{A}_{12}$            & $p$                        & $\hat{A}_{12}$            & $p$                        & $\hat{A}_{12}$            & $P$                        \\ \midrule
                  & \multicolumn{16}{c}{$R1$}                                                                                                                                                                                                                                                                                                                                       \\ \cmidrule{2-19} 
\rewardTTCRD    & 0.425          & 0.164                    & \textbf{0.805} & \textbf{\textless{}0.05} & \textbf{0.174} & \textbf{\textless{}0.05} & \textbf{0.150} & \textbf{\textless{}0.05} & \textbf{0.150} & \textbf{\textless{}0.05} & 0.255×         & \textless{}0.05          & \textbf{1.000} & \textbf{\textless{}0.05} & \textbf{0.000} & \textbf{\textless{}0.05} & \textbf{0.000} & \textbf{\textless{}0.05} \\

\rewardTTCRN    & 0.425          & 0.164                    & \textbf{0.805} & \textbf{\textless{}0.05} & \textbf{0.188} & \textbf{\textless{}0.05} & \textbf{0.150} & \textbf{\textless{}0.05} & \textbf{0.150} & \textbf{\textless{}0.05} & 0.285×         & \textless{}0.05          & \textbf{1.000} & \textbf{\textless{}0.05} & \textbf{0.000} & \textbf{\textless{}0.05} & \textbf{0.000} & \textbf{\textless{}0.05} \\

\rewardTTCSD    & 0.425          & 0.164                    & \textbf{0.805} & \textbf{\textless{}0.05} & \textbf{0.201} & \textbf{\textless{}0.05} & \textbf{0.150} & \textbf{\textless{}0.05} & \textbf{0.150} & \textbf{\textless{}0.05} & 0.172×         & \textless{}0.05          & \textbf{1.000} & \textbf{\textless{}0.05} & \textbf{0.000} & \textbf{\textless{}0.05} & \textbf{0.000} & \textbf{\textless{}0.05} \\

\rewardTTCSN    & 0.450          & 0.310                    & \textbf{0.854} & \textbf{\textless{}0.05} & \textbf{0.130} & \textbf{\textless{}0.05} & \textbf{0.150} & \textbf{\textless{}0.05} & \textbf{0.150} & \textbf{\textless{}0.05} & 0.203×         & \textless{}0.05          & \textbf{1.000} & \textbf{\textless{}0.05} & \textbf{0.000} & \textbf{\textless{}0.05} & \textbf{0.000} & \textbf{\textless{}0.05} \\ \midrule

                  & \multicolumn{16}{c}{$R2$}                                                                                                                                                                                                                                                                                                                                       \\ \cmidrule{2-19} 
\rewardTTCRD    & 0.550          & 0.506                    & \textbf{0.794} & \textbf{\textless{}0.05} & \textbf{0.340} & \textbf{\textless{}0.05} & \textbf{0.300} & \textbf{\textless{}0.05} & \textbf{0.300} & \textbf{\textless{}0.05} & 0.016×         & \textless{}0.05          & 0.635 & 0.141           & \textbf{0.075} & \textbf{\textless{}0.05} & \textbf{0.075} & \textbf{\textless{}0.05} \\

\rewardTTCRN    & 0.600 & 0.154           & \textbf{0.876} & \textbf{\textless{}0.05} & \textbf{0.151} & \textbf{\textless{}0.05} & \textbf{0.300} & \textbf{\textless{}0.05} & \textbf{0.300} & \textbf{\textless{}0.05} & 0.018×         & \textless{}0.05          & 0.629 & 0.164           & \textbf{0.075} & \textbf{\textless{}0.05} & \textbf{0.075} & \textbf{\textless{}0.05} \\

\rewardTTCSD    & 0.575          & 0.302                    & \textbf{0.835} & \textbf{\textless{}0.05} & \textbf{0.194} & \textbf{\textless{}0.05} & \textbf{0.300} & \textbf{\textless{}0.05} & \textbf{0.300} & \textbf{\textless{}0.05} & 0.015×         & \textless{}0.05          & \textbf{0.701} & \textbf{\textless{}0.05} & \textbf{0.075} & \textbf{\textless{}0.05} & \textbf{0.075} & \textbf{\textless{}0.05} \\

\rewardTTCSN    & 0.600 & 0.154           & \textbf{0.876} & \textbf{\textless{}0.05} & \textbf{0.138} & \textbf{\textless{}0.05} & \textbf{0.300} & \textbf{\textless{}0.05} & \textbf{0.300} & \textbf{\textless{}0.05} & 0.010×         & \textless{}0.05          & \textbf{0.751} & \textbf{\textless{}0.05} & \textbf{0.075} & \textbf{\textless{}0.05} & \textbf{0.075} & \textbf{\textless{}0.05} \\ \midrule

                  & \multicolumn{16}{c}{$R3$}                                                                                                                                                                                                                                                                                                                                       \\ \cmidrule{2-19} 
\rewardTTCRD    & \textbf{0.850} & \textbf{\textless{}0.05} & \textbf{0.850} & \textbf{\textless{}0.05} & \textbf{0.275} & \textbf{\textless{}0.05} & \textbf{0.000} & \textbf{\textless{}0.05} & \textbf{0.000} & \textbf{\textless{}0.05} & \textbf{0.800} & \textbf{\textless{}0.05} & \textbf{1.000} & \textbf{\textless{}0.05} & \textbf{0.000} & \textbf{\textless{}0.05} & \textbf{0.000} & \textbf{\textless{}0.05} \\

\rewardTTCRN    & \textbf{0.825} & \textbf{\textless{}0.05} & \textbf{0.825} & \textbf{\textless{}0.05} & \textbf{0.325} & \textbf{\textless{}0.05} & \textbf{0.000} & \textbf{\textless{}0.05} & \textbf{0.000} & \textbf{\textless{}0.05} & \textbf{0.723} & \textbf{\textless{}0.05} & \textbf{1.000} & \textbf{\textless{}0.05} & \textbf{0.000} & \textbf{\textless{}0.05} & \textbf{0.000} & \textbf{\textless{}0.05} \\

\rewardTTCSD    & \textbf{0.950} & \textbf{\textless{}0.05} & \textbf{0.950} & \textbf{\textless{}0.05} & \textbf{0.350} & \textbf{\textless{}0.05} & \textbf{0.000} & \textbf{\textless{}0.05} & \textbf{0.000} & \textbf{\textless{}0.05} & \textbf{0.738} & \textbf{\textless{}0.05} & \textbf{1.000} & \textbf{\textless{}0.05} & \textbf{0.000} & \textbf{\textless{}0.05} & \textbf{0.000} & \textbf{\textless{}0.05} \\

\rewardTTCSN    & \textbf{0.900} & \textbf{\textless{}0.05} & \textbf{0.900} & \textbf{\textless{}0.05} & \textbf{0.275} & \textbf{\textless{}0.05} & \textbf{0.000} & \textbf{\textless{}0.05} & \textbf{0.000} & \textbf{\textless{}0.05} & \textbf{0.723} & \textbf{\textless{}0.05} & \textbf{1.000} & \textbf{\textless{}0.05} & \textbf{0.000} & \textbf{\textless{}0.05} & \textbf{0.000} & \textbf{\textless{}0.05} \\ \midrule

                  & \multicolumn{16}{c}{$R4$}                                                                                                                                                                                                                                                                                                                                       \\ \cmidrule{2-19} 
\rewardTTCRD    & \textbf{0.925} & \textbf{\textless{}0.05} & \textbf{0.925} & \textbf{\textless{}0.05} & \textbf{0.375} & \textbf{\textless{}0.05} & \textbf{0.000} & \textbf{\textless{}0.05} & \textbf{0.000} & \textbf{\textless{}0.05} & \textbf{0.850} & \textbf{\textless{}0.05} & \textbf{1.000} & \textbf{\textless{}0.05} & \textbf{0.000} & \textbf{\textless{}0.05} & \textbf{0.000} & \textbf{\textless{}0.05} \\

\rewardTTCRN    & \textbf{0.875} & \textbf{\textless{}0.05} & \textbf{0.875} & \textbf{\textless{}0.05} & \textbf{0.175} & \textbf{\textless{}0.05} & \textbf{0.000} & \textbf{\textless{}0.05} & \textbf{0.000} & \textbf{\textless{}0.05} & \textbf{0.900} & \textbf{\textless{}0.05} & \textbf{1.000} & \textbf{\textless{}0.05} & \textbf{0.000} & \textbf{\textless{}0.05} & \textbf{0.000} & \textbf{\textless{}0.05} \\

\rewardTTCSD    & \textbf{0.925} & \textbf{\textless{}0.05} & \textbf{0.925} & \textbf{\textless{}0.05} & \textbf{0.375} & \textbf{\textless{}0.05} & \textbf{0.000} & \textbf{\textless{}0.05} & \textbf{0.000} & \textbf{\textless{}0.05} & \textbf{0.805} & \textbf{\textless{}0.05} & \textbf{1.000} & \textbf{\textless{}0.05} & \textbf{0.000} & \textbf{\textless{}0.05} & \textbf{0.000} & \textbf{\textless{}0.05} \\

\rewardTTCSN    & \textbf{0.900} & \textbf{\textless{}0.05} & \textbf{0.900} & \textbf{\textless{}0.05} & \textbf{0.275} & \textbf{\textless{}0.05} & \textbf{0.000} & \textbf{\textless{}0.05} & \textbf{0.000} & \textbf{\textless{}0.05} & \textbf{0.835} & \textbf{\textless{}0.05} & \textbf{1.000} & \textbf{\textless{}0.05} & \textbf{0.000} & \textbf{\textless{}0.05} & \textbf{0.000} & \textbf{\textless{}0.05} \\ \bottomrule
\end{tabular}

%% file: figs/unrealistic.tex
	\begin{subfigure}[b]{0.242\textwidth}
		\centering
		\includegraphics[width=\textwidth]{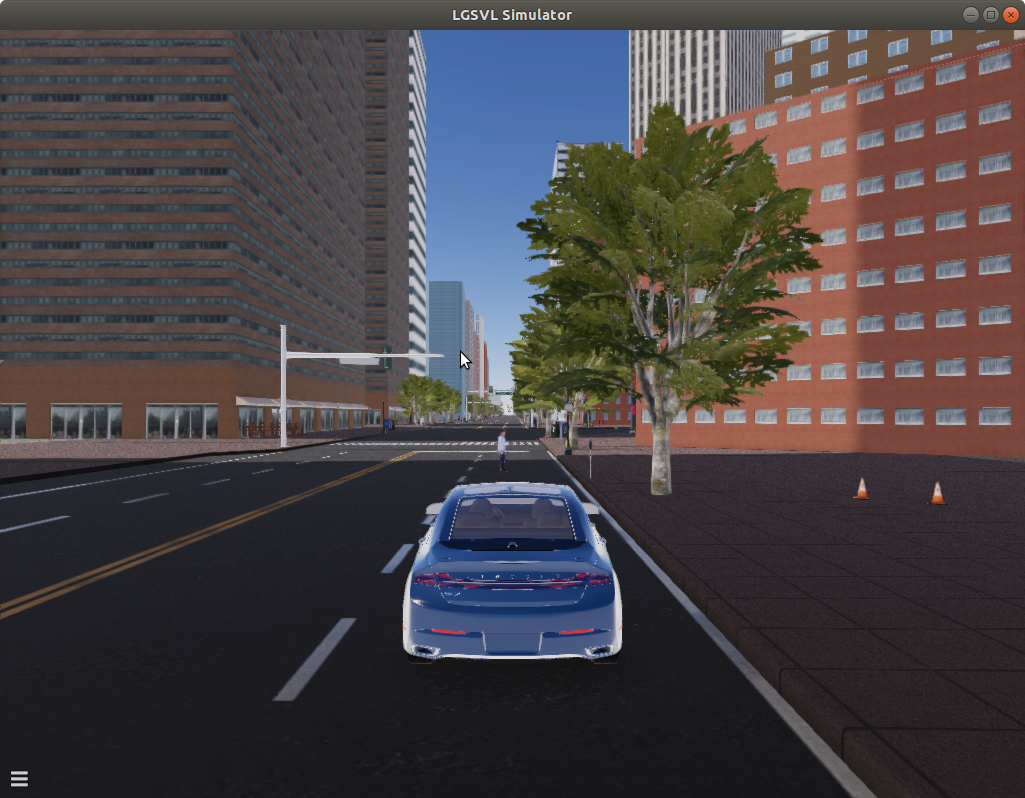}
            \includegraphics[width=\textwidth]{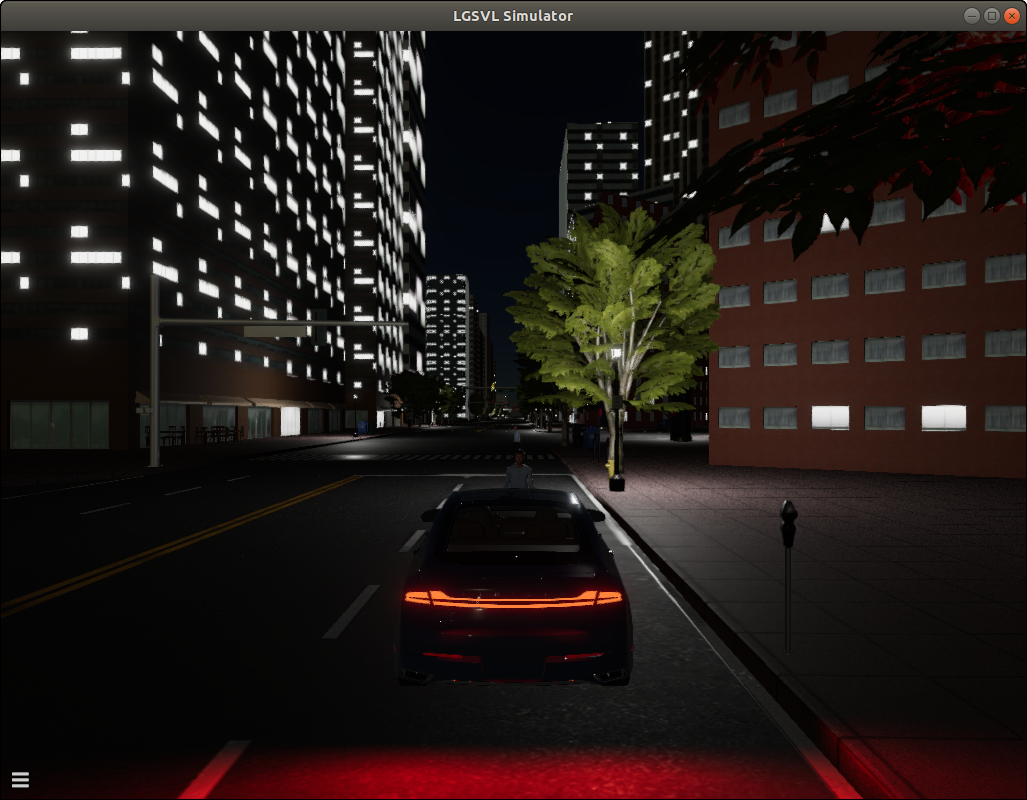}
		\caption{\centering {Unrealistic Time Changes (UTC)}}
		\label{fig:u1}
	\end{subfigure}
 	\hfill
	\begin{subfigure}[b]{0.242\textwidth}
		\centering
		\includegraphics[width=\textwidth]{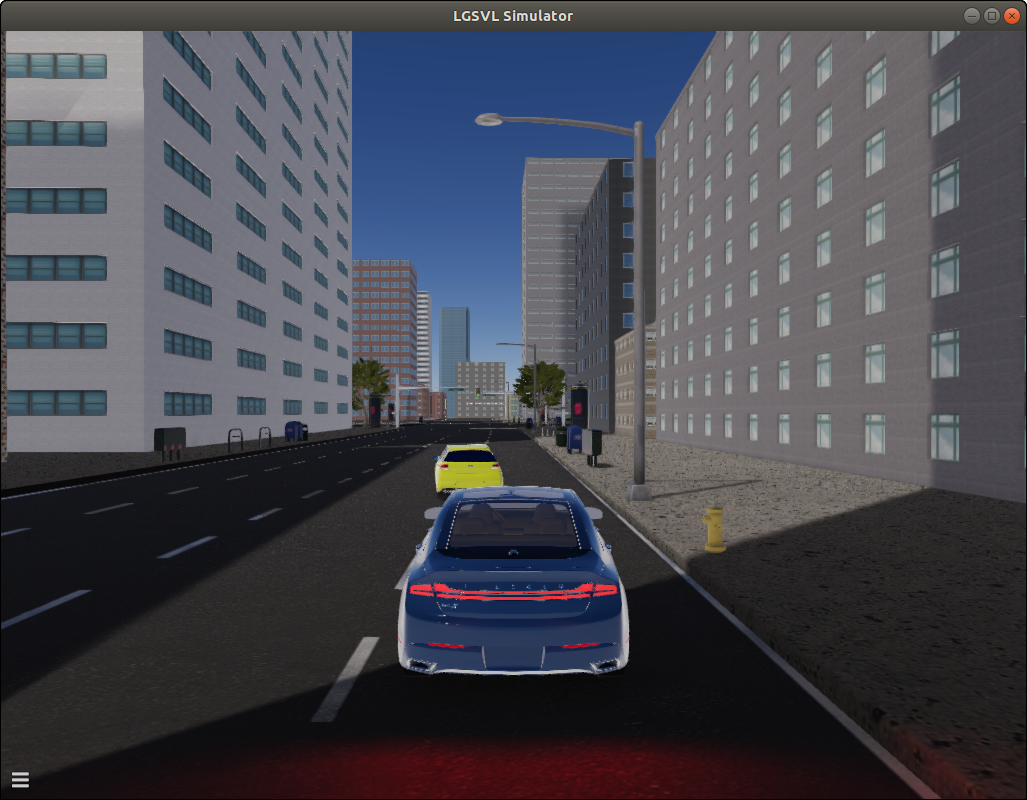}
            \includegraphics[width=\textwidth]{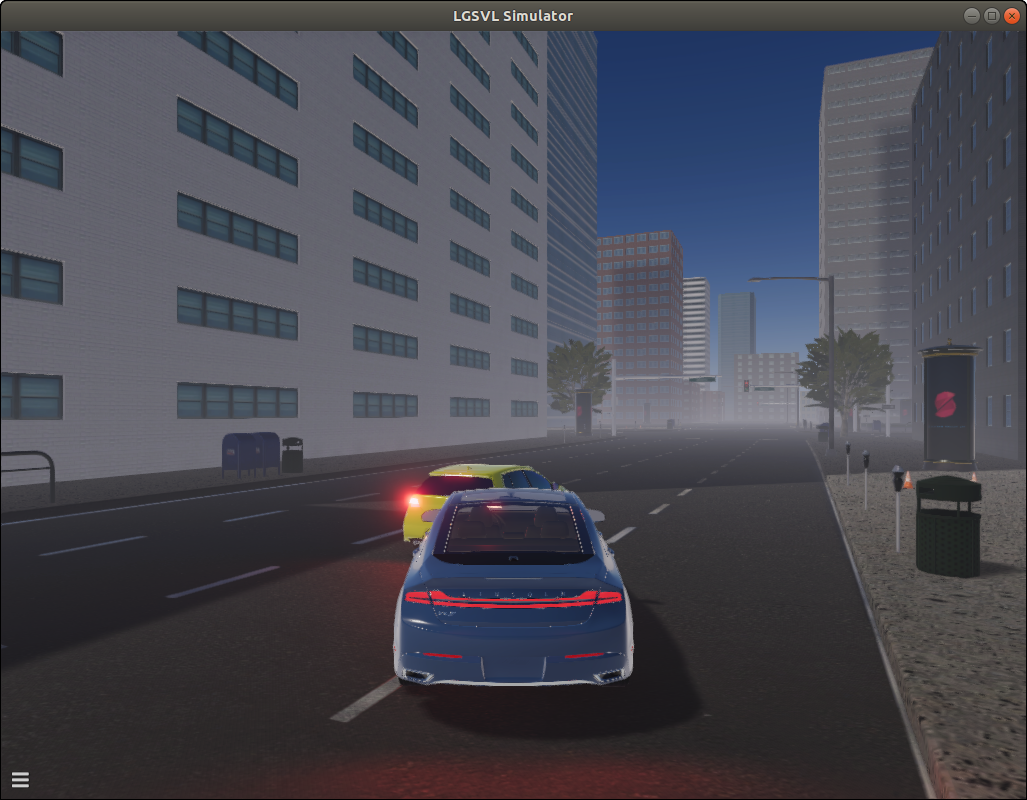}
		\caption{\centering {Unrealistic Weather Changes (UWC)}}
		\label{fig:u5}
	\end{subfigure}
        \hfill
	\begin{subfigure}[b]{0.242\textwidth}
		\centering
		\includegraphics[width=\textwidth]{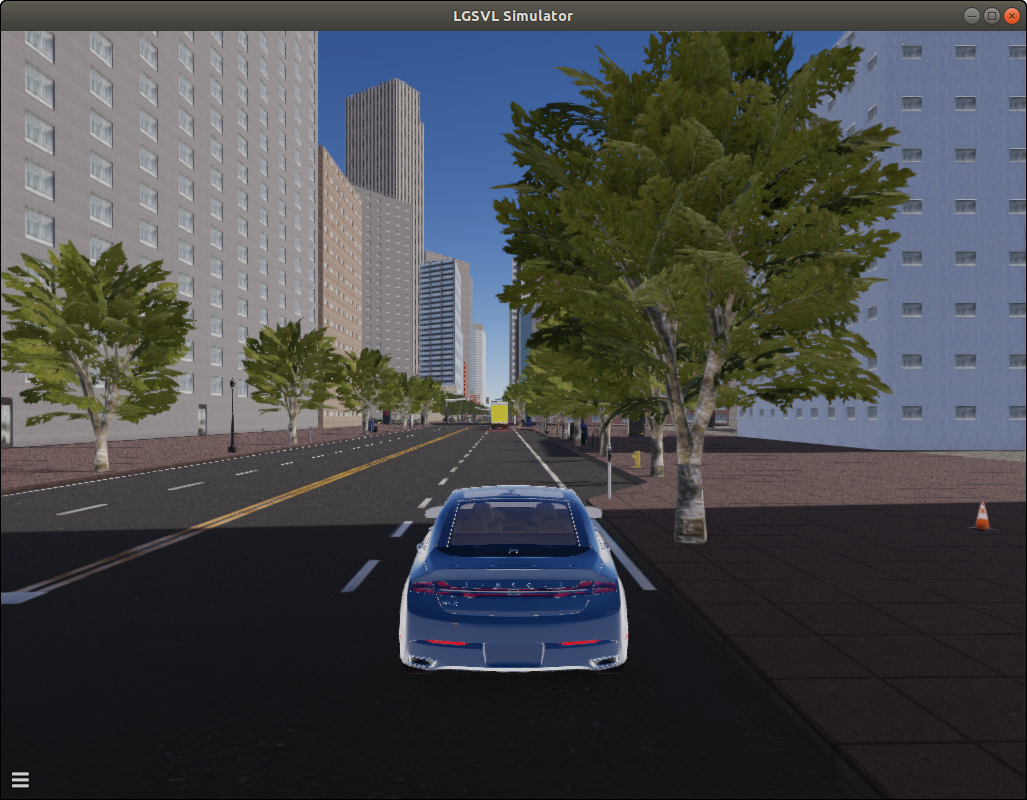}
            \includegraphics[width=\textwidth]{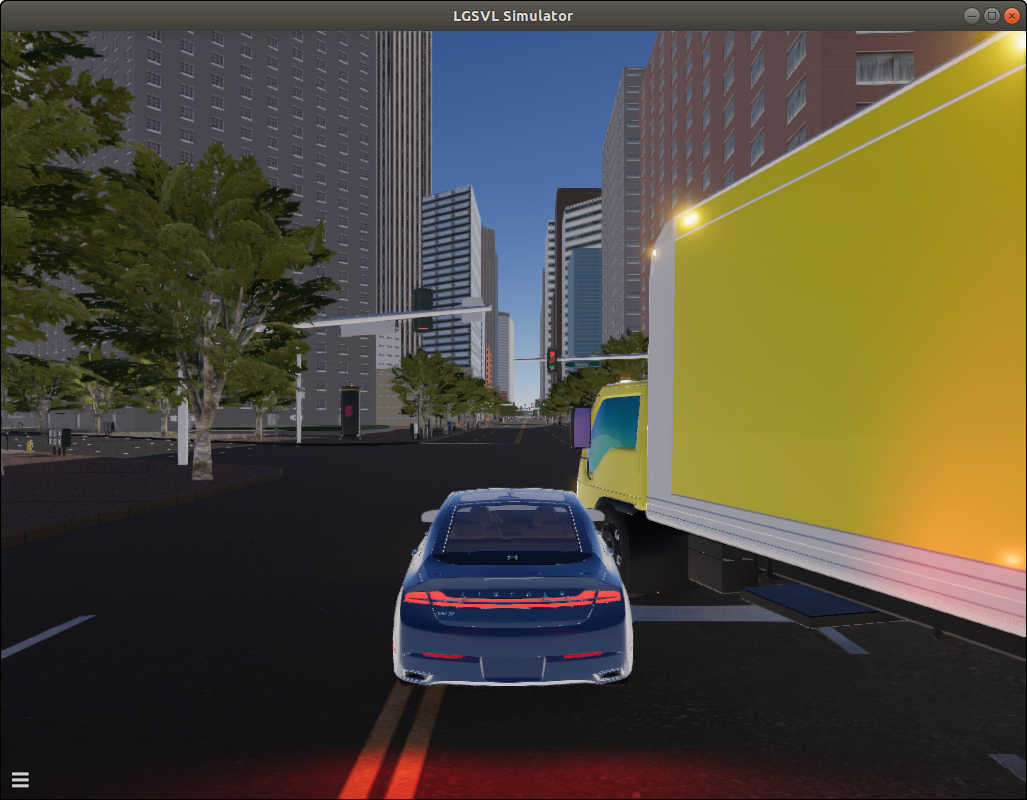}
		\caption{\centering {Violation of Safety Distance (VSD)}}
		\label{fig:u9}
	\end{subfigure}
         \hfill
	\begin{subfigure}[b]{0.242\textwidth}
		\centering
		\includegraphics[width=\textwidth]{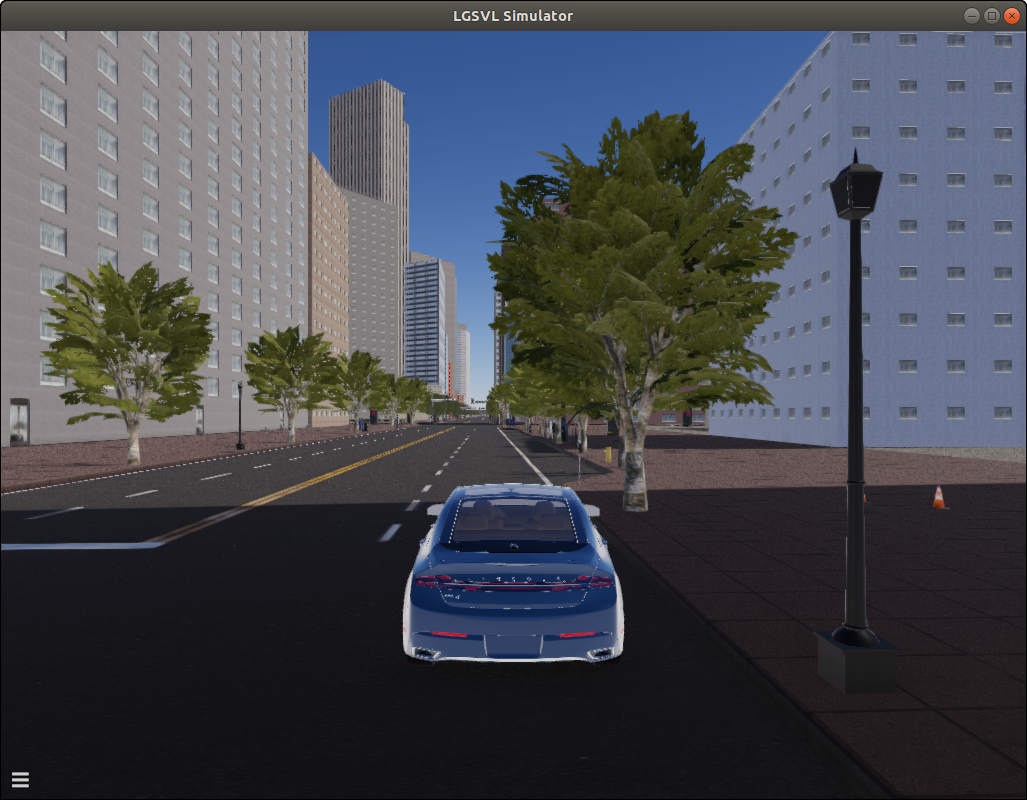}
            \includegraphics[width=\textwidth]{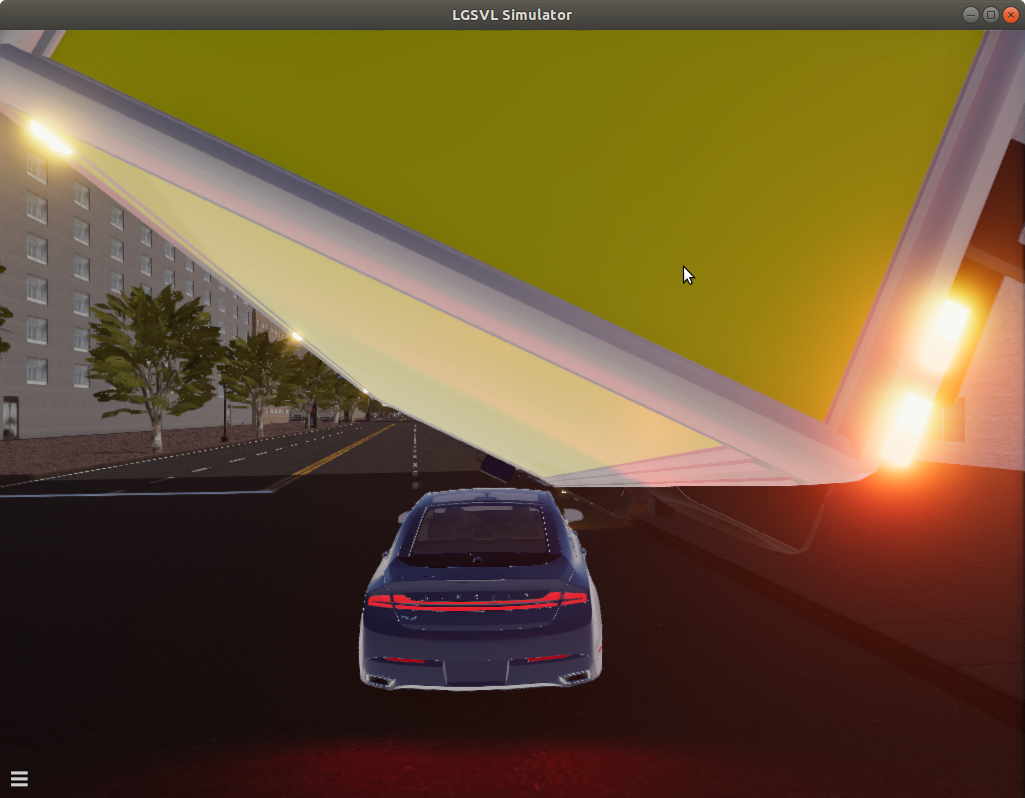}
		\caption{\tabular[t]{@{}c@{}}Overlapping Areas \\ (OA)\endtabular
  }
		\label{fig:u11}
	\end{subfigure}

%% file: tables/RQ2.2_Diversity_All_Realistic.tex
\begin{tabular}{ccccccccccccccccc}
\toprule 
                                                                                       & \multicolumn{4}{c}{$R1$}   & \multicolumn{4}{c}{$R2$}   & \multicolumn{4}{c}{$R3$}      & \multicolumn{4}{c}{$R4$}    \\ \cmidrule(r){1-1} \cmidrule(r){2-5} \cmidrule(r){6-9} \cmidrule(r){10-13} \cmidrule(r){14-17}
\rewardTTC \textit{vs.}& \multicolumn{2}{c}{\metricDiversityAPI}                   & \multicolumn{2}{c}{\metricDiversityScenario}              & \multicolumn{2}{c}{\metricDiversityAPI}                   & \multicolumn{2}{c}{\metricDiversityScenario}  & \multicolumn{2}{c}{\metricDiversityAPI}                   & \multicolumn{2}{c}{\metricDiversityScenario}              & \multicolumn{2}{c}{\metricDiversityAPI}                   & \multicolumn{2}{c}{\metricDiversityScenario}                   \\
                                                                                       \deepcollision & $\hat{A}_{12}$            & $p$                        & $\hat{A}_{12}$            & $p$                        & $\hat{A}_{12}$            & $p$                        & $\hat{A}_{12}$            & $p$      & $\hat{A}_{12}$            & $p$                        & $\hat{A}_{12}$            & $p$                        & $\hat{A}_{12}$            & $p$                        & $\hat{A}_{12}$            & $p$                  \\ \cmidrule(r){1-1} \cmidrule(r){2-5} \cmidrule(r){6-9} \cmidrule(r){10-13} \cmidrule(r){14-17}

\rewardTTCRD & \textbf{0.880} & \textbf{\textless{}0.05} & \textbf{0.909} & \textbf{\textless{}0.05} & \textbf{0.750} & \textbf{\textless{}0.05} & \textbf{0.739} & \textbf{\textless{}0.05} & \textbf{0.786} & \textbf{\textless{}0.05} & \textbf{0.767} & \textbf{\textless{}0.05} & \textbf{0.761} & \textbf{\textless{}0.05} & 0.570          & 0.455\\
\rewardTTCRN & \textbf{0.839} & \textbf{\textless{}0.05} & \textbf{0.877} & \textbf{\textless{}0.05} & \textbf{0.730} & \textbf{\textless{}0.05} & \textbf{0.820} & \textbf{\textless{}0.05} & \textbf{0.774} & \textbf{\textless{}0.05} & \textbf{0.749} & \textbf{\textless{}0.05} & \textbf{0.838} & \textbf{\textless{}0.05} & \textbf{0.762} & \textbf{\textless{}0.05}\\
\rewardTTCSD& \textbf{0.870} & \textbf{\textless{}0.05} & \textbf{0.811} & \textbf{\textless{}0.05} & \textbf{0.805} & \textbf{\textless{}0.05} & \textbf{0.860} & \textbf{\textless{}0.05} & \textbf{0.785} & \textbf{\textless{}0.05} & \textbf{0.728} & \textbf{\textless{}0.05} & \textbf{0.719} & \textbf{\textless{}0.05} & 0.453          & 0.615\\
\rewardTTCSN& \textbf{0.795} & \textbf{\textless{}0.05} & \textbf{0.844} & \textbf{\textless{}0.05} & \textbf{0.844} & \textbf{\textless{}0.05} & \textbf{0.799} & \textbf{\textless{}0.05} & \textbf{0.746} & \textbf{\textless{}0.05} & 0.679          & 0.054                    & \textbf{0.765} & \textbf{\textless{}0.05} & 0.556          & 0.549
\\ \bottomrule
\end{tabular}

%% file: algos/algo-deepqtest.tex
\begin{algorithmic}[1]
\Require
  \Statex Let \textit{TEnv} be the test environment; \textit{AS} be the environment configuration action space; \textit{terminate} be the stopping criteria.
\Ensure
  \Statex $Act_{list}$: a list of actions selected from \textit{AS}
  \Statex

\State $Act_{list}$ $\leftarrow$ []
\State Initialize replay memory $D$ with capacity N; Initialize Q-network $Q$ with random weights $\theta$; Initialize target network $\hat{Q}$ with weights $\theta^- = \theta$
\For{episode = 1, M}
\State $s_t$ $\leftarrow$ observe a state of \textit{TEnv}

\While{$\neg$ \textit{terminate}}
\Statex
\State $act_{t}=\left\{\begin{aligned}
& arg\max_aQ(s_t,a),  & \text{with probability 1-}{\epsilon}\\
& \text{randomly select an action}, & {\epsilon}
\end{aligned}
\right.$
\Statex
\State \textit{terminate}, $r_t$, $s_{t+1}$ $\leftarrow$ step($act_{t}$, $s_t$) 
\State Store transition $<s_t, a_t, r_t, s_{t+1}>$ in $D$
\If{$D$ is full}
\State Sample random minibatch of transition $<s_t, a_t, r_t, s_{t+1}>$ from $D$
\State Update Q-network
\EndIf
\State Every $C$ steps reset $\hat{Q} = Q$ 
\State $s_t$ $\leftarrow$ $s_{t+1}$
\State Put $act_t$ into $Act_{list}$

\EndWhile
\EndFor
\State \Return $Act_{list}$

\end{algorithmic}

%% file: algos/algo-calculate_diversity.tex
\begin{algorithmic}[1]
\Require
  \Statex Let $S_{a}=<scene_{a}^{1}, scene_{a}^{2},...,scene_{a}^{n}>$, $S_{b}=<scene_{b}^{1}, scene_{b}^{2},...,scene_{b}^{n}>$ be two driving scenarios; 
  \Statex A $scene$ is defined as $scene=<property^1, property^2, ..., property^{np}>$;
  
\Ensure
  \Statex $similarity_{scenario}$: $similarity$ between $S_a$ and $S_b$.

\Function{scenarioSimilarity}{$S_a$, $S_b$}
  \State $similarity_{scenario} \leftarrow 0$
  \For{$interval$ $\leftarrow$ $0$ $to$ $n - 1$}
    \State $similarity_{ab}$, $similarity_{ba}$ $\leftarrow$ 0, 0
    \For{i $\leftarrow$ $1$ $to$ $n - interval$}
    \State $similarity_{ab}$ $\leftarrow$ $similarity_{ab}$ + $\textsc{sceneSimilarity}(scene_{a}^{i}, scene_{b}^{i + interval})$
    \State $similarity_{ba}$ $\leftarrow$ $similarity_{ba}$ + $\textsc{sceneSimilarity}(scene_{b}^{i}, scene_{a}^{i + interval})$
    \EndFor
    \State $similarity_{scenario}$ $\leftarrow$ $\max$ \{$similarity_{scenario}$, $similarity_{ab} / n$, $similarity_{ba} / n$\}
  \EndFor
  \State \Return $similarity_{scenario}$
\EndFunction

\Statex

\Function{sceneSimilarity}{$scene_a$, $scene_b$}
  \State $similarity_{scene} \leftarrow 0$
\For{i $\leftarrow$ $1$ $to$ $np$}
    \If{$property_{a}^{i}$ $\notin$ $Obstacles$} \algorithmiccomment{$Obstacles$ is \{$NPC Vehicles$, $Pedestrians$, $Static Obstacles$\}}
    \If{$property_{a}^{i}$ = $property_{b}^{i}$}
    \State $similarity_{scene} \leftarrow similarity_{scene} + 1$
    \Else
    \State $similarity_{scene} \leftarrow similarity_{scene} + \textsc{obstacleSimilarity}(property_{a}^{i}, property_{b}^{i})$
    \EndIf
    \EndIf
\EndFor
\State \Return $similarity_{scene} / np$
\EndFunction

\Statex

\Function{obstacleSimilarity}{$obstacles_{a}$, $obstacles_{b}$}
\State $similarity$ $\leftarrow$ $0$
\State $number_a$, $number_b$ = $\textsc{countObstacles}(obstacles_{a})$, $\textsc{countObstacles}(obstacles_{b})$
\State $outer_{obstacles}$, $inner_{obstacles}$ $\leftarrow$ $obstacle_{a}$, $obstacles_{b}$
\If{$number_a$ $\leq$ $number_b$}
\State $outer_{obstacles}$, $inner_{obstacles}$ $\leftarrow$ $obstacle_{b}$, $obstacles_{a}$
\EndIf

\For{$i$ $\leftarrow$ $1$ $to$ $\min\{number_a, number_b\}$}
    \State $similarity_i$ $\leftarrow$ $0$
    \For{$j$ $\leftarrow$ $1$ $to$ $\max\{number_a, number_b\}$}
        \State $similarity_i$ $\leftarrow$ $\max \{similarity_i, \textsc{countIdenticalValues}(outer_{obstacles}(i), inner_{obstacles}(j))\}$
    \EndFor
    \State $similarity$ $\leftarrow$ $similarity$ + $similarity_i$
\EndFor
\State \Return $similarity$ / $(number_a + number_b)$
\EndFunction



\end{algorithmic}

%% file: tables/Appendix_RQ1_TTCReward.tex
\begin{tabular}{cccccccccccc}
\toprule
\multicolumn{2}{c}{\multirow{2}{*}{\begin{tabular}[c]{@{}c@{}}\deepqtest\\ \textit{vs. (\randomStrategy, \greedyStrategy)}\end{tabular}}} & \multicolumn{2}{c}{\metricTTC}                   & \multicolumn{2}{c}{\metricDTO}                   & \multicolumn{2}{c}{\metricJerk}                  & \multicolumn{2}{c}{\metricCollisonNum}             & \multicolumn{2}{c}{\metricCollisionTime}         \\
\multicolumn{2}{c}{}                                                                                  & $\hat{A}_{12}$            & $p$                        & $\hat{A}_{12}$            & $p$                        & $\hat{A}_{12}$            & $p$                        & $\hat{A}_{12}$            & $p$                        & $\hat{A}_{12}$            & $p$                        \\ \midrule
                                                           &                                          & \multicolumn{10}{c}{\rewardTTCRD}                                                                                                                                                                                                   \\ \cmidrule{3-12} 
\multirow{2}{*}{\textit{R1}}                                        & \textit{RS}                                       & \textbf{0.117} & \textbf{\textless{}0.05} & \textbf{0.090} & \textbf{\textless{}0.05} & \textbf{0.820} & \textbf{\textless{}0.05} & \textbf{0.825} & \textbf{\textless{}0.05} & \textbf{0.155} & \textbf{\textless{}0.05} \\
                                                           & \textit{GS}                                       & \textbf{0.058} & \textbf{\textless{}0.05} & \textbf{0.168} & \textbf{\textless{}0.05} & \textbf{0.795} & \textbf{\textless{}0.05} & \textbf{0.825} & \textbf{\textless{}0.05} & \textbf{0.233} & \textbf{\textless{}0.05} \\
\multirow{2}{*}{\textit{R2}}                                        & \textit{RS}                                       & \textbf{0.133} & \textbf{\textless{}0.05} & \textbf{0.117} & \textbf{\textless{}0.05} & \textbf{0.713} & \textbf{\textless{}0.05} & \textbf{0.775} & \textbf{\textless{}0.05} & \textbf{0.190} & \textbf{\textless{}0.05} \\
                                                           & \textit{GS}                                       & \textbf{0.195} & \textbf{\textless{}0.05} & \textbf{0.138} & \textbf{\textless{}0.05} & \textbf{0.716} & \textbf{\textless{}0.05} & \textbf{0.700} & \textbf{\textless{}0.05} & \textbf{0.226} & \textbf{\textless{}0.05} \\
\multirow{2}{*}{\textit{R3}}                                        & \textit{RS}                                       & \textbf{0.158} & \textbf{\textless{}0.05} & \textbf{0.155} & \textbf{\textless{}0.05} & 0.552          & 0.579                    & \textbf{0.750} & \textbf{\textless{}0.05} & \textbf{0.228} & \textbf{\textless{}0.05} \\
                                                           & \textit{GS}                                       & \textbf{0.160} & \textbf{\textless{}0.05} & \textbf{0.128} & \textbf{\textless{}0.05} & 0.535          & 0.715                    & \textbf{0.675} & \textbf{\textless{}0.05} & \textbf{0.270} & \textbf{\textless{}0.05} \\
\multirow{2}{*}{\textit{R4}}                                        & \textit{RS}                                       & \textbf{0.005} & \textbf{\textless{}0.05} & \textbf{0.113} & \textbf{\textless{}0.05} & \textbf{0.895} & \textbf{\textless{}0.05} & \textbf{0.900} & \textbf{\textless{}0.05} & \textbf{0.109} & \textbf{\textless{}0.05} \\
                                                           & \textit{GS}                                       & \textbf{0.010} & \textbf{\textless{}0.05} & \textbf{0.068} & \textbf{\textless{}0.05} & \textbf{0.848} & \textbf{\textless{}0.05} & \textbf{0.900} & \textbf{\textless{}0.05} & \textbf{0.090} & \textbf{\textless{}0.05} \\ \midrule
                                                           &                                          & \multicolumn{10}{c}{\rewardTTCRN}                                                                                                                                                                                                   \\ \cmidrule{3-12} 
\multirow{2}{*}{\textit{R1}}                                        & \textit{RS}                                       & \textbf{0.075} & \textbf{\textless{}0.05} & \textbf{0.045} & \textbf{\textless{}0.05} & \textbf{0.897} & \textbf{\textless{}0.05} & \textbf{0.900} & \textbf{\textless{}0.05} & \textbf{0.100} & \textbf{\textless{}0.05} \\
                                                           & \textit{GS}                                       & \textbf{0.043} & \textbf{\textless{}0.05} & \textbf{0.087} & \textbf{\textless{}0.05} & \textbf{0.843} & \textbf{\textless{}0.05} & \textbf{0.775} & \textbf{\textless{}0.05} & \textbf{0.185} & \textbf{\textless{}0.05} \\
\multirow{2}{*}{\textit{R2}}                                        & \textit{RS}                                       & \textbf{0.170} & \textbf{\textless{}0.05} & \textbf{0.070} & \textbf{\textless{}0.05} & \textbf{0.698} & \textbf{\textless{}0.05} & \textbf{0.850} & \textbf{\textless{}0.05} & \textbf{0.131} & \textbf{\textless{}0.05} \\
                                                           & \textit{GS}                                       & \textbf{0.107} & \textbf{\textless{}0.05} & \textbf{0.052} & \textbf{\textless{}0.05} & \textbf{0.748} & \textbf{\textless{}0.05} & \textbf{0.750} & \textbf{\textless{}0.05} & \textbf{0.119} & \textbf{\textless{}0.05} \\
\multirow{2}{*}{\textit{R3}}                                        & \textit{RS}                                       & \textbf{0.068} & \textbf{\textless{}0.05} & \textbf{0.152} & \textbf{\textless{}0.05} & \textbf{0.618} & \textbf{0.208}           & \textbf{0.725} & \textbf{\textless{}0.05} & \textbf{0.241} & \textbf{\textless{}0.05} \\
                                                           & \textit{GS}                                       & \textbf{0.087} & \textbf{\textless{}0.05} & \textbf{0.085} & \textbf{\textless{}0.05} & \textbf{0.512} & \textbf{0.903}           & \textbf{0.625} & \textbf{0.122}           & \textbf{0.316} & \textbf{\textless{}0.05} \\
\multirow{2}{*}{\textit{R4}}                                        & \textit{RS}                                       & \textbf{0.072} & \textbf{\textless{}0.05} & \textbf{0.043} & \textbf{\textless{}0.05} & \textbf{0.848} & \textbf{\textless{}0.05} & \textbf{0.800} & \textbf{\textless{}0.05} & \textbf{0.203} & \textbf{\textless{}0.05} \\
                                                           & \textit{GS}                                       & \textbf{0.077} & \textbf{\textless{}0.05} & \textbf{0.068} & \textbf{\textless{}0.05} & \textbf{0.840} & \textbf{\textless{}0.05} & \textbf{0.850} & \textbf{\textless{}0.05} & \textbf{0.134} & \textbf{\textless{}0.05} \\ \midrule
                                                           &                                          & \multicolumn{10}{c}{\rewardTTCSD}                                                                                                                                                                                                   \\ \cmidrule{3-12} 
\multirow{2}{*}{\textit{R1}}                                        & \textit{RS}                                       & \textbf{0.045} & \textbf{\textless{}0.05} & \textbf{0.165} & \textbf{\textless{}0.05} & \textbf{0.860} & \textbf{\textless{}0.05} & \textbf{0.850} & \textbf{\textless{}0.05} & \textbf{0.111} & \textbf{\textless{}0.05} \\
                                                           & \textit{GS}                                       & \textbf{0.092} & \textbf{\textless{}0.05} & \textbf{0.185} & \textbf{\textless{}0.05} & \textbf{0.833} & \textbf{\textless{}0.05} & \textbf{0.825} & \textbf{\textless{}0.05} & \textbf{0.150} & \textbf{\textless{}0.05} \\
\multirow{2}{*}{\textit{R2}}                                        & \textit{RS}                                       & \textbf{0.102} & \textbf{\textless{}0.05} & \textbf{0.070} & \textbf{\textless{}0.05} & 0.637          & 0.140                    & \textbf{0.775} & \textbf{\textless{}0.05} & \textbf{0.188} & \textbf{\textless{}0.05} \\
                                                           & \textit{GS}                                       & \textbf{0.070} & \textbf{\textless{}0.05} & \textbf{0.098} & \textbf{\textless{}0.05} & 0.652          & 0.102                    & \textbf{0.775} & \textbf{\textless{}0.05} & \textbf{0.206} & \textbf{\textless{}0.05} \\
\multirow{2}{*}{\textit{R3}}                                        & \textit{RS}                                       & \textbf{0.163} & \textbf{\textless{}0.05} & \textbf{0.050} & \textbf{\textless{}0.05} & 0.610          & 0.239                    & \textbf{0.850} & \textbf{\textless{}0.05} & \textbf{0.126} & \textbf{\textless{}0.05} \\
                                                           & \textit{GS}                                       & \textbf{0.195} & \textbf{\textless{}0.05} & \textbf{0.092} & \textbf{\textless{}0.05} & 0.557          & 0.543                    & \textbf{0.775} & \textbf{\textless{}0.05} & \textbf{0.150} & \textbf{\textless{}0.05} \\
\multirow{2}{*}{\textit{R4}}                                        & \textit{RS}                                       & \textbf{0.025} & \textbf{\textless{}0.05} & \textbf{0.052} & \textbf{\textless{}0.05} & \textbf{0.868} & \textbf{\textless{}0.05} & \textbf{0.925} & \textbf{\textless{}0.05} & \textbf{0.075} & \textbf{\textless{}0.05} \\
                                                           & \textit{GS}                                       & \textbf{0.077} & \textbf{\textless{}0.05} & \textbf{0.055} & \textbf{\textless{}0.05} & \textbf{0.802} & \textbf{\textless{}0.05} & \textbf{0.800} & \textbf{\textless{}0.05} & \textbf{0.121} & \textbf{\textless{}0.05} \\ \midrule
                                                           &                                          & \multicolumn{10}{c}{\rewardTTCSN}                                                                                                                                                                                                   \\ \cmidrule{3-12} 
\multirow{2}{*}{\textit{R1}}                                        & \textit{RS}                                       & \textbf{0.035} & \textbf{\textless{}0.05} & \textbf{0.175} & \textbf{\textless{}0.05} & \textbf{0.757} & \textbf{\textless{}0.05} & \textbf{0.900} & \textbf{\textless{}0.05} & \textbf{0.084} & \textbf{\textless{}0.05} \\
                                                           & \textit{GS}                                       & \textbf{0.048} & \textbf{\textless{}0.05} & \textbf{0.158} & \textbf{\textless{}0.05} & \textbf{0.740} & \textbf{\textless{}0.05} & \textbf{0.900} & \textbf{\textless{}0.05} & \textbf{0.084} & \textbf{\textless{}0.05} \\
\multirow{2}{*}{\textit{R2}}                                        & \textit{RS}                                       & \textbf{0.045} & \textbf{\textless{}0.05} & \textbf{0.010} & \textbf{\textless{}0.05} & 0.630          & 0.164                    & \textbf{0.825} & \textbf{\textless{}0.05} & \textbf{0.151} & \textbf{\textless{}0.05} \\
                                                           & \textit{GS}                                       & \textbf{0.058} & \textbf{\textless{}0.05} & \textbf{0.102} & \textbf{\textless{}0.05} & \textbf{0.740} & \textbf{\textless{}0.05} & \textbf{0.800} & \textbf{\textless{}0.05} & \textbf{0.163} & \textbf{\textless{}0.05} \\
\multirow{2}{*}{\textit{R3}}                                        & \textit{RS}                                       & \textbf{0.270} & \textbf{\textless{}0.05} & \textbf{0.102} & \textbf{\textless{}0.05} & \textbf{0.718} & \textbf{\textless{}0.05} & \textbf{0.825} & \textbf{\textless{}0.05} & \textbf{0.139} & \textbf{\textless{}0.05} \\
                                                           & \textit{GS}                                       & \textbf{0.098} & \textbf{\textless{}0.05} & \textbf{0.122} & \textbf{\textless{}0.05} & 0.650          & 0.107                    & \textbf{0.700} & \textbf{\textless{}0.05} & \textbf{0.287} & \textbf{\textless{}0.05} \\
\multirow{2}{*}{\textit{R4}}                                        & \textit{RS}                                       & \textbf{0.035} & \textbf{\textless{}0.05} & \textbf{0.052} & \textbf{\textless{}0.05} & \textbf{0.815} & \textbf{\textless{}0.05} & \textbf{0.875} & \textbf{\textless{}0.05} & \textbf{0.138} & \textbf{\textless{}0.05} \\
                                                           & \textit{GS}                                       & \textbf{0.052} & \textbf{\textless{}0.05} & \textbf{0.058} & \textbf{\textless{}0.05} & \textbf{0.748} & \textbf{\textless{}0.05} & \textbf{0.875} & \textbf{\textless{}0.05} & \textbf{0.138} & \textbf{\textless{}0.05} \\ \bottomrule
\end{tabular}

%% file: tables/Appendix_RQ1_DistanceReward.tex
\begin{tabular}{cccccccccccc}
\toprule
\multicolumn{2}{c}{\multirow{2}{*}{\begin{tabular}[c]{@{}c@{}}\deepqtest\\ \textit{vs. (\randomStrategy, \greedyStrategy)}\end{tabular}}} & \multicolumn{2}{c}{\metricTTC}                   & \multicolumn{2}{c}{\metricDTO}                   & \multicolumn{2}{c}{\metricJerk}                  & \multicolumn{2}{c}{\metricCollisonNum}             & \multicolumn{2}{c}{\metricCollisionTime}         \\
\multicolumn{2}{c}{}                                                                                  & $\hat{A}_{12}$            & $p$                        & $\hat{A}_{12}$            & $p$                        & $\hat{A}_{12}$            & $p$                        & $\hat{A}_{12}$            & $p$                        & $\hat{A}_{12}$            & $p$                        \\ \midrule
                                                           &                                          & \multicolumn{10}{c}{\rewardDTORD}                                                                                                                                                                                                   \\ \cmidrule{3-12} 
\multirow{2}{*}{\textit{R1}}                                   & \textit{RS}                                   & \textbf{0.198} & \textbf{\textless{}0.05} & \textbf{0.055} & \textbf{\textless{}0.05} & \textbf{0.713} & \textbf{\textless{}0.05} & \textbf{0.750} & \textbf{\textless{}0.05} & \textbf{0.203} & \textbf{\textless{}0.05} \\
                                                               & \textit{GS}                                   & \textbf{0.098} & \textbf{\textless{}0.05} & \textbf{0.060} & \textbf{\textless{}0.05} & \textbf{0.719} & \textbf{\textless{}0.05} & \textbf{0.825} & \textbf{\textless{}0.05} & \textbf{0.158} & \textbf{\textless{}0.05} \\
\multirow{2}{*}{\textit{R2}}                                   & \textit{RS}                                   & \textbf{0.133} & \textbf{\textless{}0.05} & \textbf{0.072} & \textbf{\textless{}0.05} & 0.540          & 0.675                    & \textbf{0.850} & \textbf{\textless{}0.05} & \textbf{0.083} & \textbf{\textless{}0.05} \\
                                                               & \textit{GS}                                   & \textbf{0.085} & \textbf{\textless{}0.05} & \textbf{0.058} & \textbf{\textless{}0.05} & 0.675          & 0.060                    & \textbf{0.875} & \textbf{\textless{}0.05} & \textbf{0.095} & \textbf{\textless{}0.05} \\
\multirow{2}{*}{\textit{R3}}                                   & \textit{RS}                                   & \textbf{0.170} & \textbf{\textless{}0.05} & \textbf{0.075} & \textbf{\textless{}0.05} & 0.502          & 0.989                    & 0.575          & 0.356                    & 0.340          & 0.063                    \\
                                                               & \textit{GS}                                   & \textbf{0.163} & \textbf{\textless{}0.05} & \textbf{0.068} & \textbf{\textless{}0.05} & 0.547          & 0.617                    & \textbf{0.700} & \textbf{\textless{}0.05} & \textbf{0.287} & \textbf{\textless{}0.05} \\
\multirow{2}{*}{\textit{R4}}                                   & \textit{RS}                                   & \textbf{0.212} & \textbf{\textless{}0.05} & \textbf{0.075} & \textbf{\textless{}0.05} & \textbf{0.823} & \textbf{\textless{}0.05} & \textbf{0.750} & \textbf{\textless{}0.05} & \textbf{0.229} & \textbf{\textless{}0.05} \\
                                                               & \textit{GS}                                   & \textbf{0.268} & \textbf{\textless{}0.05} & \textbf{0.065} & \textbf{\textless{}0.05} & \textbf{0.858} & \textbf{\textless{}0.05} & \textbf{0.825} & \textbf{\textless{}0.05} & \textbf{0.175} & \textbf{\textless{}0.05} \\\midrule
                                                           &                                          & \multicolumn{10}{c}{\rewardDTORN}                                                                                                                                                                                                   \\ \cmidrule{3-12} 
\multirow{2}{*}{\textit{R1}}                                   & \textit{RS}                                   & \textbf{0.125} & \textbf{\textless{}0.05} & \textbf{0.055} & \textbf{\textless{}0.05} & \textbf{0.685} & \textbf{\textless{}0.05} & \textbf{0.875} & \textbf{\textless{}0.05} & \textbf{0.105} & \textbf{\textless{}0.05} \\
                                                               & \textit{GS}                                   & \textbf{0.130} & \textbf{\textless{}0.05} & \textbf{0.080} & \textbf{\textless{}0.05} & 0.665          & 0.076                    & \textbf{0.875} & \textbf{\textless{}0.05} & \textbf{0.107} & \textbf{\textless{}0.05} \\
\multirow{2}{*}{\textit{R2}}                                   & \textit{RS}                                   & \textbf{0.080} & \textbf{\textless{}0.05} & \textbf{0.060} & \textbf{\textless{}0.05} & 0.455          & 0.636                    & \textbf{0.775} & \textbf{\textless{}0.05} & \textbf{0.161} & \textbf{\textless{}0.05} \\
                                                               & \textit{GS}                                   & \textbf{0.115} & \textbf{\textless{}0.05} & \textbf{0.055} & \textbf{\textless{}0.05} & 0.497          & 0.989                    & \textbf{0.800} & \textbf{\textless{}0.05} & \textbf{0.163} & \textbf{\textless{}0.05} \\
\multirow{2}{*}{\textit{R3}}                                   & \textit{RS}                                   & \textbf{0.147} & \textbf{\textless{}0.05} & \textbf{0.010} & \textbf{\textless{}0.05} & \textbf{0.748} & \textbf{\textless{}0.05} & \textbf{0.850} & \textbf{\textless{}0.05} & \textbf{0.135} & \textbf{\textless{}0.05} \\
                                                               & \textit{GS}                                   & \textbf{0.150} & \textbf{\textless{}0.05} & \textbf{0.018} & \textbf{\textless{}0.05} & \textbf{0.770} & \textbf{\textless{}0.05} & \textbf{0.775} & \textbf{\textless{}0.05} & \textbf{0.171} & \textbf{\textless{}0.05} \\
\multirow{2}{*}{\textit{R4}}                                   & \textit{RS}                                   & \textbf{0.282} & \textbf{\textless{}0.05} & \textbf{0.010} & \textbf{\textless{}0.05} & \textbf{0.850} & \textbf{\textless{}0.05} & \textbf{0.900} & \textbf{\textless{}0.05} & \textbf{0.085} & \textbf{\textless{}0.05} \\
                                                               & \textit{GS}                                   & \textbf{0.328} & \textbf{\textless{}0.05} & \textbf{0.033} & \textbf{\textless{}0.05} & \textbf{0.772} & \textbf{\textless{}0.05} & \textbf{0.875} & \textbf{\textless{}0.05} & \textbf{0.087} & \textbf{\textless{}0.05} \\\midrule
                                                           &                                          & \multicolumn{10}{c}{\rewardDTOSD}                                                                                                                                                                                                   \\ \cmidrule{3-12} 
\multirow{2}{*}{\textit{R1}}                                   & \textit{RS}                                   & \textbf{0.168} & \textbf{\textless{}0.05} & \textbf{0.043} & \textbf{\textless{}0.05} & \textbf{0.723} & \textbf{\textless{}0.05} & \textbf{0.750} & \textbf{\textless{}0.05} & \textbf{0.261} & \textbf{\textless{}0.05} \\
                                                               & \textit{GS}                                   & \textbf{0.160} & \textbf{\textless{}0.05} & \textbf{0.040} & \textbf{\textless{}0.05} & \textbf{0.723} & \textbf{\textless{}0.05} & \textbf{0.675} & \textbf{\textless{}0.05} & \textbf{0.254} & \textbf{\textless{}0.05} \\
\multirow{2}{*}{\textit{R2}}                                   & \textit{RS}                                   & \textbf{0.048} & \textbf{\textless{}0.05} & \textbf{0.068} & \textbf{\textless{}0.05} & 0.378          & 0.190                    & \textbf{0.825} & \textbf{\textless{}0.05} & \textbf{0.099} & \textbf{\textless{}0.05} \\
                                                               & \textit{GS}                                   & \textbf{0.087} & \textbf{\textless{}0.05} & \textbf{0.080} & \textbf{\textless{}0.05} & 0.438          & 0.508                    & \textbf{0.850} & \textbf{\textless{}0.05} & \textbf{0.109} & \textbf{\textless{}0.05} \\
\multirow{2}{*}{\textit{R3}}                                   & \textit{RS}                                   & \textbf{0.100} & \textbf{\textless{}0.05} & \textbf{0.048} & \textbf{\textless{}0.05} & \textbf{0.730} & \textbf{\textless{}0.05} & \textbf{0.800} & \textbf{\textless{}0.05} & \textbf{0.158} & \textbf{\textless{}0.05} \\
                                                               & \textit{GS}                                   & \textbf{0.145} & \textbf{\textless{}0.05} & \textbf{0.055} & \textbf{\textless{}0.05} & \textbf{0.698} & \textbf{\textless{}0.05} & \textbf{0.750} & \textbf{\textless{}0.05} & \textbf{0.209} & \textbf{\textless{}0.05} \\
\multirow{2}{*}{\textit{R4}}                                   & \textit{RS}                                   & \textbf{0.282} & \textbf{\textless{}0.05} & \textbf{0.055} & \textbf{\textless{}0.05} & \textbf{0.830} & \textbf{\textless{}0.05} & \textbf{0.825} & \textbf{\textless{}0.05} & \textbf{0.144} & \textbf{\textless{}0.05} \\
                                                               & \textit{GS}                                   & \textbf{0.388} & \textbf{0.059}           & \textbf{0.065} & \textbf{\textless{}0.05} & \textbf{0.828} & \textbf{\textless{}0.05} & \textbf{0.800} & \textbf{\textless{}0.05} & \textbf{0.179} & \textbf{\textless{}0.05} \\\midrule
                                                           &                                          & \multicolumn{10}{c}{\rewardDTOSN}                                                                                                                                                                                                   \\ \cmidrule{3-12} 
\multirow{2}{*}{\textit{R1}}                                   & \textit{RS}                                   & \textbf{0.233} & \textbf{\textless{}0.05} & \textbf{0.070} & \textbf{\textless{}0.05} & \textbf{0.797} & \textbf{\textless{}0.05} & \textbf{0.825} & \textbf{\textless{}0.05} & \textbf{0.168} & \textbf{\textless{}0.05} \\
                                                               & \textit{GS}                                   & \textbf{0.278} & \textbf{\textless{}0.05} & \textbf{0.045} & \textbf{\textless{}0.05} & \textbf{0.834} & \textbf{\textless{}0.05} & \textbf{0.725} & \textbf{\textless{}0.05} & \textbf{0.259} & \textbf{\textless{}0.05} \\
\multirow{2}{*}{\textit{R2}}                                   & \textit{RS}                                   & \textbf{0.193} & \textbf{\textless{}0.05} & \textbf{0.090} & \textbf{\textless{}0.05} & 0.398          & 0.273                    & \textbf{0.825} & \textbf{\textless{}0.05} & \textbf{0.130} & \textbf{\textless{}0.05} \\
                                                               & \textit{GS}                                   & \textbf{0.165} & \textbf{\textless{}0.05} & \textbf{0.090} & \textbf{\textless{}0.05} & 0.470          & 0.756                    & \textbf{0.775} & \textbf{\textless{}0.05} & \textbf{0.269} & \textbf{\textless{}0.05} \\
\multirow{2}{*}{\textit{R3}}                                   & \textit{RS}                                   & \textbf{0.102} & \textbf{\textless{}0.05} & \textbf{0.037} & \textbf{\textless{}0.05} & 0.490          & 0.925                    & \textbf{0.800} & \textbf{\textless{}0.05} & \textbf{0.142} & \textbf{\textless{}0.05} \\
                                                               & \textit{GS}                                   & \textbf{0.105} & \textbf{\textless{}0.05} & \textbf{0.025} & \textbf{\textless{}0.05} & 0.472          & 0.776                    & \textbf{0.800} & \textbf{\textless{}0.05} & \textbf{0.138} & \textbf{\textless{}0.05} \\
\multirow{2}{*}{\textit{R4}}                                   & \textit{RS}                                   & \textbf{0.228} & \textbf{\textless{}0.05} & \textbf{0.035} & \textbf{\textless{}0.05} & \textbf{0.688} & \textbf{\textless{}0.05} & \textbf{0.850} & \textbf{\textless{}0.05} & \textbf{0.146} & \textbf{\textless{}0.05} \\
                                                               & \textit{GS}                                   & \textbf{0.205} & \textbf{\textless{}0.05} & \textbf{0.043} & \textbf{\textless{}0.05} & \textbf{0.730} & \textbf{\textless{}0.05} & \textbf{0.825} & \textbf{\textless{}0.05} & \textbf{0.165} & \textbf{\textless{}0.05} \\\bottomrule
\end{tabular}

%% file: tables/Appendix_RQ1_JerkReward.tex
\begin{tabular}{cccccccccccc}
\toprule
\multicolumn{2}{c}{\multirow{2}{*}{\begin{tabular}[c]{@{}c@{}}\deepqtest\\ \textit{vs. (\randomStrategy, \greedyStrategy)}\end{tabular}}} & \multicolumn{2}{c}{\metricTTC}                   & \multicolumn{2}{c}{\metricDTO}                   & \multicolumn{2}{c}{\metricJerk}                  & \multicolumn{2}{c}{\metricCollisonNum}             & \multicolumn{2}{c}{\metricCollisionTime}         \\
\multicolumn{2}{c}{}                                                                                  & $\hat{A}_{12}$            & $p$                        & $\hat{A}_{12}$            & $p$                        & $\hat{A}_{12}$            & $p$                        & $\hat{A}_{12}$            & $p$                        & $\hat{A}_{12}$            & $p$                        \\ \midrule
                                                           &                                          & \multicolumn{10}{c}{\rewardJerkRD}                                                                                                                                                                                                   \\ \cmidrule{3-12} 
\multirow{2}{*}{\textit{R1}}                                   & \textit{RS}                                   & 0.618          & 0.208                    & 0.338          & 0.081                    & \textbf{0.682} & \textbf{\textless{}0.05} & \textbf{0.675} & \textbf{\textless{}0.05} & \textbf{0.315} & \textbf{\textless{}0.05} \\
                                                               & \textit{GS}                                   & 0.623          & 0.190                    & 0.422          & 0.409                    & 0.595          & 0.310                    & 0.650          & 0.058                    & 0.354          & 0.076                    \\
\multirow{2}{*}{\textit{R2}}                                   & \textit{RS}                                   & 0.405          & 0.182                    & 0.470          & 0.756                    & 0.432          & 0.473                    & \textbf{0.750} & \textbf{\textless{}0.05} & \textbf{0.244} & \textbf{\textless{}0.05} \\
                                                               & \textit{GS}                                   & \textbf{0.263} & \textbf{\textless{}0.05} & 0.472          & 0.776                    & 0.422          & 0.409                    & \textbf{0.825} & \textbf{\textless{}0.05} & \textbf{0.136} & \textbf{\textless{}0.05} \\
\multirow{2}{*}{\textit{R3}}                                   & \textit{RS}                                   & \textbf{0.160} & \textbf{\textless{}0.05} & 0.610          & 0.239                    & 0.390          & 0.239                    & 0.575          & 0.356                    & 0.420          & 0.356                    \\
                                                               & \textit{GS}                                   & \textbf{0.152} & \textbf{\textless{}0.05} & 0.578          & 0.409                    & 0.405          & 0.310                    & 0.650          & 0.058                    & \textbf{0.335} & \textbf{\textless{}0.05} \\
\multirow{2}{*}{\textit{R4}}                                   & \textit{RS}                                   & \textbf{0.290} & \textbf{\textless{}0.05} & \textbf{0.228} & \textbf{\textless{}0.05} & \textbf{0.855} & \textbf{\textless{}0.05} & \textbf{0.800} & \textbf{\textless{}0.05} & \textbf{0.190} & \textbf{\textless{}0.05} \\
                                                               & \textit{GS}                                   & \textbf{0.180} & \textbf{\textless{}0.05} & \textbf{0.180} & \textbf{\textless{}0.05} & \textbf{0.815} & \textbf{\textless{}0.05} & \textbf{0.775} & \textbf{\textless{}0.05} & \textbf{0.229} & \textbf{\textless{}0.05} \\ \midrule
                                                           &                                          & \multicolumn{10}{c}{\rewardJerkRN}                                                                                                                                                                                                   \\ \cmidrule{3-12} 
\multirow{2}{*}{\textit{R1}}                                   & \textit{RS}                                   & \textbf{0.290} & \textbf{\textless{}0.05} & 0.435          & 0.490                    & \textbf{0.782} & \textbf{\textless{}0.05} & \textbf{0.675} & \textbf{\textless{}0.05} & 0.360          & 0.104                    \\
                                                               & \textit{GS}                                   & \textbf{0.225} & \textbf{\textless{}0.05} & 0.340          & 0.086                    & \textbf{0.782} & \textbf{\textless{}0.05} & \textbf{0.725} & \textbf{\textless{}0.05} & \textbf{0.274} & \textbf{\textless{}0.05} \\
\multirow{2}{*}{\textit{R2}}                                   & \textit{RS}                                   & \textbf{0.177} & \textbf{\textless{}0.05} & 0.535          & 0.715                    & 0.547          & 0.617                    & \textbf{0.675} & \textbf{\textless{}0.05} & \textbf{0.286} & \textbf{\textless{}0.05} \\
                                                               & \textit{GS}                                   & \textbf{0.263} & \textbf{\textless{}0.05} & 0.512          & 0.903                    & 0.560          & 0.525                    & \textbf{0.700} & \textbf{\textless{}0.05} & \textbf{0.282} & \textbf{\textless{}0.05} \\
\multirow{2}{*}{\textit{R3}}                                   & \textit{RS}                                   & \textbf{0.280} & \textbf{\textless{}0.05} & 0.738×         & \textless{}0.05          & 0.487          & 0.903                    & 0.600          & 0.196                    & 0.389          & 0.162                    \\
                                                               & \textit{GS}                                   & \textbf{0.255} & \textbf{\textless{}0.05} & 0.642          & 0.126                    & 0.510          & 0.925                    & 0.550          & 0.534                    & 0.424          & 0.359                    \\
\multirow{2}{*}{\textit{R4}}                                   & \textit{RS}                                   & \textbf{0.152} & \textbf{\textless{}0.05} & \textbf{0.275} & \textbf{\textless{}0.05} & \textbf{0.690} & \textbf{\textless{}0.05} & \textbf{0.700} & \textbf{\textless{}0.05} & \textbf{0.304} & \textbf{\textless{}0.05} \\
                                                               & \textit{GS}                                   & \textbf{0.145} & \textbf{\textless{}0.05} & \textbf{0.215} & \textbf{\textless{}0.05} & \textbf{0.770} & \textbf{\textless{}0.05} & \textbf{0.775} & \textbf{\textless{}0.05} & \textbf{0.210} & \textbf{\textless{}0.05} \\ \midrule
                                                           &                                          & \multicolumn{10}{c}{\rewardJerkSD}                                                                                                                                                                                                   \\ \cmidrule{3-12} 
\multirow{2}{*}{\textit{R1}}                                   & \textit{RS}                                   & \textbf{0.237} & \textbf{\textless{}0.05} & 0.427          & 0.441                    & 0.650          & 0.107                    & \textbf{0.675} & \textbf{\textless{}0.05} & \textbf{0.273} & \textbf{\textless{}0.05} \\
                                                               & \textit{GS}                                   & \textbf{0.253} & \textbf{\textless{}0.05} & 0.502          & 0.989                    & 0.568          & 0.473                    & \textbf{0.700} & \textbf{\textless{}0.05} & \textbf{0.271} & \textbf{\textless{}0.05} \\
\multirow{2}{*}{\textit{R2}}                                   & \textit{RS}                                   & \textbf{0.273} & \textbf{\textless{}0.05} & 0.360          & 0.133                    & 0.585          & 0.365                    & \textbf{0.725} & \textbf{\textless{}0.05} & \textbf{0.240} & \textbf{\textless{}0.05} \\
                                                               & \textit{GS}                                   & \textbf{0.292} & \textbf{\textless{}0.05} & 0.482          & 0.860                    & 0.625          & 0.181                    & \textbf{0.700} & \textbf{\textless{}0.05} & \textbf{0.300} & \textbf{\textless{}0.05} \\
\multirow{2}{*}{\textit{R3}}                                   & \textit{RS}                                   & \textbf{0.085} & \textbf{\textless{}0.05} & 0.603          & 0.273                    & \textbf{0.682} & \textbf{\textless{}0.05} & \textbf{0.750} & \textbf{\textless{}0.05} & \textbf{0.239} & \textbf{\textless{}0.05} \\
                                                               & \textit{GS}                                   & \textbf{0.237} & \textbf{\textless{}0.05} & 0.485          & 0.882                    & 0.593          & 0.323                    & \textbf{0.700} & \textbf{\textless{}0.05} & \textbf{0.297} & \textbf{\textless{}0.05} \\
\multirow{2}{*}{\textit{R4}}                                   & \textit{RS}                                   & \textbf{0.365} & \textbf{0.074}           & \textbf{0.145} & \textbf{\textless{}0.05} & \textbf{0.720} & \textbf{\textless{}0.05} & \textbf{0.825} & \textbf{\textless{}0.05} & \textbf{0.165} & \textbf{\textless{}0.05} \\
                                                               & \textit{GS}                                   & \textbf{0.325} & \textbf{\textless{}0.05} & \textbf{0.158} & \textbf{\textless{}0.05} & \textbf{0.734} & \textbf{\textless{}0.05} & \textbf{0.825} & \textbf{\textless{}0.05} & \textbf{0.171} & \textbf{\textless{}0.05} \\ \midrule
                                                           &                                          & \multicolumn{10}{c}{\rewardJerkSN}                                                                                                                                                                                                   \\ \cmidrule{3-12} 
\multirow{2}{*}{\textit{R1}}                               & \textit{RS}                              & 0.445          & 0.561                    & 0.393          & 0.250                    & 0.680          & 0.053                    & 0.625          & 0.110                    & 0.390          & 0.176                    \\
                                                           & \textit{GS}                              & 0.365          & 0.148                    & 0.557          & 0.543                    & 0.593          & 0.323                    & 0.650          & 0.051                    & \textbf{0.333} & \textbf{\textless{}0.05} \\
\multirow{2}{*}{\textit{R2}}                               & \textit{RS}                              & \textbf{0.223} & \textbf{\textless{}0.05} & 0.485          & 0.882                    & \textbf{0.695} & \textbf{\textless{}0.05} & \textbf{0.725} & \textbf{\textless{}0.05} & \textbf{0.297} & \textbf{\textless{}0.05} \\
                                                           & \textit{GS}                              & \textbf{0.235} & \textbf{\textless{}0.05} & 0.455          & 0.636                    & 0.660          & 0.086                    & \textbf{0.700} & \textbf{\textless{}0.05} & \textbf{0.270} & \textbf{\textless{}0.05} \\
\multirow{2}{*}{\textit{R3}}                               & \textit{RS}                              & \textbf{0.110} & \textbf{\textless{}0.05} & 0.610          & 0.239                    & \textbf{0.698} & \textbf{\textless{}0.05} & \textbf{0.750} & \textbf{\textless{}0.05} & \textbf{0.235} & \textbf{\textless{}0.05} \\
                                                           & \textit{GS}                              & \textbf{0.182} & \textbf{\textless{}0.05} & 0.795×         & \textless{}0.05          & \textbf{0.682} & \textbf{\textless{}0.05} & 0.625          & 0.118                    & 0.366          & 0.135                    \\
\multirow{2}{*}{\textit{R4}}                               & \textit{RS}                              & \textbf{0.355} & \textbf{\textless{}0.05} & \textbf{0.235} & \textbf{\textless{}0.05} & 0.647          & 0.114                    & \textbf{0.825} & \textbf{\textless{}0.05} & \textbf{0.181} & \textbf{\textless{}0.05} \\
                                                           & \textit{GS}                              & \textbf{0.320} & \textbf{\textless{}0.05} & \textbf{0.215} & \textbf{\textless{}0.05} & 0.650          & 0.107                    & \textbf{0.825} & \textbf{\textless{}0.05} & \textbf{0.190} & \textbf{\textless{}0.05} \\ \bottomrule
\end{tabular}

%% file: tables/RQ2_Appendix.tex
\begin{tabular}{ccccccccccccccccc}
\toprule
\multicolumn{3}{c}{\multirow{2}{*}{Reward Comparison}}             & \multicolumn{2}{c}{\metricTTC}                   & \multicolumn{2}{c}{\metricDTO}                 & \multicolumn{2}{c}{\metricJerk}                  & \multicolumn{2}{c}{\metricCollisonNum}             & \multicolumn{2}{c}{\metricCollisionTime}         & \multicolumn{2}{c}{\metricDiversityAPI}              & \multicolumn{2}{c}{\metricDiversityScenario} \\
                             &                              &             & $\hat{A}_{12}$            & $p$                        & $\hat{A}_{12}$            & $p$                        & $\hat{A}_{12}$            & $p$                        & $\hat{A}_{12}$            & $p$                        & $\hat{A}_{12}$            & $p$                        & $\hat{A}_{12}$            & $p$                        & $\hat{A}_{12}$      & $p$                \\ \midrule
                             &                              &             & \multicolumn{14}{c}{\textit{RD}}                                                                                                                                                                                                                                                                    \\ \cmidrule{4-17} 
\multirow{3}{*}{\textit{R1}} & \multirow{2}{*}{\rewardTTC} & \rewardDTO & 0.495          & 0.833                    & 0.838×         & \textless{}0.05          & \textbf{0.695} & \textbf{\textless{}0.05} & 0.525          & 0.722                    & 0.496          & 0.978                    & 0.551          & 0.583                    & 0.521    & 0.828            \\
                             &                              & \rewardJerk & \textbf{0.133} & \textbf{\textless{}0.05} & \textbf{0.250} & \textbf{\textless{}0.05} & \textbf{0.695} & \textbf{\textless{}0.05} & 0.625          & 0.099                    & 0.364          & 0.134                    & 0.338          & 0.079                    & 0.334    & 0.073            \\
                             & \rewardDTO                  & \rewardJerk & \textbf{0.128} & \textbf{\textless{}0.05} & \textbf{0.100} & \textbf{\textless{}0.05} & 0.542          & 0.655                    & 0.600          & 0.196                    & 0.354          & 0.108                    & 0.276×         & \textless{}0.05          & 0.302×   & \textless{}0.05  \\
\multirow{3}{*}{\textit{R2}} & \multirow{2}{*}{\rewardTTC} & \rewardDTO & \textbf{0.125} & \textbf{\textless{}0.05} & 0.850×         & \textless{}0.05          & \textbf{0.693} & \textbf{\textless{}0.05} & 0.425          & 0.225                    & 0.535          & 0.692                    & 0.514          & 0.884                    & 0.492    & 0.946            \\
                             &                              & \rewardJerk & \textbf{0.217} & \textbf{\textless{}0.05} & 0.323          & 0.056                    & \textbf{0.710} & \textbf{\textless{}0.05} & 0.475          & 0.722                    & 0.459          & 0.650                    & 0.460          & 0.662                    & 0.445    & 0.560            \\
                             & \rewardDTO                  & \rewardJerk & 0.440          & 0.704                    & \textbf{0.170} & \textbf{\textless{}0.05} & 0.552          & 0.579                    & 0.550          & 0.394                    & 0.453          & 0.604                    & 0.471          & 0.758                    & 0.476    & 0.807            \\
\multirow{3}{*}{\textit{R3}} & \multirow{2}{*}{\rewardTTC} & \rewardDTO & 0.458          & 0.640                    & 0.855×         & \textless{}0.05          & 0.522          & 0.818                    & 0.550          & 0.534                    & 0.471          & 0.756                    & 0.565          & 0.484                    & 0.547    & 0.421            \\
                             &                              & \rewardJerk & \textbf{0.278} & \textbf{\textless{}0.05} & \textbf{0.113} & \textbf{\textless{}0.05} & 0.562          & 0.507                    & 0.550          & 0.534                    & 0.386          & 0.207                    & 0.393          & 0.244                    & 0.422    & 0.407            \\
                             & \rewardDTO                  & \rewardJerk & \textbf{0.347} & \textbf{\textless{}0.05} & \textbf{0.055} & \textbf{\textless{}0.05} & 0.545          & 0.636                    & 0.500          & 1.000                    & 0.431          & 0.441                    & 0.329          & 0.063                    & 0.371    & 0.085            \\
\multirow{3}{*}{\textit{R4}} & \multirow{2}{*}{\rewardTTC} & \rewardDTO & \textbf{0.158} & \textbf{\textless{}0.05} & 0.868×         & \textless{}0.05          & 0.578          & 0.409                    & 0.600          & 0.154                    & 0.461          & 0.678                    & 0.499          & 1.000                    & 0.284×   & \textless{}0.05  \\
                             &                              & \rewardJerk & \textbf{0.122} & \textbf{\textless{}0.05} & \textbf{0.188} & \textbf{\textless{}0.05} & 0.583          & 0.379                    & 0.600          & 0.154                    & 0.378          & 0.183                    & 0.407          & 0.318                    & 0.296×   & \textless{}0.05  \\
                             & \rewardDTO                  & \rewardJerk & 0.420          & 0.113                    & \textbf{0.100} & \textbf{\textless{}0.05} & 0.530          & 0.756                    & 0.500          & 1.000                    & 0.419          & 0.373                    & 0.420          & 0.390                    & 0.454    & 0.625            \\ \midrule
                             &                              &             & \multicolumn{14}{c}{\textit{RN}}                                                                                                                                                                                                                                                                    \\ \cmidrule{4-17} 
\multirow{3}{*}{\textit{R1}} & \multirow{2}{*}{\rewardTTC} & \rewardDTO & 0.525          & 0.584                    & 0.880×         & \textless{}0.05          & 0.620          & 0.199                    & 0.475          & 0.696                    & 0.667          & 0.068                    & \textbf{0.703} & \textbf{\textless{}0.05} & 0.686    & \textless{}0.05  \\
                             &                              & \rewardJerk & \textbf{0.188} & \textbf{\textless{}0.05} & \textbf{0.295} & \textbf{\textless{}0.05} & 0.578          & 0.409                    & 0.575          & 0.302                    & 0.396          & 0.258                    & 0.396          & 0.263                    & 0.380    & 0.197            \\
                             & \rewardDTO                  & \rewardJerk & \textbf{0.290} & \textbf{\textless{}0.05} & \textbf{0.117} & \textbf{\textless{}0.05} & 0.455          & 0.636                    & 0.600          & 0.154                    & \textbf{0.264} & \textbf{\textless{}0.05} & 0.255×         & \textless{}0.05          & 0.231×   & \textless{}0.05  \\
\multirow{3}{*}{\textit{R2}} & \multirow{2}{*}{\rewardTTC} & \rewardDTO & \textbf{0.170} & \textbf{\textless{}0.05} & 0.855×         & \textless{}0.05          & \textbf{0.757} & \textbf{\textless{}0.05} & 0.550          & 0.447                    & 0.476          & 0.797                    & 0.527          & 0.765                    & 0.540    & 0.640            \\
                             &                              & \rewardJerk & \textbf{0.225} & \textbf{\textless{}0.05} & \textbf{0.145} & \textbf{\textless{}0.05} & \textbf{0.685} & \textbf{\textless{}0.05} & \textbf{0.650} & \textbf{\textless{}0.05} & \textbf{0.295} & \textbf{\textless{}0.05} & 0.320×         & \textless{}0.05          & 0.338×   & \textless{}0.05  \\
                             & \rewardDTO                  & \rewardJerk & 0.443          & 0.521                    & \textbf{0.048} & \textbf{\textless{}0.05} & 0.445          & 0.561                    & 0.600          & 0.196                    & 0.333          & 0.060                    & 0.292×         & \textless{}0.05          & 0.352    & 0.054            \\
\multirow{3}{*}{\textit{R3}} & \multirow{2}{*}{\rewardTTC} & \rewardDTO & 0.378          & 0.189                    & 0.820×         & \textless{}0.05          & 0.472          & 0.776                    & 0.425          & 0.302                    & 0.547          & 0.603                    & 0.565          & 0.476                    & 0.501    & 1.000            \\
                             &                              & \rewardJerk & \textbf{0.168} & \textbf{\textless{}0.05} & \textbf{0.072} & \textbf{\textless{}0.05} & 0.588          & 0.351                    & 0.600          & 0.215                    & 0.352          & 0.094                    & 0.443          & 0.539                    & 0.326    & 0.061            \\
                             & \rewardDTO                  & \rewardJerk & \textbf{0.307} & \textbf{\textless{}0.05} & \textbf{0.007} & \textbf{\textless{}0.05} & 0.650          & 0.107                    & \textbf{0.675} & \textbf{\textless{}0.05} & \textbf{0.307} & \textbf{\textless{}0.05} & 0.379          & 0.189                    & 0.241×   & \textless{}0.05  \\
\multirow{3}{*}{\textit{R4}} & \multirow{2}{*}{\rewardTTC} & \rewardDTO & \textbf{0.233} & \textbf{\textless{}0.05} & 0.880×         & \textless{}0.05          & 0.562          & 0.507                    & 0.450          & 0.447                    & 0.613          & 0.219                    & 0.664          & 0.074                    & 0.453    & 0.811            \\
                             &                              & \rewardJerk & 0.335          & 0.076                    & \textbf{0.177} & \textbf{\textless{}0.05} & 0.502          & 0.989                    & 0.575          & 0.325                    & 0.465          & 0.708                    & 0.509          & 0.935                    & 0.431    & 0.464            \\
                             & \rewardDTO                  & \rewardJerk & 0.652          & 0.514                    & \textbf{0.072} & \textbf{\textless{}0.05} & 0.475          & 0.797                    & 0.625          & 0.083                    & 0.343          & 0.085                    & 0.331          & 0.067                    & 0.459    & 0.464            \\ \midrule
                             &                              &             & \multicolumn{14}{c}{\textit{SD}}                                                                                                                                                                                                                                                                    \\ \cmidrule{4-17} 
\multirow{3}{*}{\textit{R1}} & \multirow{2}{*}{\rewardTTC} & \rewardDTO & 0.512          & 0.903                    & 0.912×         & \textless{}0.05          & 0.675          & 0.060                    & 0.550          & 0.482                    & 0.549          & 0.602                    & 0.603          & 0.269                    & 0.544    & 0.645            \\
                             &                              & \rewardJerk & \textbf{0.260} & \textbf{\textless{}0.05} & \textbf{0.245} & \textbf{\textless{}0.05} & \textbf{0.823} & \textbf{\textless{}0.05} & 0.625          & 0.099                    & 0.431          & 0.453                    & 0.500          & 1.000                    & 0.361    & 0.136            \\
                             & \rewardDTO                  & \rewardJerk & \textbf{0.270} & \textbf{\textless{}0.05} & \textbf{0.050} & \textbf{\textless{}0.05} & \textbf{0.698} & \textbf{\textless{}0.05} & 0.575          & 0.341                    & 0.383          & 0.194                    & 0.390          & 0.234                    & 0.371    & 0.166            \\
\multirow{3}{*}{\textit{R2}} & \multirow{2}{*}{\rewardTTC} & \rewardDTO & \textbf{0.087} & \textbf{\textless{}0.05} & 0.865×         & \textless{}0.05          & \textbf{0.760} & \textbf{\textless{}0.05} & 0.450          & 0.394                    & 0.568          & 0.466                    & 0.550          & 0.592                    & 0.690    & 0.103            \\
                             &                              & \rewardJerk & \textbf{0.113} & \textbf{\textless{}0.05} & \textbf{0.147} & \textbf{\textless{}0.05} & 0.642          & 0.126                    & 0.600          & 0.178                    & 0.331          & 0.065                    & 0.341          & 0.087                    & 0.534    & 0.860            \\
                             & \rewardDTO                  & \rewardJerk & 0.345          & 0.337                    & \textbf{0.100} & \textbf{\textless{}0.05} & 0.328          & 0.064                    & \textbf{0.650} & \textbf{\textless{}0.05} & \textbf{0.259} & \textbf{\textless{}0.05} & 0.295×         & \textless{}0.05          & 0.302×   & \textless{}0.05  \\
\multirow{3}{*}{\textit{R3}} & \multirow{2}{*}{\rewardTTC} & \rewardDTO & 0.522          & 0.603                    & 0.703×         & \textless{}0.05          & 0.492          & 0.946                    & 0.550          & 0.394                    & 0.466          & 0.718                    & 0.506          & 0.955                    & 0.480    & 0.966            \\
                             &                              & \rewardJerk & \textbf{0.240} & \textbf{\textless{}0.05} & \textbf{0.048} & \textbf{\textless{}0.05} & 0.510          & 0.925                    & 0.625          & 0.064                    & \textbf{0.274} & \textbf{\textless{}0.05} & 0.285×         & \textless{}0.05          & 0.256×   & \textless{}0.05  \\
                             & \rewardDTO                  & \rewardJerk & \textbf{0.275} & \textbf{\textless{}0.05} & \textbf{0.040} & \textbf{\textless{}0.05} & 0.502          & 0.989                    & 0.575          & 0.302                    & 0.330          & 0.063                    & 0.328          & 0.061                    & 0.338×   & \textless{}0.05  \\
\multirow{3}{*}{\textit{R4}} & \multirow{2}{*}{\rewardTTC} & \rewardDTO & \textbf{0.205} & \textbf{\textless{}0.05} & 0.885×         & \textless{}0.05          & 0.605          & 0.262                    & 0.550          & 0.447                    & 0.551          & 0.571                    & 0.552          & 0.563                    & 0.329    & 0.065            \\
                             &                              & \rewardJerk & \textbf{0.120} & \textbf{\textless{}0.05} & \textbf{0.240} & \textbf{\textless{}0.05} & 0.613          & 0.229                    & 0.575          & 0.270                    & 0.346          & 0.094                    & 0.364          & 0.140                    & 0.300×   & \textless{}0.05  \\
                             & \rewardDTO                  & \rewardJerk & 0.450          & 0.386                    & \textbf{0.092} & \textbf{\textless{}0.05} & 0.535          & 0.715                    & 0.525          & 0.740                    & \textbf{0.321} & \textbf{\textless{}0.05} & 0.334          & 0.069                    & 0.431    & 0.464            \\ \midrule
                             &                              &             & \multicolumn{14}{c}{\textit{SN}}                                                                                                                                                                                                                                                                    \\ \cmidrule{4-17} 
\multirow{3}{*}{\textit{R1}} & \multirow{2}{*}{\rewardTTC} & \rewardDTO & 0.323          & 0.056                    & 0.815×         & \textless{}0.05          & 0.522          & 0.818                    & 0.575          & 0.270                    & 0.430          & 0.449                    & 0.441          & 0.529                    & 0.406    & 0.315            \\
                             &                              & \rewardJerk & \textbf{0.152} & \textbf{\textless{}0.05} & \textbf{0.180} & \textbf{\textless{}0.05} & 0.630          & 0.164                    & \textbf{0.675} & \textbf{\textless{}0.05} & \textbf{0.250} & \textbf{\textless{}0.05} & 0.251×         & \textless{}0.05          & 0.273×   & \textless{}0.05  \\
                             & \rewardDTO                  & \rewardJerk & 0.335          & 0.076                    & \textbf{0.055} & \textbf{\textless{}0.05} & \textbf{0.713} & \textbf{\textless{}0.05} & 0.600          & 0.208                    & 0.380          & 0.183                    & 0.417          & 0.378                    & 0.479    & 0.828            \\
\multirow{3}{*}{\textit{R2}} & \multirow{2}{*}{\rewardTTC} & \rewardDTO & \textbf{0.168} & \textbf{\textless{}0.05} & 0.848×         & \textless{}0.05          & \textbf{0.790} & \textbf{\textless{}0.05} & 0.500          & 1.000                    & 0.646          & 0.105                    & 0.626          & 0.161                    & 0.650    & 0.106            \\
                             &                              & \rewardJerk & \textbf{0.080} & \textbf{\textless{}0.05} & \textbf{0.135} & \textbf{\textless{}0.05} & 0.580          & 0.394                    & 0.625          & 0.083                    & 0.326          & 0.058                    & 0.318×         & \textless{}0.05          & 0.458    & 0.654            \\
                             & \rewardDTO                  & \rewardJerk & 0.390          & 0.133                    & \textbf{0.077} & \textbf{\textless{}0.05} & 0.235×         & \textless{}0.05          & 0.625          & 0.083                    & \textbf{0.239} & \textbf{\textless{}0.05} & 0.261×         & \textless{}0.05          & 0.302×   & \textless{}0.05  \\
\multirow{3}{*}{\textit{R3}} & \multirow{2}{*}{\rewardTTC} & \rewardDTO & 0.688×         & \textless{}0.05          & 0.750×         & \textless{}0.05          & 0.640          & 0.133                    & 0.500          & 1.000                    & 0.499          & 1.000                    & 0.468          & 0.729                    & 0.390    & 0.237            \\
                             &                              & \rewardJerk & 0.325          & 0.060                    & \textbf{0.010} & \textbf{\textless{}0.05} & 0.500          & 1.000                    & 0.550          & 0.482                    & 0.383          & 0.202                    & 0.321          & 0.053                    & 0.155×   & \textless{}0.05  \\
                             & \rewardDTO                  & \rewardJerk & \textbf{0.287} & \textbf{\textless{}0.05} & \textbf{0.005} & \textbf{\textless{}0.05} & 0.323          & 0.056                    & 0.550          & 0.482                    & 0.372          & 0.165                    & 0.356          & 0.120                    & 0.292×   & \textless{}0.05  \\
\multirow{3}{*}{\textit{R4}} & \multirow{2}{*}{\rewardTTC} & \rewardDTO & 0.355          & 0.273                    & 0.895×         & \textless{}0.05          & 0.552          & 0.579                    & 0.500          & 1.000                    & 0.574          & 0.417                    & 0.576          & 0.403                    & 0.416    & 0.371            \\
                             &                              & \rewardJerk & \textbf{0.198} & \textbf{\textless{}0.05} & 0.405          & 0.310                    & 0.583          & 0.379                    & 0.550          & 0.482                    & 0.395          & 0.255                    & 0.393          & 0.245                    & 0.315×   & \textless{}0.05  \\
                             & \rewardDTO                  & \rewardJerk & 0.347          & 0.064                    & \textbf{0.098} & \textbf{\textless{}0.05} & 0.542          & 0.655                    & 0.550          & 0.482                    & 0.335          & 0.071                    & 0.333          & 0.067                    & 0.341    & 0.087            \\ \bottomrule
\end{tabular}

%% file: main.bbl
\begin{thebibliography}{10}

\bibitem{abdessalem2018testing}
Raja~Ben Abdessalem, Shiva Nejati, Lionel~C Briand, and Thomas Stifter.
\newblock Testing vision-based control systems using learnable evolutionary algorithms.
\newblock In {\em 2018 IEEE/ACM 40th International Conference on Software Engineering (ICSE)}, pages 1016--1026. IEEE, 2018.

\bibitem{abdessalem2018ASE}
Raja~Ben Abdessalem, Annibale Panichella, Shiva Nejati, Lionel~C Briand, and Thomas Stifter.
\newblock Testing autonomous cars for feature interaction failures using many-objective search.
\newblock In {\em 2018 33rd IEEE/ACM International Conference on Automated Software Engineering (ASE)}, pages 143--154. IEEE, 2018.

\bibitem{abedin2020data}
Sarder~Fakhrul Abedin, Md~Shirajum Munir, Nguyen~H Tran, Zhu Han, and Choong~Seon Hong.
\newblock Data freshness and energy-efficient uav navigation optimization: A deep reinforcement learning approach.
\newblock {\em IEEE Transactions on Intelligent Transportation Systems}, 2020.

\bibitem{akiba2019optuna}
Takuya Akiba, Shotaro Sano, Toshihiko Yanase, Takeru Ohta, and Masanori Koyama.
\newblock Optuna: A next-generation hyperparameter optimization framework.
\newblock In {\em Proceedings of the 25th ACM SIGKDD international conference on knowledge discovery \& data mining}, pages 2623--2631, 2019.

\bibitem{arcuri2011practical}
Andrea Arcuri and Lionel Briand.
\newblock A practical guide for using statistical tests to assess randomized algorithms in software engineering.
\newblock In {\em 2011 33rd International Conference on Software Engineering (ICSE)}, pages 1--10. IEEE, 2011.

\bibitem{bae2020self}
Il~Bae, Jaeyoung Moon, Junekyo Jhung, H~Suk, Taewoo Kim, Hyunbin Park, Jaekwang Cha, Jincheol Kim, Dohyun Kim, and Shiho Kim.
\newblock Self-driving like a human driver instead of a robocar: Personalized comfortable driving experience for autonomous vehicles.
\newblock {\em arXiv preprint arXiv:2001.03908}, 2020.

\bibitem{bella2011collision}
Francesco Bella and Roberta Russo.
\newblock A collision warning system for rear-end collision: a driving simulator study.
\newblock {\em Procedia-social and behavioral sciences}, 20:676--686, 2011.

\bibitem{ben2016testing}
Raja Ben~Abdessalem, Shiva Nejati, Lionel~C Briand, and Thomas Stifter.
\newblock Testing advanced driver assistance systems using multi-objective search and neural networks.
\newblock In {\em Proceedings of the 31st IEEE/ACM International Conference on Automated Software Engineering}, pages 63--74, 2016.

\bibitem{bijelic2018benchmarking}
Mario Bijelic, Tobias Gruber, and Werner Ritter.
\newblock Benchmarking image sensors under adverse weather conditions for autonomous driving.
\newblock In {\em 2018 IEEE Intelligent Vehicles Symposium (IV)}, pages 1773--1779. IEEE, 2018.

\bibitem{string_sim}
Paul~E. Black.
\newblock Ratcliff/obershelp pattern recognition.
\newblock In {\em Dictionary of Algorithms and Data Structures [online]}. Available from: \url{https://www.nist.gov/dads/HTML/ratcliffObershelp.html}, 2021.

\bibitem{buhler2008evolutionary}
Oliver B{\"u}hler and Joachim Wegener.
\newblock Evolutionary functional testing.
\newblock {\em Computers \& Operations Research}, 35(10):3144--3160, 2008.

\bibitem{calo2020generating}
Alessandro Cal{\`o}, Paolo Arcaini, Shaukat Ali, Florian Hauer, and Fuyuki Ishikawa.
\newblock Generating avoidable collision scenarios for testing autonomous driving systems.
\newblock In {\em 2020 IEEE 13th International Conference on Software Testing, Validation and Verification (ICST)}, pages 375--386. IEEE, 2020.

\bibitem{KVtest}
Indra~M Chakravarty, JD~Roy, and Radha~Govind Laha.
\newblock Handbook of methods of applied statistics.
\newblock 1967.

\bibitem{chen2021adversarial}
Baiming Chen, Xiang Chen, Qiong Wu, and Liang Li.
\newblock Adversarial evaluation of autonomous vehicles in lane-change scenarios.
\newblock {\em IEEE Transactions on Intelligent Transportation Systems}, 2021.

\bibitem{chen2019model}
Jianyu Chen, Bodi Yuan, and Masayoshi Tomizuka.
\newblock Model-free deep reinforcement learning for urban autonomous driving.
\newblock In {\em 2019 IEEE intelligent transportation systems conference (ITSC)}, pages 2765--2771. IEEE, 2019.

\bibitem{chen2020conditional}
Long Chen, Xuemin Hu, Bo~Tang, and Yu~Cheng.
\newblock Conditional dqn-based motion planning with fuzzy logic for autonomous driving.
\newblock {\em IEEE Transactions on Intelligent Transportation Systems}, 2020.

\bibitem{chen2017multi}
Xiaozhi Chen, Huimin Ma, Ji~Wan, Bo~Li, and Tian Xia.
\newblock Multi-view 3d object detection network for autonomous driving.
\newblock In {\em Proceedings of the IEEE conference on Computer Vision and Pattern Recognition}, pages 1907--1915, 2017.

\bibitem{codevilla2018offline}
Felipe Codevilla, Antonio~M Lopez, Vladlen Koltun, and Alexey Dosovitskiy.
\newblock On offline evaluation of vision-based driving models.
\newblock In {\em Proceedings of the European Conference on Computer Vision (ECCV)}, pages 236--251, 2018.

\bibitem{codevilla2018end}
Felipe Codevilla, Matthias M{\"u}ller, Antonio L{\'o}pez, Vladlen Koltun, and Alexey Dosovitskiy.
\newblock End-to-end driving via conditional imitation learning.
\newblock In {\em 2018 IEEE International Conference on Robotics and Automation (ICRA)}, pages 4693--4700. IEEE, 2018.

\bibitem{corso2019adaptive}
Anthony Corso, Peter Du, Katherine Driggs-Campbell, and Mykel~J Kochenderfer.
\newblock Adaptive stress testing with reward augmentation for autonomous vehicle validatio.
\newblock In {\em 2019 IEEE Intelligent Transportation Systems Conference (ITSC)}, pages 163--168. IEEE, 2019.

\bibitem{czarnecki2018operational}
Krzysztof Czarnecki.
\newblock Operational world model ontology for automated driving systems--part 1: Road structure.
\newblock {\em Waterloo Intelligent Systems Engineering Lab (WISE) Report, University of Waterloo}, 2018.

\bibitem{dargan2020survey}
Shaveta Dargan, Munish Kumar, Maruthi~Rohit Ayyagari, and Gulshan Kumar.
\newblock A survey of deep learning and its applications: a new paradigm to machine learning.
\newblock {\em Archives of Computational Methods in Engineering}, 27(4):1071--1092, 2020.

\bibitem{deb2002fast}
Kalyanmoy Deb, Amrit Pratap, Sameer Agarwal, and TAMT Meyarivan.
\newblock A fast and elitist multiobjective genetic algorithm: Nsga-ii.
\newblock {\em IEEE transactions on evolutionary computation}, 6(2):182--197, 2002.

\bibitem{dosovitskiy2017carla}
Alexey Dosovitskiy, German Ros, Felipe Codevilla, Antonio Lopez, and Vladlen Koltun.
\newblock Carla: An open urban driving simulator.
\newblock In {\em Conference on robot learning}, pages 1--16. PMLR, 2017.

\bibitem{dulac2015deep}
Gabriel Dulac-Arnold, Richard Evans, Hado van Hasselt, Peter Sunehag, Timothy Lillicrap, Jonathan Hunt, Timothy Mann, Theophane Weber, Thomas Degris, and Ben Coppin.
\newblock Deep reinforcement learning in large discrete action spaces.
\newblock {\em arXiv preprint arXiv:1512.07679}, 2015.

\bibitem{dulac2019challenges}
Gabriel Dulac-Arnold, Daniel Mankowitz, and Todd Hester.
\newblock Challenges of real-world reinforcement learning.
\newblock {\em arXiv preprint arXiv:1904.12901}, 2019.

\bibitem{fan2018baidu}
Haoyang Fan, Fan Zhu, Changchun Liu, Liangliang Zhang, Li~Zhuang, Dong Li, Weicheng Zhu, Jiangtao Hu, Hongye Li, and Qi~Kong.
\newblock Baidu apollo em motion planner.
\newblock {\em arXiv preprint arXiv:1807.08048}, 2018.

\bibitem{feng2020deep}
Di~Feng, Christian Haase-Sch{\"u}tz, Lars Rosenbaum, Heinz Hertlein, Claudius Glaeser, Fabian Timm, Werner Wiesbeck, and Klaus Dietmayer.
\newblock Deep multi-modal object detection and semantic segmentation for autonomous driving: Datasets, methods, and challenges.
\newblock {\em IEEE Transactions on Intelligent Transportation Systems}, 22(3):1341--1360, 2020.

\bibitem{ferlisi2021quantitative}
Settimio Ferlisi, Antonio Marchese, and Dario Peduto.
\newblock Quantitative analysis of the risk to road networks exposed to slow-moving landslides: a case study in the campania region (southern italy).
\newblock {\em Landslides}, 18:303--319, 2021.

\bibitem{furda2011enabling}
Andrei Furda and Ljubo Vlacic.
\newblock Enabling safe autonomous driving in real-world city traffic using multiple criteria decision making.
\newblock {\em IEEE Intelligent Transportation Systems Magazine}, 3(1):4--17, 2011.

\bibitem{greer2004software}
Des Greer and Guenther Ruhe.
\newblock Software release planning: an evolutionary and iterative approach.
\newblock {\em Information and software technology}, 46(4):243--253, 2004.

\bibitem{guo2018dlfuzz}
Jianmin Guo, Yu~Jiang, Yue Zhao, Quan Chen, and Jiaguang Sun.
\newblock Dlfuzz: Differential fuzzing testing of deep learning systems.
\newblock In {\em Proceedings of the 2018 26th ACM Joint Meeting on European Software Engineering Conference and Symposium on the Foundations of Software Engineering}, pages 739--743, 2018.

\bibitem{guo2019safe}
Junyao Guo, Unmesh Kurup, and Mohak Shah.
\newblock Is it safe to drive? an overview of factors, metrics, and datasets for driveability assessment in autonomous driving.
\newblock {\em IEEE Transactions on Intelligent Transportation Systems}, 21(8):3135--3151, 2019.

\bibitem{ha2017mfnet}
Qishen Ha, Kohei Watanabe, Takumi Karasawa, Yoshitaka Ushiku, and Tatsuya Harada.
\newblock Mfnet: Towards real-time semantic segmentation for autonomous vehicles with multi-spectral scenes.
\newblock In {\em 2017 IEEE/RSJ International Conference on Intelligent Robots and Systems (IROS)}, pages 5108--5115. IEEE, 2017.

\bibitem{haq2022many}
Fitash~Ul Haq, Donghwan Shin, and Lionel Briand.
\newblock Many-objective reinforcement learning for online testing of dnn-enabled systems.
\newblock {\em arXiv preprint arXiv:2210.15432}, 2022.

\bibitem{haq2020comparing}
Fitash~Ul Haq, Donghwan Shin, Shiva Nejati, and Lionel~C Briand.
\newblock Comparing offline and online testing of deep neural networks: An autonomous car case study.
\newblock In {\em 2020 IEEE 13th International Conference on Software Testing, Validation and Verification (ICST)}, pages 85--95. IEEE, 2020.

\bibitem{harel2020neuron}
Fabrice Harel-Canada, Lingxiao Wang, Muhammad~Ali Gulzar, Quanquan Gu, and Miryung Kim.
\newblock Is neuron coverage a meaningful measure for testing deep neural networks?
\newblock In {\em Proceedings of the 28th ACM Joint Meeting on European Software Engineering Conference and Symposium on the Foundations of Software Engineering}, pages 851--862, 2020.

\bibitem{harman2010search}
Mark Harman, Phil McMinn, Jerffeson Teixeira~de Souza, and Shin Yoo.
\newblock Search based software engineering: Techniques, taxonomy, tutorial.
\newblock In {\em Empirical software engineering and verification}, pages 1--59. Springer, 2010.

\bibitem{hochreiter1997long}
Sepp Hochreiter and J{\"u}rgen Schmidhuber.
\newblock Long short-term memory.
\newblock {\em Neural computation}, 9(8):1735--1780, 1997.

\bibitem{hu2019deepmutation++}
Qiang Hu, Lei Ma, Xiaofei Xie, Bing Yu, Yang Liu, and Jianjun Zhao.
\newblock Deepmutation++: A mutation testing framework for deep learning systems.
\newblock In {\em 2019 34th IEEE/ACM International Conference on Automated Software Engineering (ASE)}, pages 1158--1161. IEEE, 2019.

\bibitem{itkonen2017trade}
Teemu~H Itkonen, Jami Pekkanen, Otto Lappi, Iisakki Kosonen, Tapio Luttinen, and Heikki Summala.
\newblock Trade-off between jerk and time headway as an indicator of driving style.
\newblock {\em PloS one}, 12(10):e0185856, 2017.

\bibitem{jin2011surrogate}
Yaochu Jin.
\newblock Surrogate-assisted evolutionary computation: Recent advances and future challenges.
\newblock {\em Swarm and Evolutionary Computation}, 1(2):61--70, 2011.

\bibitem{kato2018autoware}
Shinpei Kato, Shota Tokunaga, Yuya Maruyama, Seiya Maeda, Manato Hirabayashi, Yuki Kitsukawa, Abraham Monrroy, Tomohito Ando, Yusuke Fujii, and Takuya Azumi.
\newblock Autoware on board: Enabling autonomous vehicles with embedded systems.
\newblock In {\em 2018 ACM/IEEE 9th International Conference on Cyber-Physical Systems (ICCPS)}, pages 287--296. IEEE, 2018.

\bibitem{keller2013will}
Christoph~G Keller and Dariu~M Gavrila.
\newblock Will the pedestrian cross? a study on pedestrian path prediction.
\newblock {\em IEEE Transactions on Intelligent Transportation Systems}, 15(2):494--506, 2013.

\bibitem{kim2019guiding}
Jinhan Kim, Robert Feldt, and Shin Yoo.
\newblock Guiding deep learning system testing using surprise adequacy.
\newblock In {\em 2019 IEEE/ACM 41st International Conference on Software Engineering (ICSE)}, pages 1039--1049. IEEE, 2019.

\bibitem{kiran2021deep}
B~Ravi Kiran, Ibrahim Sobh, Victor Talpaert, Patrick Mannion, Ahmad~A Al~Sallab, Senthil Yogamani, and Patrick P{\'e}rez.
\newblock Deep reinforcement learning for autonomous driving: A survey.
\newblock {\em IEEE Transactions on Intelligent Transportation Systems}, 2021.

\bibitem{kiran2020deep}
B~Ravi Kiran, Ibrahim Sobh, Victor Talpaert, Patrick Mannion, Ahmad A~Al Sallab, Senthil Yogamani, and Patrick P{\'e}rez.
\newblock Deep reinforcement learning for autonomous driving: A survey.
\newblock {\em arXiv preprint arXiv:2002.00444}, 2020.

\bibitem{RobustTest}
Barbara Kitchenham, Lech Madeyski, David Budgen, Jacky Keung, Pearl Brereton, Stuart Charters, Shirley Gibbs, and Amnart Pohthong.
\newblock Robust statistical methods for empirical software engineering.
\newblock {\em Empirical Software Engineering}, 22(2):579--630, 2017.

\bibitem{dm2022}
Mykel~J. Kochenderfer and Kyle H.~Wray Tim A.~Wheeler.
\newblock {\em Algorithms for Decision Making}.
\newblock MIT press, 2022.

\bibitem{kolchin2014template}
Maxim Kolchin and Fedor Kozlov.
\newblock A template-based information extraction from web sites with unstable markup.
\newblock In {\em Semantic Web Evaluation Challenge: SemWebEval 2014 at ESWC 2014, Anissaras, Crete, Greece, May 25-29, 2014, Revised Selected Papers}, pages 89--94. Springer, 2014.

\bibitem{konak2006multi}
Abdullah Konak, David~W Coit, and Alice~E Smith.
\newblock Multi-objective optimization using genetic algorithms: A tutorial.
\newblock {\em Reliability engineering \& system safety}, 91(9):992--1007, 2006.

\bibitem{koonce2008traffic}
Peter Koonce and Lee Rodegerdts.
\newblock Traffic signal timing manual.
\newblock Technical report, United States. Federal Highway Administration, 2008.

\bibitem{krizhevsky2012imagenet}
Alex Krizhevsky, Ilya Sutskever, and Geoffrey~E Hinton.
\newblock Imagenet classification with deep convolutional neural networks.
\newblock {\em Advances in neural information processing systems}, 25:1097--1105, 2012.

\bibitem{li2016vehicle}
Bo~Li, Tianlei Zhang, and Tian Xia.
\newblock Vehicle detection from 3d lidar using fully convolutional network.
\newblock {\em arXiv preprint arXiv:1608.07916}, 2016.

\bibitem{li2020av}
Guanpeng Li, Yiran Li, Saurabh Jha, Timothy Tsai, Michael Sullivan, Siva Kumar~Sastry Hari, Zbigniew Kalbarczyk, and Ravishankar Iyer.
\newblock Av-fuzzer: Finding safety violations in autonomous driving systems.
\newblock In {\em 2020 IEEE 31st International Symposium on Software Reliability Engineering (ISSRE)}, pages 25--36. IEEE, 2020.

\bibitem{liaw2018tune}
Richard Liaw, Eric Liang, Robert Nishihara, Philipp Moritz, Joseph~E Gonzalez, and Ion Stoica.
\newblock Tune: A research platform for distributed model selection and training.
\newblock {\em arXiv preprint arXiv:1807.05118}, 2018.

\bibitem{liu2014multiobjective}
Chunming Liu, Xin Xu, and Dewen Hu.
\newblock Multiobjective reinforcement learning: A comprehensive overview.
\newblock {\em IEEE Transactions on Systems, Man, and Cybernetics: Systems}, 45(3):385--398, 2014.

\bibitem{lu2022learning}
Chengjie Lu, Yize Shi, Huihui Zhang, Man Zhang, Tiexin Wang, Tao Yue, and Shaukat Ali.
\newblock Learning configurations of operating environment of autonomous vehicles to maximize their collisions.
\newblock {\em IEEE Transactions on Software Engineering}, 49(1):384--402, 2022.

\bibitem{10174023}
Chengjie Lu, Tao Yue, and Shaukat Ali.
\newblock Deepscenario: An open driving scenario dataset for autonomous driving system testing.
\newblock In {\em 2023 IEEE/ACM 20th International Conference on Mining Software Repositories (MSR)}, pages 52--56, 2023.

\bibitem{ma2018deepmutation}
Lei Ma, Fuyuan Zhang, Jiyuan Sun, Minhui Xue, Bo~Li, Felix Juefei-Xu, Chao Xie, Li~Li, Yang Liu, Jianjun Zhao, et~al.
\newblock Deepmutation: Mutation testing of deep learning systems.
\newblock In {\em 2018 IEEE 29th International Symposium on Software Reliability Engineering (ISSRE)}, pages 100--111. IEEE, 2018.

\bibitem{maeda2018road}
Hiroya Maeda, Yoshihide Sekimoto, Toshikazu Seto, Takehiro Kashiyama, and Hiroshi Omata.
\newblock Road damage detection using deep neural networks with images captured through a smartphone.
\newblock {\em arXiv preprint arXiv:1801.09454}, 2018.

\bibitem{mann1947test}
Henry~B Mann and Donald~R Whitney.
\newblock On a test of whether one of two random variables is stochastically larger than the other.
\newblock {\em The annals of mathematical statistics}, pages 50--60, 1947.

\bibitem{mnih2015human}
Volodymyr Mnih, Koray Kavukcuoglu, David Silver, Andrei~A Rusu, Joel Veness, Marc~G Bellemare, Alex Graves, Martin Riedmiller, Andreas~K Fidjeland, Georg Ostrovski, et~al.
\newblock Human-level control through deep reinforcement learning.
\newblock {\em nature}, 518(7540):529--533, 2015.

\bibitem{mukadam2017tactical}
Mustafa Mukadam, Akansel Cosgun, Alireza Nakhaei, and Kikuo Fujimura.
\newblock Tactical decision making for lane changing with deep reinforcement learning.
\newblock 2017.

\bibitem{openscenario}
American~Society of~Addiction~Medicine.
\newblock Asam openscenario, 2022.

\bibitem{ocl}
OMG.
\newblock Object constraint language v2.0.
\newblock Object Management Group Adopted Specification (formal/06-05-01), 2006.

\bibitem{paszke2019pytorch}
Adam Paszke, Sam Gross, Francisco Massa, Adam Lerer, James Bradbury, Gregory Chanan, Trevor Killeen, Zeming Lin, Natalia Gimelshein, Luca Antiga, et~al.
\newblock Pytorch: An imperative style, high-performance deep learning library.
\newblock {\em Advances in neural information processing systems}, 32:8026--8037, 2019.

\bibitem{pei2017deepxplore}
Kexin Pei, Yinzhi Cao, Junfeng Yang, and Suman Jana.
\newblock Deepxplore: Automated whitebox testing of deep learning systems.
\newblock In {\em proceedings of the 26th Symposium on Operating Systems Principles}, pages 1--18, 2017.

\bibitem{peng2020first}
Zi~Peng, Jinqiu Yang, Tse-Hsun Chen, and Lei Ma.
\newblock A first look at the integration of machine learning models in complex autonomous driving systems: a case study on apollo.
\newblock In {\em Proceedings of the 28th ACM Joint Meeting on European Software Engineering Conference and Symposium on the Foundations of Software Engineering}, pages 1240--1250, 2020.

\bibitem{pfeuffer2018optimal}
Andreas Pfeuffer and Klaus Dietmayer.
\newblock Optimal sensor data fusion architecture for object detection in adverse weather conditions.
\newblock In {\em 2018 21st International Conference on Information Fusion (FUSION)}, pages 1--8. IEEE, 2018.

\bibitem{piewak2018improved}
Florian Piewak, Peter Pinggera, Markus Enzweiler, David Pfeiffer, and Marius Z{\"o}llner.
\newblock Improved semantic stixels via multimodal sensor fusion.
\newblock In {\em German Conference on Pattern Recognition}, pages 447--458. Springer, 2018.

\bibitem{queiroz2019geoscenario}
Rodrigo Queiroz, Thorsten Berger, and Krzysztof Czarnecki.
\newblock Geoscenario: An open dsl for autonomous driving scenario representation.
\newblock In {\em 2019 IEEE Intelligent Vehicles Symposium (IV)}, pages 287--294. IEEE, 2019.

\bibitem{ro2020new}
Jin~Woo Ro, Partha~S Roop, and Avinash Malik.
\newblock A new safety distance calculation for rear-end collision avoidance.
\newblock {\em IEEE Transactions on Intelligent Transportation Systems}, 22(3):1742--1747, 2020.

\bibitem{rodriguez2016rest}
Carlos Rodr{\'\i}guez, Marcos Baez, Florian Daniel, Fabio Casati, Juan~Carlos Trabucco, Luigi Canali, and Gianraffaele Percannella.
\newblock Rest apis: a large-scale analysis of compliance with principles and best practices.
\newblock In {\em International conference on web engineering}, pages 21--39. Springer, 2016.

\bibitem{rong2020lgsvl}
Guodong Rong, Byung~Hyun Shin, Hadi Tabatabaee, Qiang Lu, Steve Lemke, M{\=a}rti{\c{n}}{\v{s}} Mo{\v{z}}eiko, Eric Boise, Geehoon Uhm, Mark Gerow, Shalin Mehta, et~al.
\newblock Lgsvl simulator: A high fidelity simulator for autonomous driving.
\newblock In {\em 2020 IEEE 23rd International Conference on Intelligent Transportation Systems (ITSC)}, pages 1--6. IEEE, 2020.

\bibitem{ruiming2018end}
Zhang Ruiming, Liu Chengju, and Chen Qijun.
\newblock End-to-end control of kart agent with deep reinforcement learning.
\newblock In {\em 2018 IEEE International Conference on Robotics and Biomimetics (ROBIO)}, pages 1688--1693. IEEE, 2018.

\bibitem{sallab2017deep}
Ahmad~EL Sallab, Mohammed Abdou, Etienne Perot, and Senthil Yogamani.
\newblock Deep reinforcement learning framework for autonomous driving.
\newblock {\em Electronic Imaging}, 2017(19):70--76, 2017.

\bibitem{schneider2017multimodal}
Lukas Schneider, Manuel Jasch, Bj{\"o}rn Fr{\"o}hlich, Thomas Weber, Uwe Franke, Marc Pollefeys, and Matthias R{\"a}tsch.
\newblock Multimodal neural networks: Rgb-d for semantic segmentation and object detection.
\newblock In {\em Scandinavian conference on image analysis}, pages 98--109. Springer, 2017.

\bibitem{shaout2011advanced}
Adnan Shaout, Dominic Colella, and Selim Awad.
\newblock Advanced driver assistance systems-past, present and future.
\newblock In {\em 2011 Seventh International Computer Engineering Conference (ICENCO'2011)}, pages 72--82. IEEE, 2011.

\bibitem{sillmann2017understanding}
Jana Sillmann, Thordis Thorarinsdottir, Noel Keenlyside, Nathalie Schaller, Lisa~V Alexander, Gabriele Hegerl, Sonia~I Seneviratne, Robert Vautard, Xuebin Zhang, and Francis~W Zwiers.
\newblock Understanding, modeling and predicting weather and climate extremes: Challenges and opportunities.
\newblock {\em Weather and climate extremes}, 18:65--74, 2017.

\bibitem{simonyan2014very}
Karen Simonyan and Andrew Zisserman.
\newblock Very deep convolutional networks for large-scale image recognition.
\newblock {\em arXiv preprint arXiv:1409.1556}, 2014.

\bibitem{UniversityofTorontoTTC}
Jonathan~Kelly Steven-Waslander.
\newblock Motion planning for self-driving cars.
\newblock {\em Coursera\&University of Toronto, https://www.coursera.org/lecture/motion-planning-self-driving-cars/lesson-3-time-to-collision-pS9zl}, 2021.

\bibitem{stocco2022mind}
Andrea Stocco, Brian Pulfer, and Paolo Tonella.
\newblock Mind the gap! a study on the transferability of virtual vs physical-world testing of autonomous driving systems.
\newblock {\em IEEE Transactions on Software Engineering}, 2022.

\bibitem{sutton2018reinforcement}
Richard~S Sutton and Andrew~G Barto.
\newblock {\em Reinforcement learning: An introduction}.
\newblock MIT press, 2018.

\bibitem{apollo}
Baidu~Apollo team.
\newblock Apollo: Open source autonomous driving, 2017.

\bibitem{tian2018deeptest}
Yuchi Tian, Kexin Pei, Suman Jana, and Baishakhi Ray.
\newblock Deeptest: Automated testing of deep-neural-network-driven autonomous cars.
\newblock In {\em Proceedings of the 40th international conference on software engineering}, pages 303--314, 2018.

\bibitem{ulbrich2015defining}
Simon Ulbrich, Till Menzel, Andreas Reschka, Fabian Schuldt, and Markus Maurer.
\newblock Defining and substantiating the terms scene, situation, and scenario for automated driving.
\newblock In {\em 2015 IEEE 18th International Conference on Intelligent Transportation Systems}, pages 982--988. IEEE, 2015.

\bibitem{vamplew2011empirical}
Peter Vamplew, Richard Dazeley, Adam Berry, Rustam Issabekov, and Evan Dekker.
\newblock Empirical evaluation methods for multiobjective reinforcement learning algorithms.
\newblock 2011.

\bibitem{vardhan2017hd}
Harsha Vardhan.
\newblock Hd maps: New age maps powering autonomous vehicles.
\newblock {\em Geospatial world}, 22, 2017.

\bibitem{vogel2003comparison}
Katja Vogel.
\newblock A comparison of headway and time to collision as safety indicators.
\newblock {\em Accident analysis \& prevention}, 35(3):427--433, 2003.

\bibitem{watkins1992q}
Christopher~JCH Watkins and Peter Dayan.
\newblock Q-learning.
\newblock {\em Machine learning}, 8(3-4):279--292, 1992.

\bibitem{Deephunter}
Xiaofei Xie, Lei Ma, Felix Juefei-Xu, Minhui Xue, Hongxu Chen, Yang Liu, Jianjun Zhao, Bo~Li, Jianxiong Yin, and Simon See.
\newblock Deephunter: A coverage-guided fuzz testing framework for deep neural networks.
\newblock In {\em Proceedings of the 28th ACM SIGSOFT International Symposium on Software Testing and Analysis}, pages 146--157, 2019.

\bibitem{zhang2018Deeproad}
Mengshi Zhang, Yuqun Zhang, Lingming Zhang, Cong Liu, and Sarfraz Khurshid.
\newblock Deeproad: Gan-based metamorphic testing and input validation framework for autonomous driving systems.
\newblock In {\em 2018 33rd IEEE/ACM International Conference on Automated Software Engineering (ASE)}, pages 132--142. IEEE, 2018.

\bibitem{zhang2018improved}
Zijun Zhang.
\newblock Improved adam optimizer for deep neural networks.
\newblock In {\em 2018 IEEE/ACM 26th International Symposium on Quality of Service (IWQoS)}, pages 1--2. IEEE, 2018.

\bibitem{zhou2020deepbillboard}
Husheng Zhou, Wei Li, Zelun Kong, Junfeng Guo, Yuqun Zhang, Bei Yu, Lingming Zhang, and Cong Liu.
\newblock Deepbillboard: Systematic physical-world testing of autonomous driving systems.
\newblock In {\em Proceedings of the ACM/IEEE 42nd International Conference on Software Engineering}, pages 347--358, 2020.

\bibitem{PyTorchTutorial}
Morvan Zhou.
\newblock pytorch tutorials.
\newblock \url{https://github.com/MorvanZhou/PyTorch-Tutorial}, 2021.

\bibitem{zhou2019automated}
Wei Zhou, Julie~Stephany Berrio, Stewart Worrall, and Eduardo Nebot.
\newblock Automated evaluation of semantic segmentation robustness for autonomous driving.
\newblock {\em IEEE Transactions on Intelligent Transportation Systems}, 21(5):1951--1963, 2019.

\bibitem{zhu2020safe}
Meixin Zhu, Yinhai Wang, Ziyuan Pu, Jingyun Hu, Xuesong Wang, and Ruimin Ke.
\newblock Safe, efficient, and comfortable velocity control based on reinforcement learning for autonomous driving.
\newblock {\em Transportation Research Part C: Emerging Technologies}, 117:102662, 2020.

\end{thebibliography}
